\documentclass[]{elsarticle}
\usepackage[T1]{fontenc}
\usepackage{amssymb,amsmath}
\usepackage{graphicx}
\usepackage{subcaption}
\usepackage{booktabs}
\usepackage{siunitx}
\usepackage{placeins}
\usepackage{lscape}
\usepackage[colorlinks=true,citecolor=blue,urlcolor=blue]{hyperref}
\usepackage{lineno}

\begin{document}

\begin{frontmatter}
	
\title{Possible quasi-periodic optical oscillations of ZTF blazars}
	
\author[inst1]{Na Wang}
\author[inst2]{Guowei Ren}
\author[inst3]{Shun Zhang}
\author[inst3]{Tingfeng Yi}
\author[inst1]{Tong Liu}
\ead{tongliu@xmu.edu.cn}
\author[inst1]{Mouyuan Sun}
\ead{msun88@xmu.edu.cn}
	
\affiliation[inst1]{organization={Department of Astronomy, Xiamen University},
	city={Xiamen}, state={Fujian}, postcode={361005}, country={China}}
\affiliation[inst2]{organization={Department of Astronomy, University of Science and Technology of China}, 
    city={Hefei}, state={Anhui}, postcode={230026}, country={China}}
\affiliation[inst3]{organization={Department of Physics, Yunnan Normal University},
	city={Kunming}, state={Yunnan}, postcode={650500}, country={China}}

\begin{abstract}
Based on the Zwicky Transient Facility (ZTF), we selected 10 blazars as our sample sources. Among these, we found four blazars (J 0923.5+4125, J 1221.3+3010, J 1503.5+4759, and J 1652.7+4024) showing possible indications of quasi periodic oscillations (QPOs) modulation. We conducted a detailed analysis of their optical light curves (g- and r-bands) over the past five years using the root mean square (RMS)-Flux relation, flux distribution, and QPO detection methods to investigate their variability characteristics. A linear RMS-Flux relation is present in both bands, and their flux distributions follow a log-normal form. This suggests that optical variability may arise from multiplicative, nonlinear processes across different timescales and flux states. Further QPO analysis using the weighted wavelet Z-transform (WWZ), Lomb-Scargle periodogram (LSP), and autoregressive integrated moving average (ARIMA) methods identified candidate periodic signals in four blazars. J 0923.5+4125 (period $\sim205$\,days) and J~1221.3+3010 ($\sim630$\,days) show local significances of $\sim3\sigma$, whereas J~1503.5+4759 ($\sim38.5$\,days) and J~1652.7+4024 ($\sim48$\,days) reach $\sim4\sigma$. After accounting for the look-elsewhere effect, the global significances for J 1503.5+4759 in the g- and r-bands are $\sim$ 2.7$\sigma$, while for J 1652.7+4024 they are approximately $\sim$ 2.5$\sigma$ in both bands. These two blazars warrant further monitoring and investigation.
\end{abstract}
	
	
\end{frontmatter}

\section{Introduction}\label{sec:introduction}

Active Galactic Nuclei (AGNs) are among the brightest sources in the universe and are powered by accreting supermassive black holes (SMBHs) at their centers. AGNs with relativistic jets oriented within $\sim 15^\circ$ of the line of sight are classified as blazars \citep{1995PASP..107..803U}. The spectral energy distribution of blazars typically exhibits a double-peaked structure. The low-energy peak, generally occurring between radio and X-ray frequencies, is attributed to synchrotron radiation from relativistic electrons, with optical emission mainly due to this process, supplemented by contributions from the accretion disk, host galaxy, and broad-line region (BLR). The high-energy peak, typically observed between X-ray and gamma-ray frequencies, is thought to result from inverse Compton scattering of low-energy photons. Two main models have been proposed to explain the source of these seed photons: the Synchrotron Self-Compton model, where the same electron population responsible for synchrotron emission also upscatters these photons to higher energies, and the External Compton model, where softer seed photons are provided by various AGN regions, such as the accretion disk, BLR, and dusty torus \citep{1993ApJ...416..458D}.

Blazars are further classified into two subclasses: Flat Spectrum Radio Quasars (FSRQs) and BL Lac objects (BL Lacs) \citep{1980ARA&A..18..321A}. FSRQs exhibit broad optical emission lines (EW $>5\,\text{\AA}$)and are typically viewed at angles of $4^\circ$, whereas BL Lacs have either no emission lines or only weak ones (EW $<5\,\text{\AA}$) with an average viewing angle of around $9^\circ$ \citep{2009ApJ...696L..22L}. Blazars can also be categorized based on the peak energy of their synchrotron radiation into Low Energy Synchrotron Peaked BL Lacs, Intermediate Synchrotron Peaked BL Lacs, and High Energy Synchrotron Peaked BL Lacs \citep{1994astro-ph..9412073}. Compared to BL Lacs, FSRQs have a more luminous BLR and a more efficient accretion process, with typically higher black hole (BH) masses and jet powers \citep{2014MNRAS.441.3375X}.  The two subclasses also differ in their redshift distribution, synchrotron peak frequency distribution, radio power, and cosmic evolution \citep{2009A&A...508..107G}.

One of the most prominent observational characteristics of blazars is their dramatic, rapid variability across all wavelengths, with timescales ranging from minutes to years \citep{1997ARA&A..35..445U, 2000ApJS..127...11G, 2014MNRAS.441.1899F, 2024MNRAS.527.6970K}. Studying this variability across different wavelengths and timescales provides insights into the internal structure of blazars, the location of emission regions, and the physical mechanisms driving the emission. Variability is generally classified into intraday variability (IDV), short-term variability (STV), and long-term variability (LTV). IDV refers to significant changes in flux or magnitude occurring within minutes to a day, with timescales typically ranging from minutes to hours \citep{1995ARA&A..33..163W}. STV involves changes on timescales of days to months and can help test the validity of AGN physical models and determine relevant parameters \citep{2004A&A...422..505G}. LTV occurs over timescales of months to years. Year-scale quasi-periodic oscillations (QPOs) have been interpreted using several models, including the supermassive binary black holes (SMBBHs) scenario \citep{2021..506..3791}, precessing jet model \citep{2014MNRAS.437.2744T}, helical jet model \citep{2008Natur..452..966M}, and Lense-Thirring precession model of the accretion disk \citep{1997ApJ...492..L59S}.

Although Very Long Baseline Interferometry (VLBI) is capable of resolving the relativistic jets launched by SMBHs, the central engine itself remains unresolved due to the extreme physical conditions in its vicinity. The unresolved region encompasses critical physical components, such as the accretion flow, the disk–jet connection, and the role of magnetic fields in initiating and collimating the jets. In this context, variability analysis serves as a powerful tool to probe the central engine's properties and the emission mechanisms of blazars. Studying variability helps constrain the size, magnetic field, and other characteristics of the regions around the dense cores of blazars.

In this paper, we utilize five years of optical observation data (g- and r-bands) from Zwicky Transient Facility (ZTF) to conduct a detailed time-series analysis of four blazars. Section \ref{Data Source and Sample Composition} discusses the selection of sample sources and the retrieval of archival optical data. Section \ref{Analysis and Results} exhibits various variability analysis methods, including the fractional variability, RMS–Flux relation, flux distribution, and QPO. In Section \ref{Discussion}, we discuss the results and their potential implications for the optical emission properties of the sample sources. Finally, brief conclusions are given in Section \ref{Conclusion}.

\section{Data}\label{Data Source and Sample Composition}

Our initial sample is derived from the Fermi 4FGL-DR3, which includes 5788 sources \citep{2023ApJS..265..31}. Subsequently, we obtained 1052 blazars after performing a cross-match with the 19th ZTF optical band data\footnote{\url{https://irsa.ipac.caltech.edu/Missions/ztf.html}}. ZTF is an optical time-domain survey that utilizes the Palomar 48-inch Schmidt telescope, equipped with a wide-field camera providing a field of view spanning 47 square degrees and an 8-second readout time \citep{2019PASP..131a8002B}. In the released data, the average observation cadence is 3 days, which greatly facilitates the study of blazar variability. To retrieve the light curve of a given source\footnote{\url{https://alerce.science/services/ztf-explorer/}}, its corresponding ZTF observation ID must first be acquired. The system initially conducts a query using the source coordinates and selects the ID closest to the original coordinates, with an angular separation of $2''$. In addition, low signal-to-noise ratios (S/N) and insufficient data points can significantly weaken significance and increase the risk of both false positives and missed detections of QPOs. To reduce uncertainties and ensure adequate sampling for reliable periodicity detection, we retained only sources with the S/N $>15$ and more than 100 valid data points over the entire observation period. Based on these criteria, a total of 10 blazars with sufficiently high-quality light curves were selected, meeting the basic requirements for QPO searches. Ultimately, after conducting periodicity analysis on these 10 blazars, 4 blazars were found to display QPO-like behaviour at tentative significance (see Section \ref{Analysis and Results} for details). We then extracted the g- and r-band light curves of these four sources (J 0923.5+4125, J 1221.3+3010, J 1503.5+4759, and J 1652.7+4024) for further variability analysis.

\section{Analysis and Results}\label{Analysis and Results}

\subsection{variability}

Fractional variability ($F_{\rm var}$) is a parameter used to quantify the variability in light curves, accounting for both intrinsic variance and measurement uncertainties. It provides a normalized measure of the variability, often utilized in studying of blazars and other astronomical objects \citep{2019Galaxies..7(2)..62}. The fractional variability is obtained as the square root of the normalized excess variance, i.e.,
\begin{equation}
	F_{\rm var} = \sqrt{\frac{S^{2}-\langle\sigma _{err}^{2}\rangle}{\langle x\rangle^{2} } },
\end{equation}
where \( x \) is the flux, $S$ is variance, and $\langle\sigma _{err}^{2}\rangle$ is square error. The uncertainty associated with the fractional variability is expressed by
\begin{equation}
	\Delta F_{\rm var}= \sqrt{F_{\rm var}^{2}+err(\sigma_{NXS}^{2})}-F_{\rm var},
\end{equation}
where the error in the normalized excess variance is computed as
\begin{equation}
	err(\sigma_{NXS}^{2} ) = \sqrt{\sqrt{\frac{2}{N}}\frac{\langle\sigma _{err}^{2}\rangle}{\langle x\rangle^{2}} +\frac{\langle\sigma_{err}^{2}\rangle}{N} \frac{4F_{\rm var}^{2} }{\langle x\rangle^2} },
\end{equation}
where \( N \) is the number of flux values. The fractional variability is an essential tool in astrophysical studies, enabling the differentiation between intrinsic variability and measurement-induced variations in observed light curves. The resulting $F_{\rm var}$ for the sample sources are listed in Table \ref{tab1}; the four blazars show evident flux variations.

\setlength{\tabcolsep}{3pt}
\begin{table*}
	\caption{Source sample of ZTF Blazars.}
	\noindent\hspace*{-8.5em}
	\label{tab1}
	\begin{tabular}{ccccccccc}
		\toprule
		4FGL Name & Source Type & R.A & Decl. & Spectral Index & ${F_{var}}$(g-band) (\%) & ${F_{var}}$(r-band) (\%) & Redshift & Color variability\\
		\midrule
		J 0923.5+4125 & FSRQ & 140.895 & 41.428 & 2.355 & 63.76 $\pm$ 0.01 & 55.47 $\pm$ 0.08 & 0.028 & BWB \\
		J 1221.3+3010 & BLL & 185.345 & 30.168 & 1.707 & 28.21 $\pm$ 0.05 & 24.02 $\pm$ 0.05 & 0.184 & RWB\\
		J 1503.5+4759 & BLL & 225.896 & 47.996 & 2.221 & 20.26 $\pm$ 0.19 & 15.97 $\pm$ 0.12 & 0.345 & BWB\\
		J 1652.7+4024 & BLL & 253.198 & 40.405 & 1.927 & 11.89 $\pm$ 0.31 & 10.57 $\pm$ 0.26 & 1.803 & BWB\\
		\bottomrule
	\end{tabular}
\end{table*}

\begin{figure*}
	\centering
	\begin{subfigure}[b]{0.24\textwidth}
		\includegraphics[width=\textwidth]{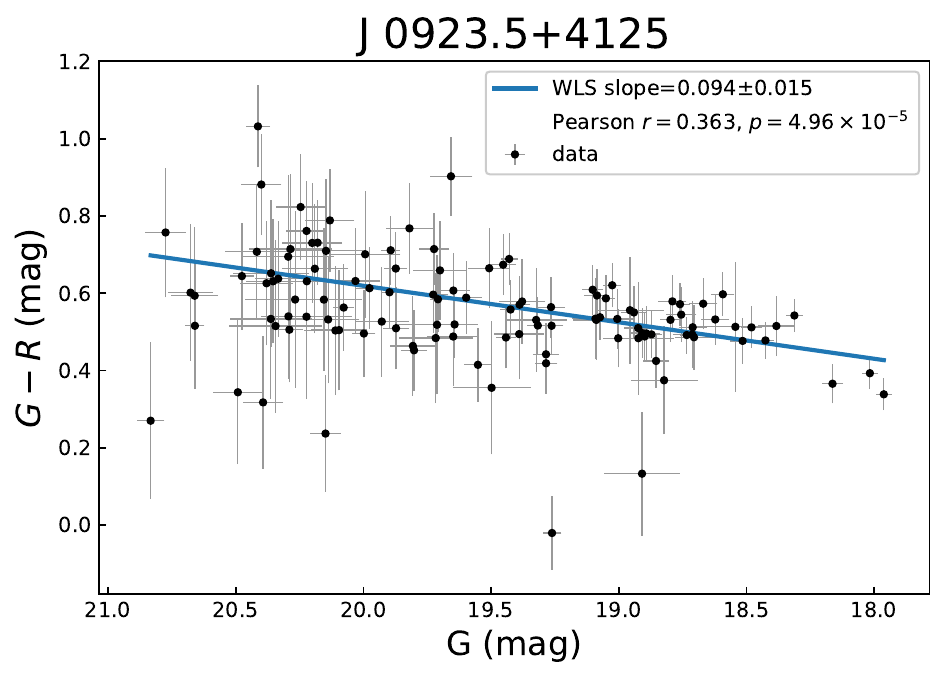}
	\end{subfigure}
	\hfill
	\begin{subfigure}[b]{0.24\textwidth}
		\includegraphics[width=\textwidth]{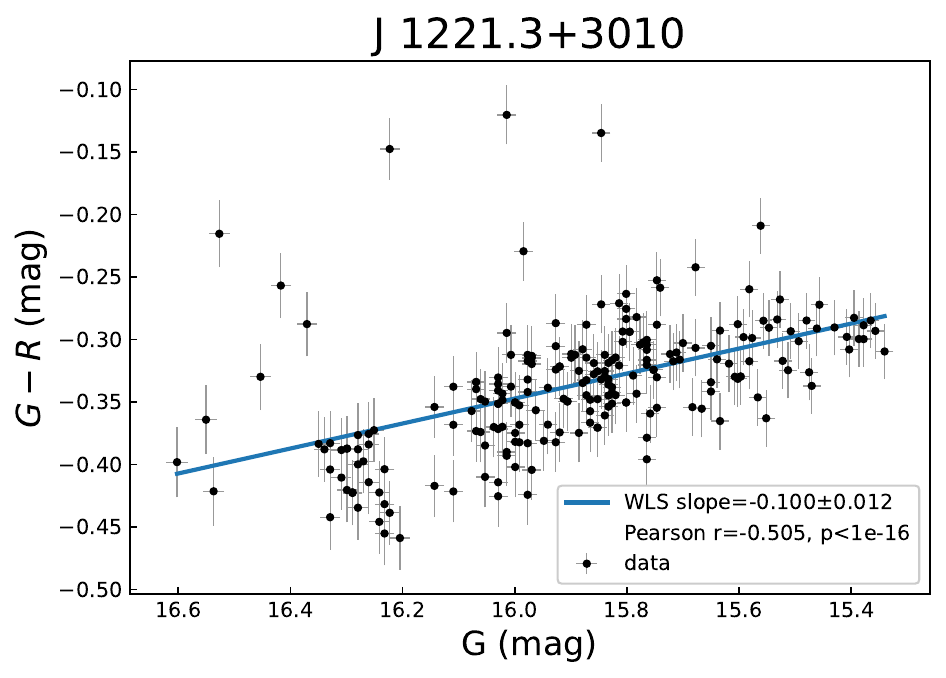}
	\end{subfigure}
	\hfill
	\begin{subfigure}[b]{0.24\textwidth}
		\includegraphics[width=\textwidth]{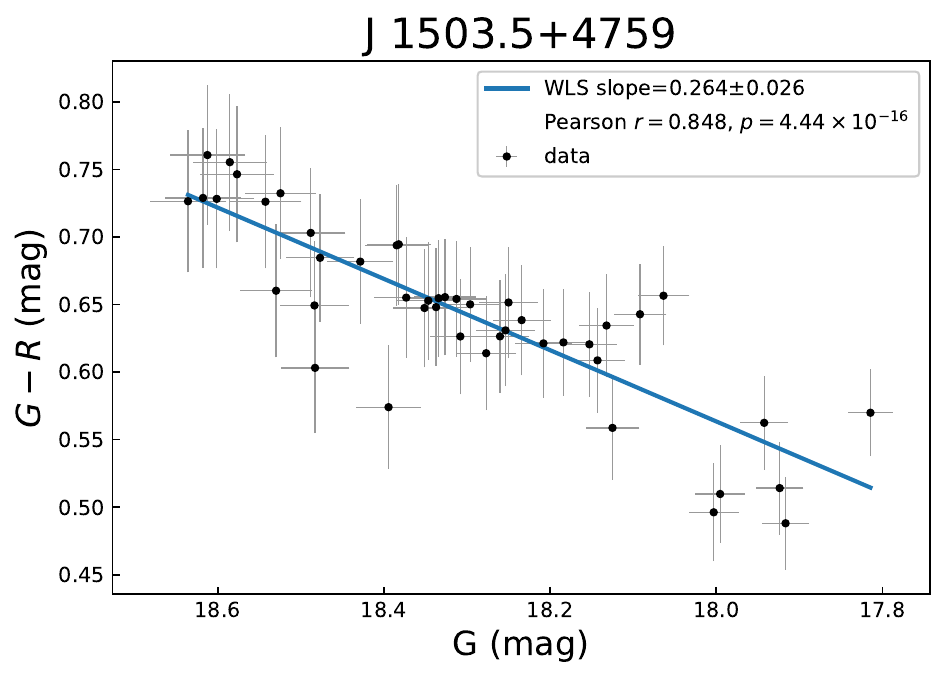}
	\end{subfigure}
	\hfill
	\begin{subfigure}[b]{0.24\textwidth}
		\includegraphics[width=\textwidth]{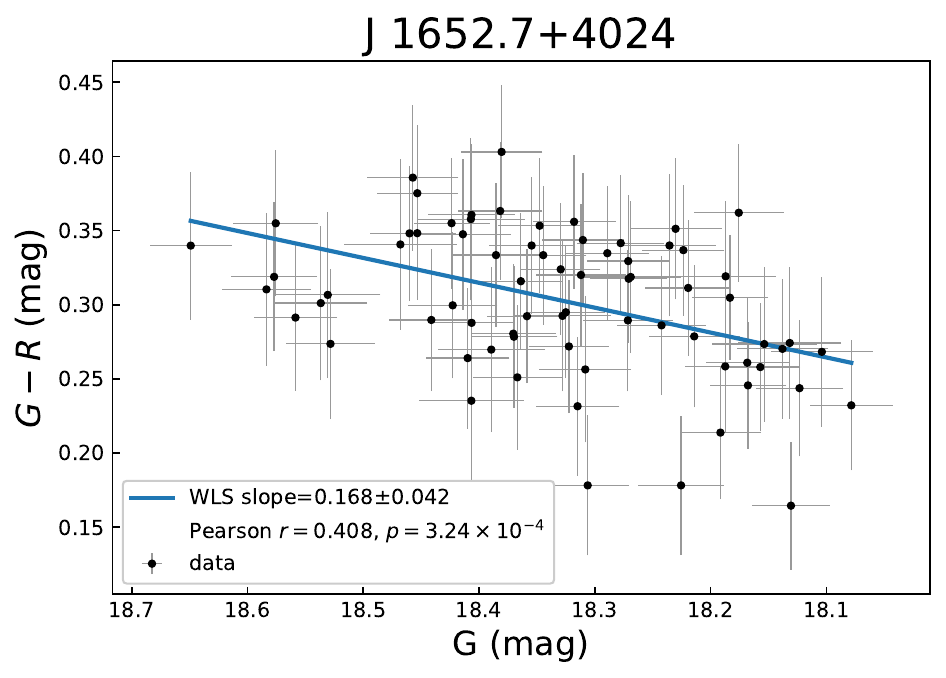}
	\end{subfigure}
	\caption{G – R vs. G-band magnitudes for the four blazars, he blue solid line represents the result of linear regression analysis.}
	\label{fig1}
\end{figure*}

\begin{figure*}
	\centering
	\begin{subfigure}[b]{0.24\textwidth}
		\includegraphics[width=\textwidth]{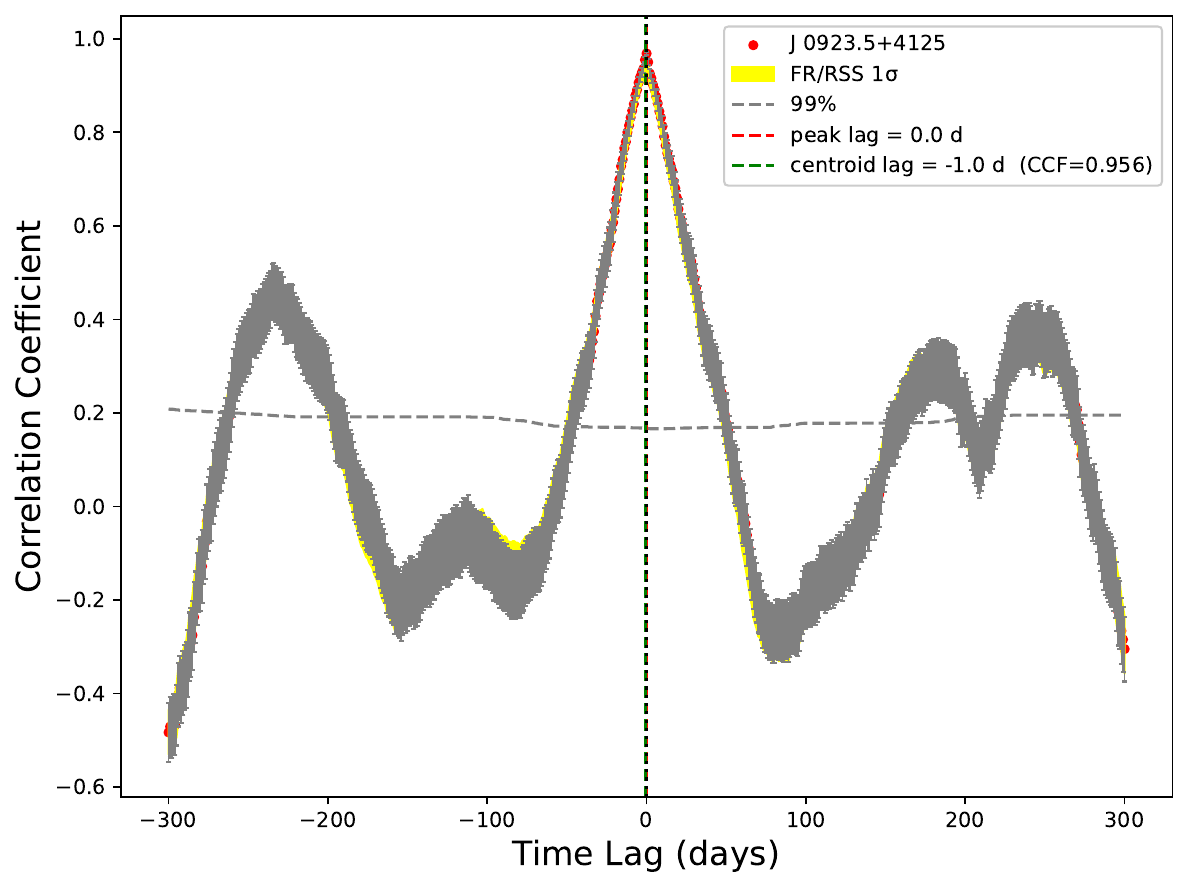}
	\end{subfigure}
	\hfill
	\begin{subfigure}[b]{0.24\textwidth}
		\includegraphics[width=\textwidth]{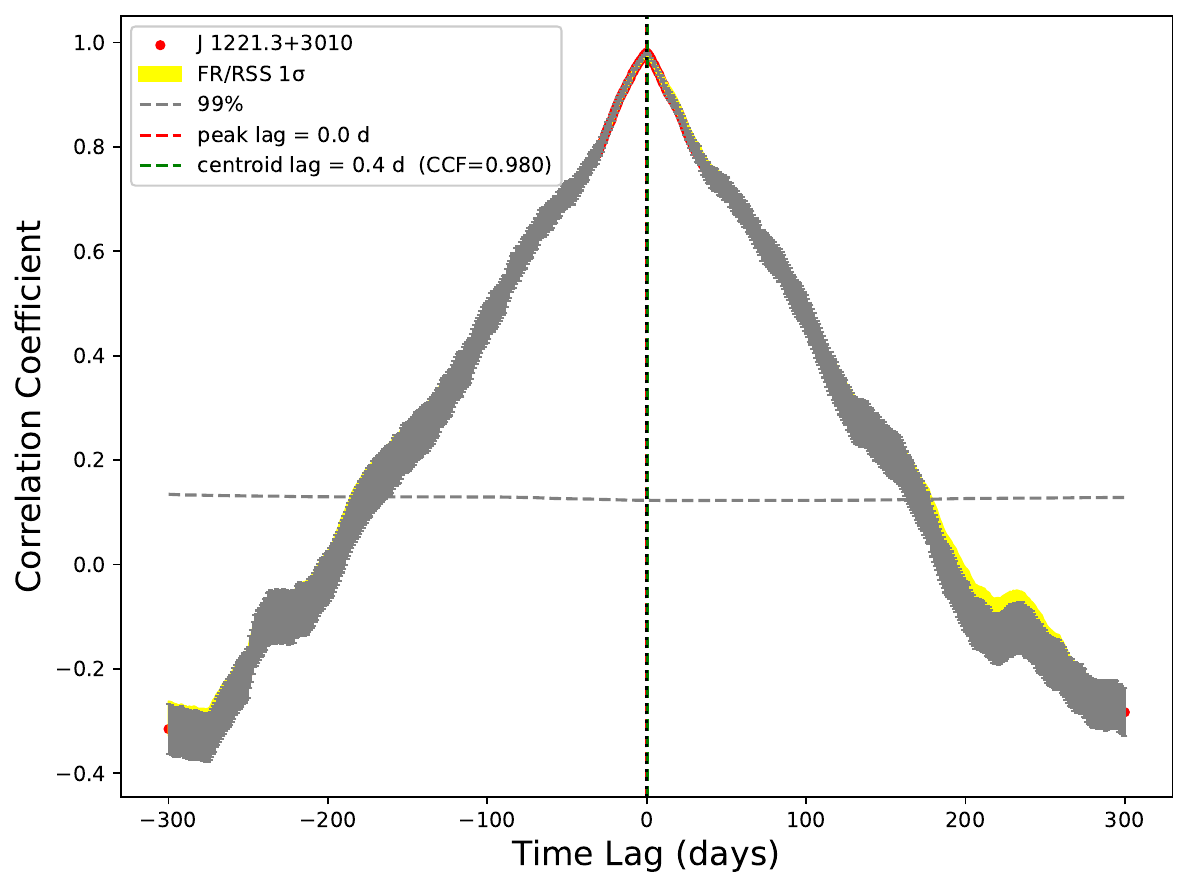}
	\end{subfigure}
	\hfill
	\begin{subfigure}[b]{0.24\textwidth}
		\includegraphics[width=\textwidth]{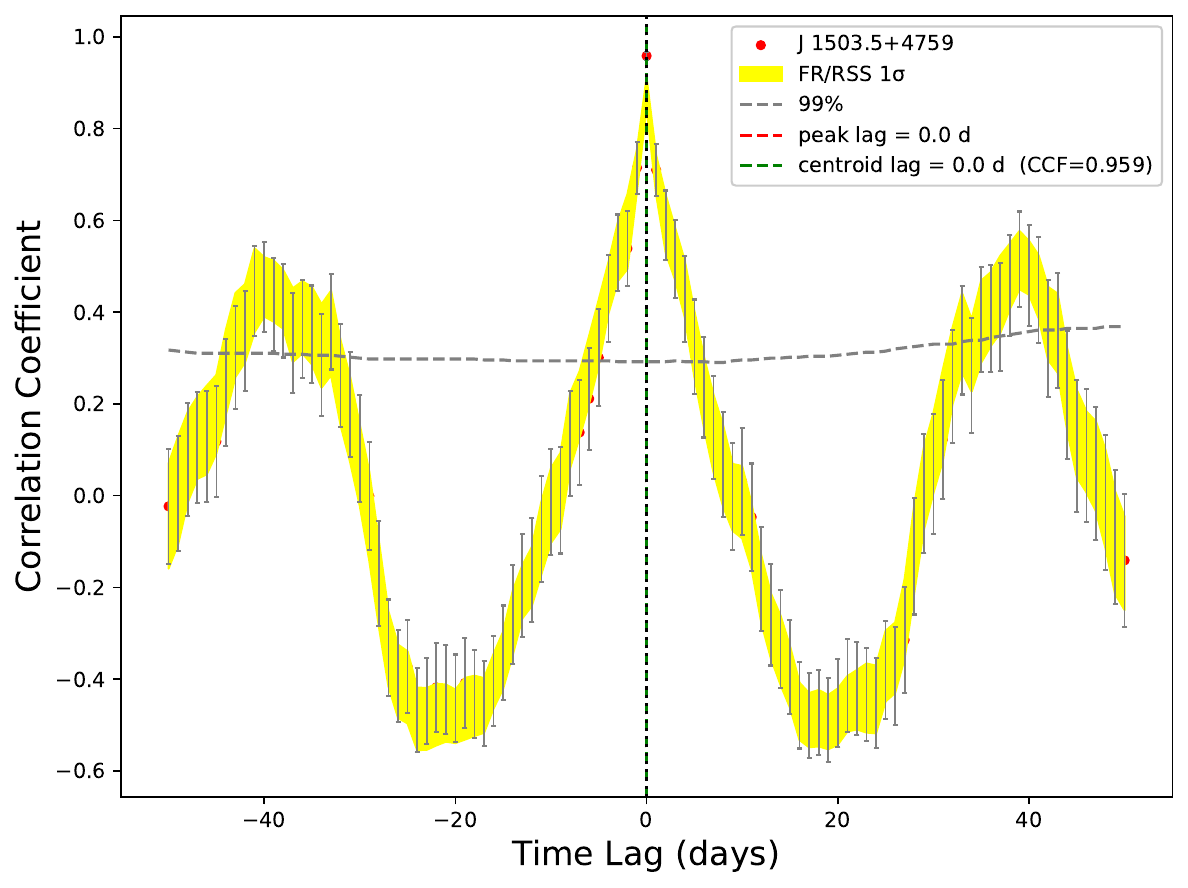}
	\end{subfigure}
	\hfill
	\begin{subfigure}[b]{0.24\textwidth}
		\includegraphics[width=\textwidth]{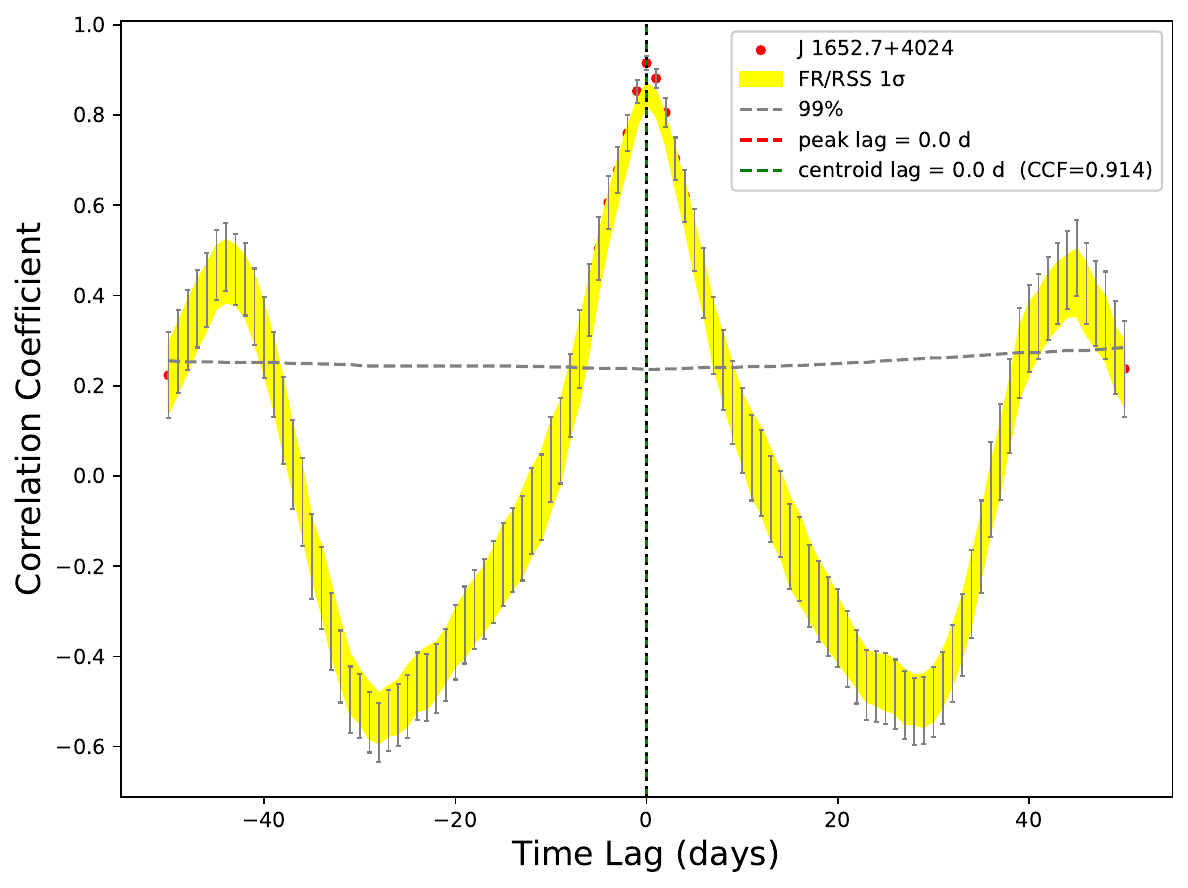}
	\end{subfigure}
	\caption{Results of the CCF analysis. The yellow shaded region represents the 1$\sigma$ uncertainty, and the gray dashed line marks the 99\% theoretical significance.}
	\label{fig2}
\end{figure*}

Figure \ref{fig1} shows the color–magnitude relation ($G-R$ vs. $G$) for the four blazars. For each source, we report the Pearson correlation coefficient $r$, the slope of the linear fit, and the null-hypothesis probability $p$. Based on the slope and $p$-value, only J 1221.3+3010 exhibits a redder-when-brighter (RWB), while the other three sources display a bluer-when-brighter (BWB). We  also performed the cross-correlation function (CCF), with the results presented in Figure \ref{fig2}. The uncertainties were estimated through 1000 Monte Carlo simulations, adopting the 16–84\% percentile range as the $1\sigma$ interval, which is shown as the yellow shaded region. The gray dashed line indicates the 99\% significance threshold for the null hypothesis of zero correlation. The results demonstrate that the variations in the g- and r-bands are synchronous, with a lag consistent with zero, suggesting that the two bands likely originate from the same emission region.

\subsection{RMS-Flux relation}

The RMS-Flux relation is a critical tool for understanding the variability properties of variable astrophysical sources. The RMS is defined as
\begin{equation}
	\text{RMS} = \sqrt{\frac{1}{N} \sum_{i=1}^{N} (x_i-\bar{x})^2}.
\end{equation}
This relationship examines the correlation between the RMS amplitude of variability on shorter timescales and the mean flux level on longer timescales. It has been reported for accreting compact objects \citep{2005MNRAS..359..345}, and is also applicable to AGNs and blazars.

\begin{figure*}
	\centering
	\begin{subfigure}[b]{0.24\textwidth}
		\includegraphics[width=\textwidth]{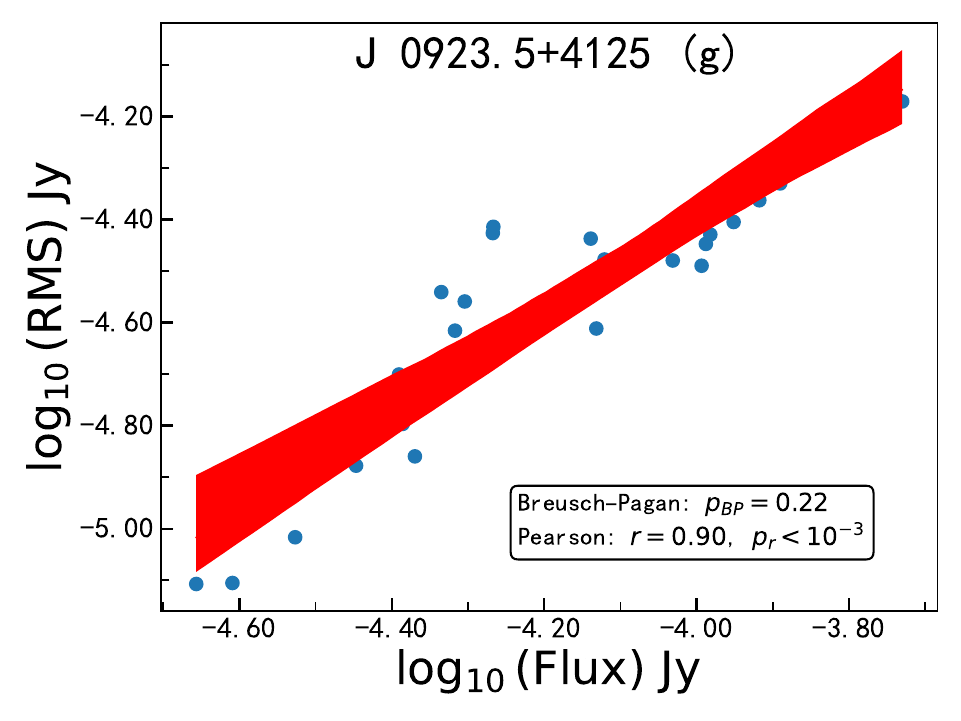}
	\end{subfigure}
	\hfill
	\begin{subfigure}[b]{0.24\textwidth}
		\includegraphics[width=\textwidth]{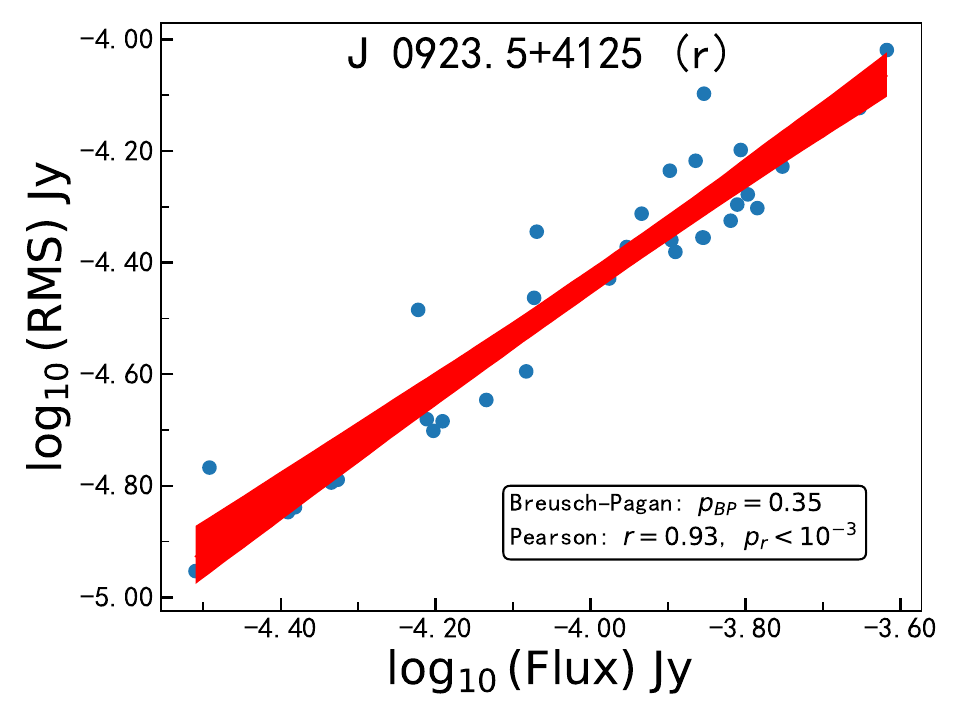}
	\end{subfigure}
	\hfill
	\begin{subfigure}[b]{0.24\textwidth}
		\includegraphics[width=\textwidth]{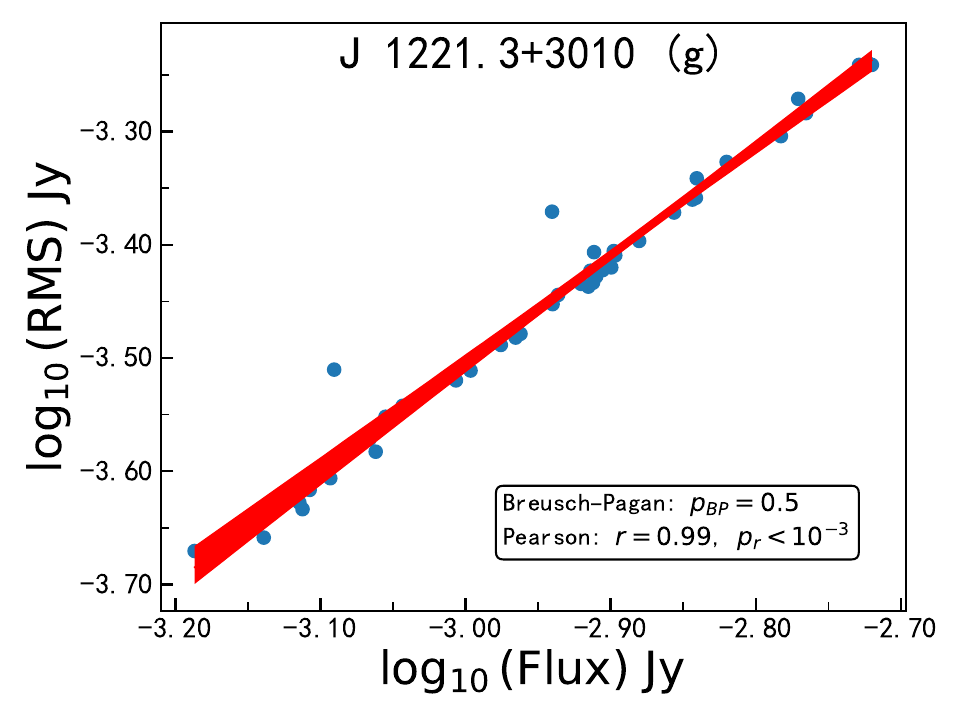}
	\end{subfigure}
	\hfill
	\begin{subfigure}[b]{0.24\textwidth}
		\includegraphics[width=\textwidth]{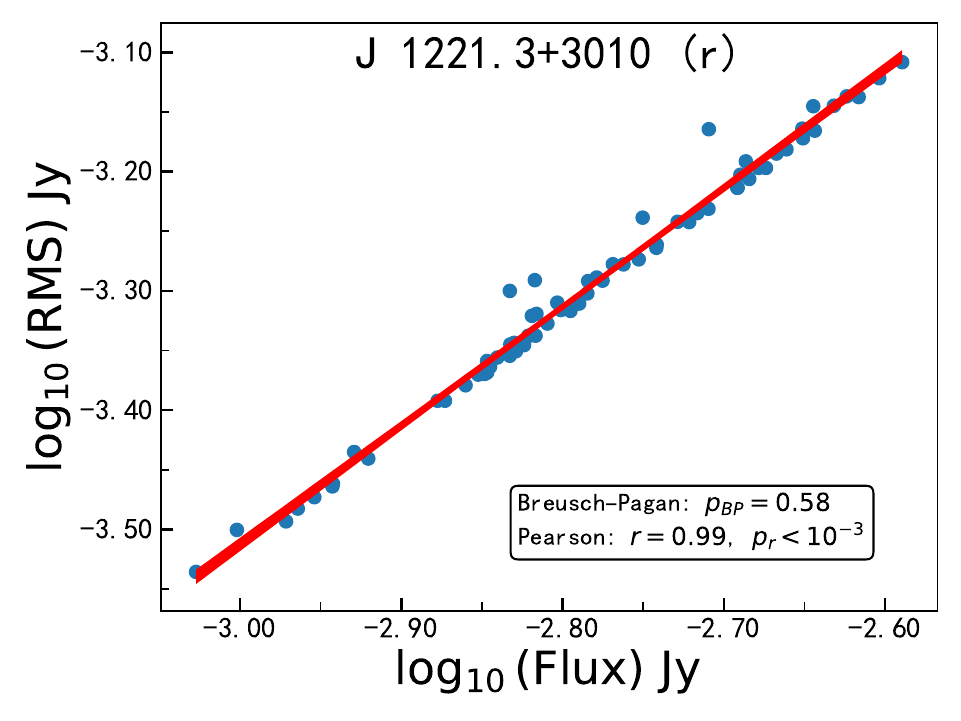}
	\end{subfigure}	
	\vspace{0.1cm}
	\begin{subfigure}[b]{0.24\textwidth}
		\includegraphics[width=\textwidth]{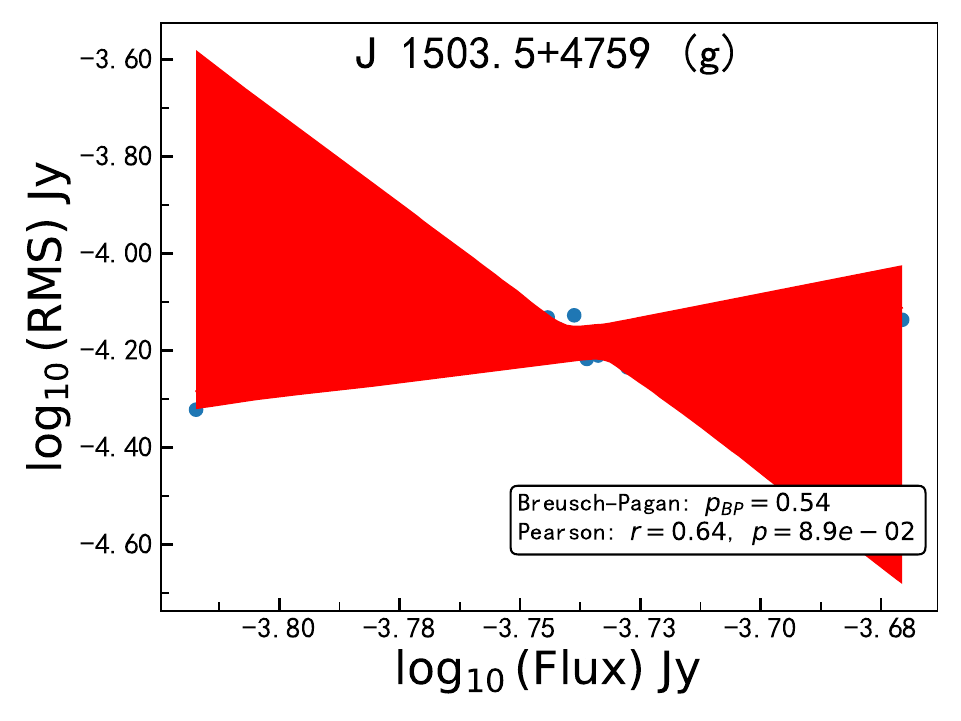}
	\end{subfigure}
	\hfill
	\begin{subfigure}[b]{0.24\textwidth}
		\includegraphics[width=\textwidth]{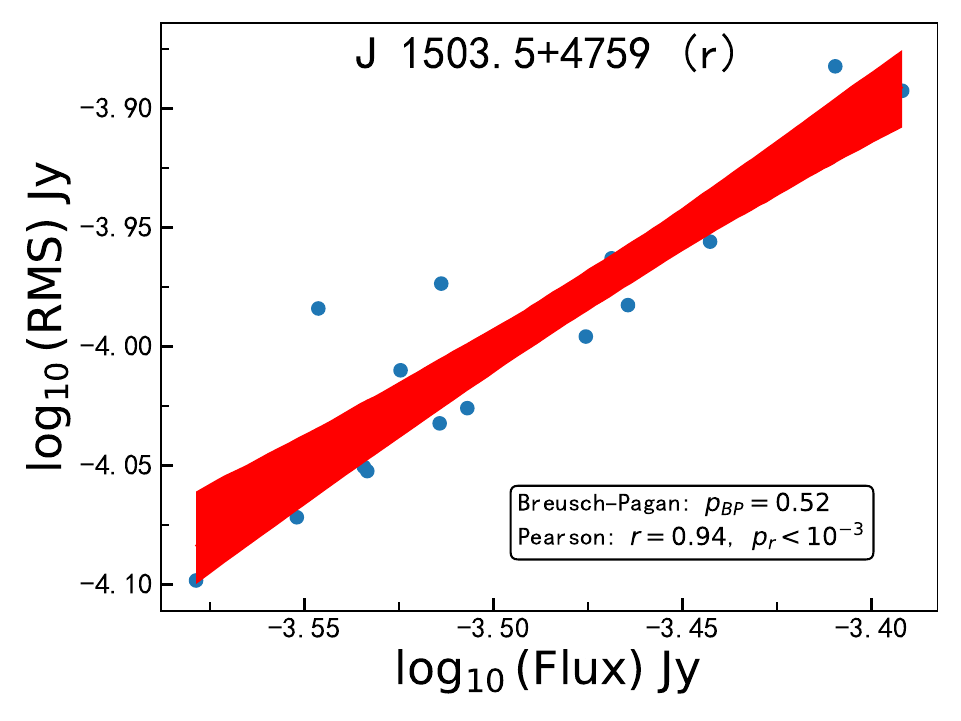}
	\end{subfigure}
	\hfill
	\begin{subfigure}[b]{0.24\textwidth}
		\includegraphics[width=\textwidth]{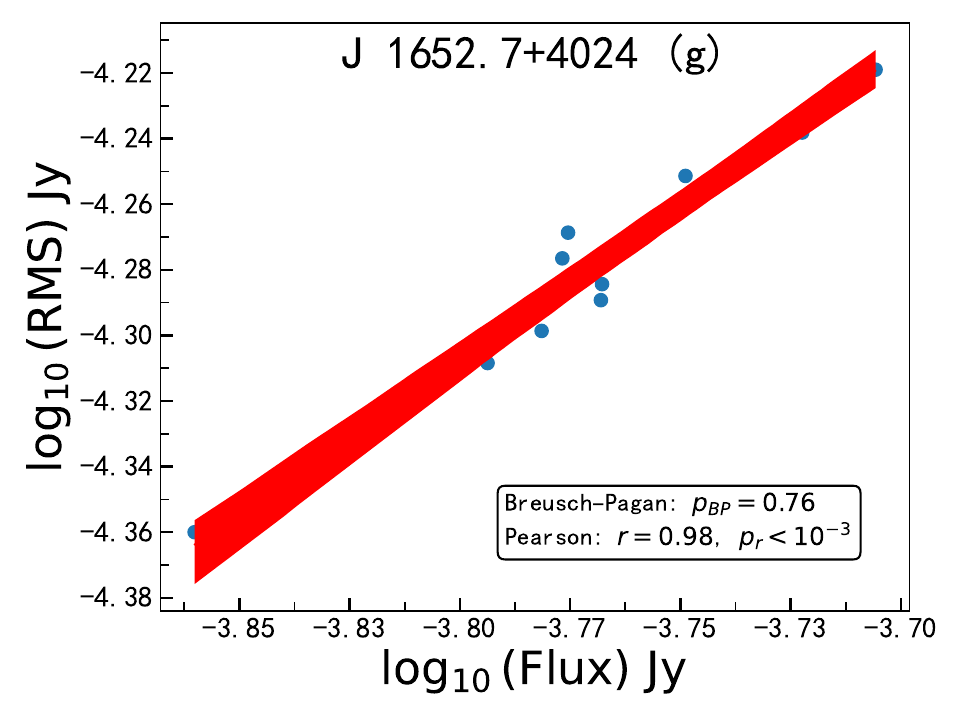}
	\end{subfigure}
	\hfill
	\begin{subfigure}[b]{0.24\textwidth}
		\includegraphics[width=\textwidth]{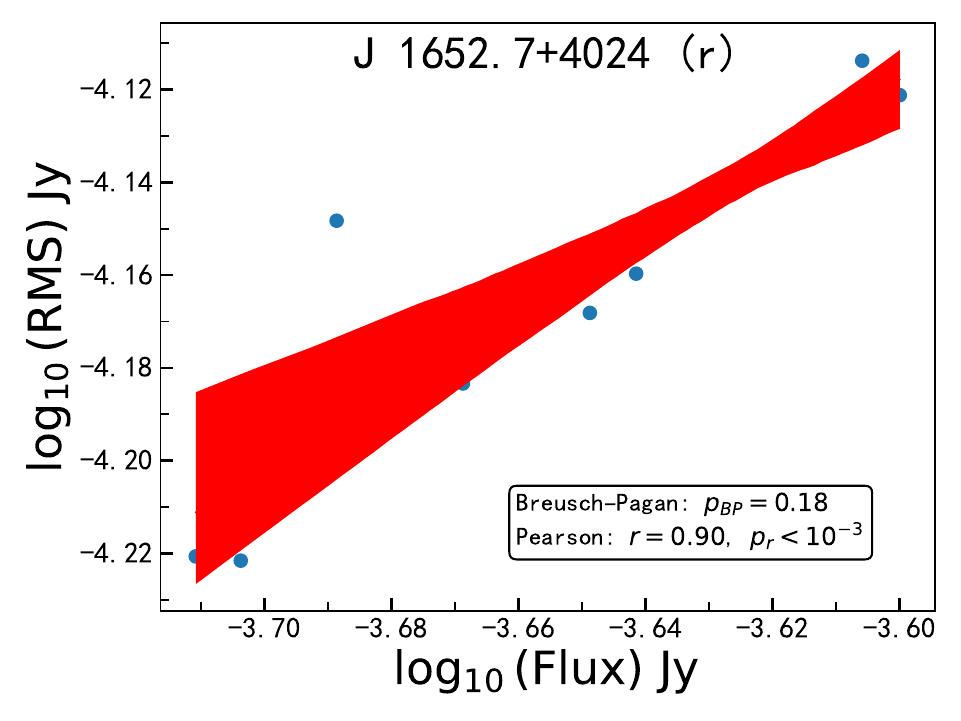}
	\end{subfigure}
	\caption{Graphs of 10 data points per bin for the RMS–Flux relation. The red-shaded area denotes the 95\% confidence interval. The four blazars exhibit a strong linear RMS–Flux relation in both the g- and r-band.}
	\label{fig3}
\end{figure*}

\begin{figure*}
	\centering
	\begin{subfigure}[b]{0.24\textwidth}
		\includegraphics[width=\textwidth]{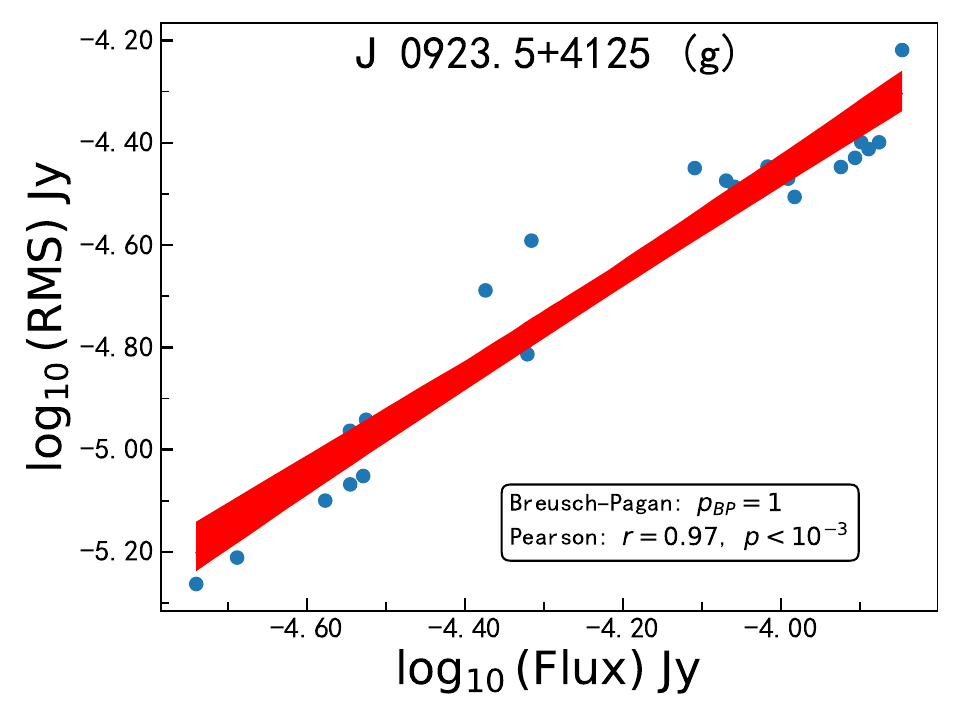}
	\end{subfigure}
	\hfill
	\begin{subfigure}[b]{0.24\textwidth}
		\includegraphics[width=\textwidth]{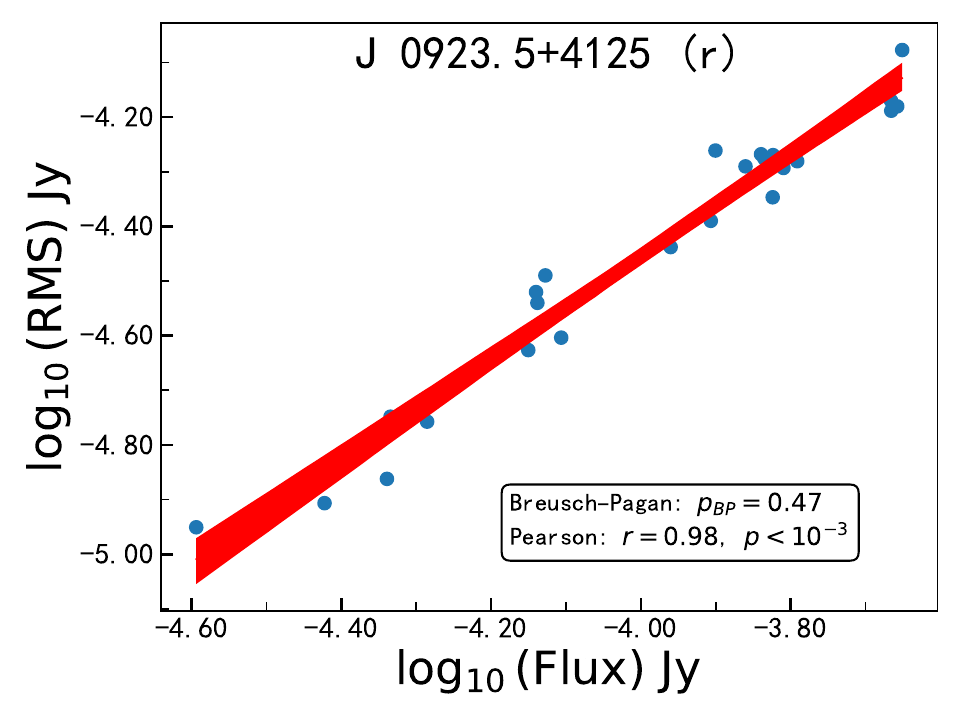}
	\end{subfigure}
	\hfill
	\begin{subfigure}[b]{0.24\textwidth}
		\includegraphics[width=\textwidth]{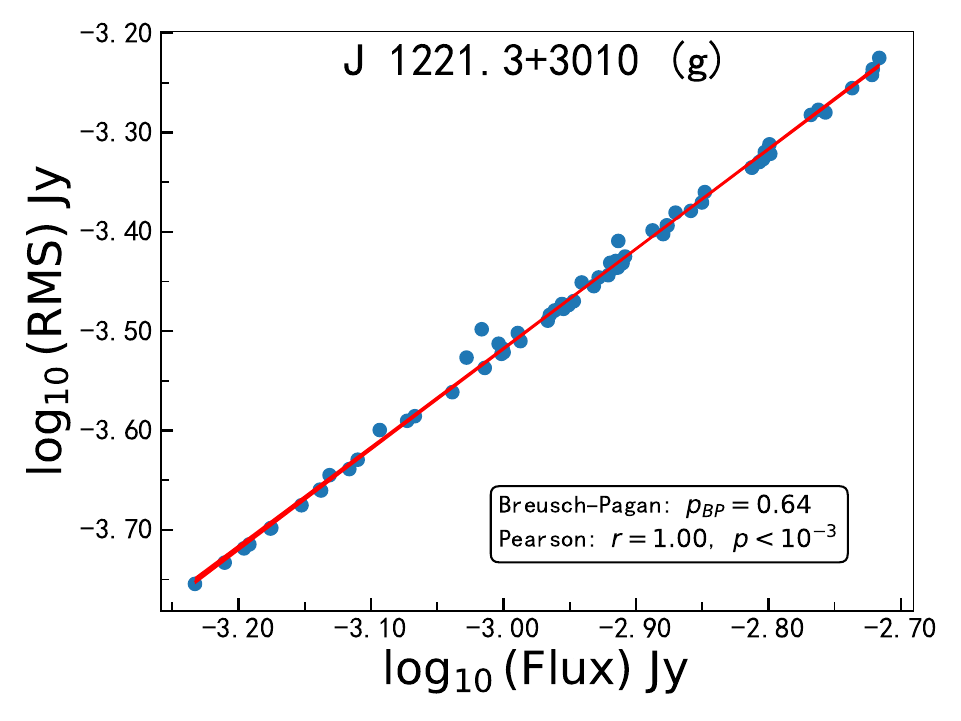}
	\end{subfigure}
	\hfill
	\begin{subfigure}[b]{0.24\textwidth}
		\includegraphics[width=\textwidth]{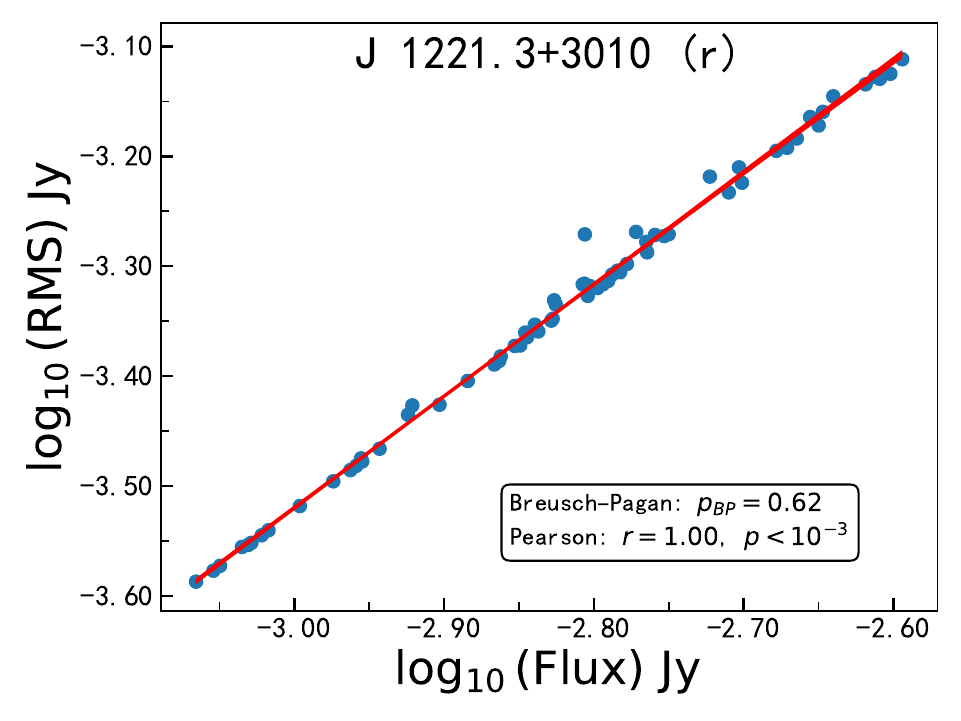}
	\end{subfigure}
	\vspace{0.1cm}
	\begin{subfigure}[b]{0.24\textwidth}
		\includegraphics[width=\textwidth]{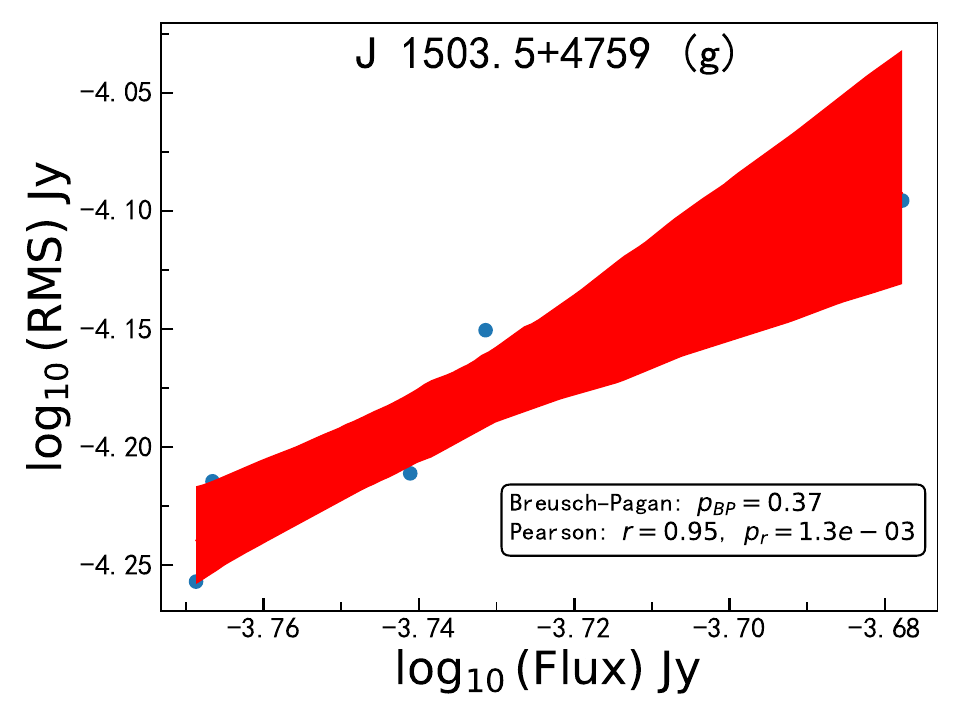}
	\end{subfigure}
	\hfill
	\begin{subfigure}[b]{0.24\textwidth}
		\includegraphics[width=\textwidth]{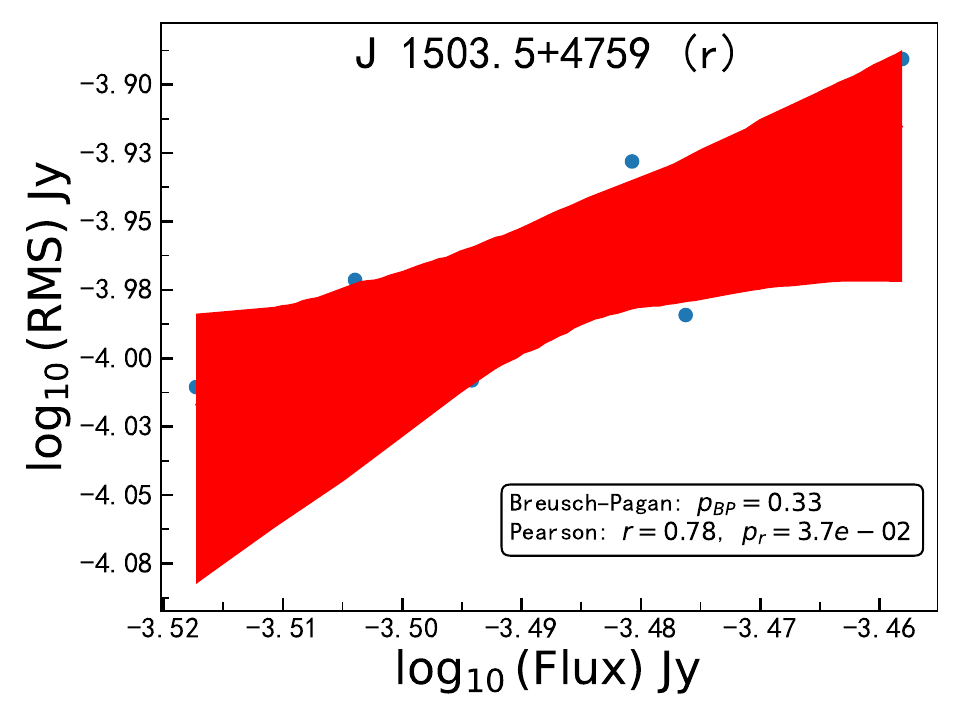}
	\end{subfigure}
	\hfill
	\begin{subfigure}[b]{0.24\textwidth}
		\includegraphics[width=\textwidth]{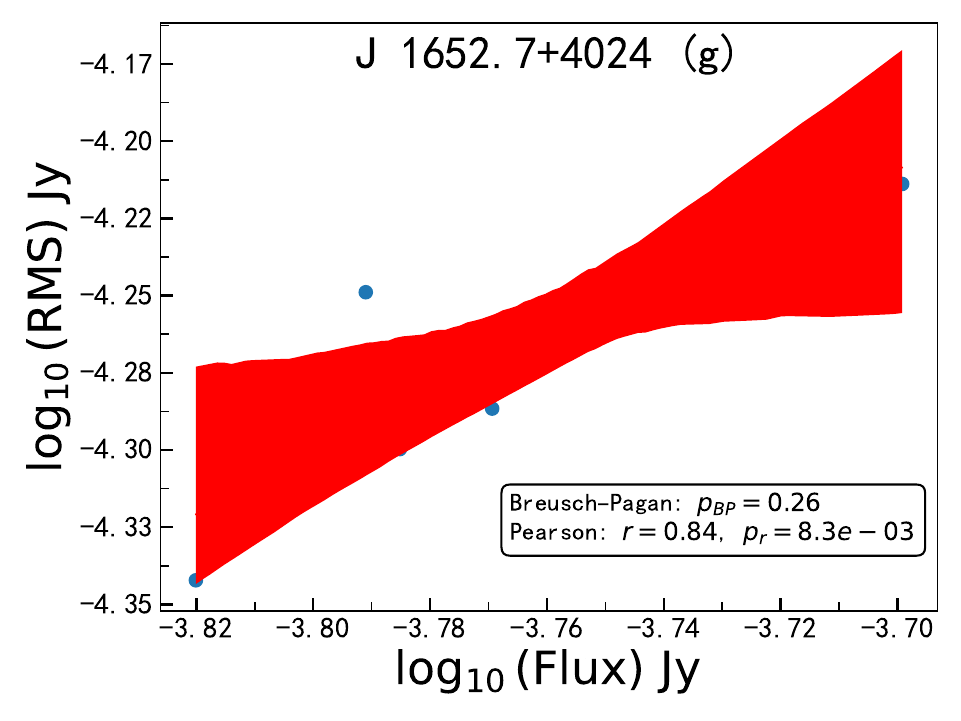}
	\end{subfigure}
	\hfill
	\begin{subfigure}[b]{0.24\textwidth}
		\includegraphics[width=\textwidth]{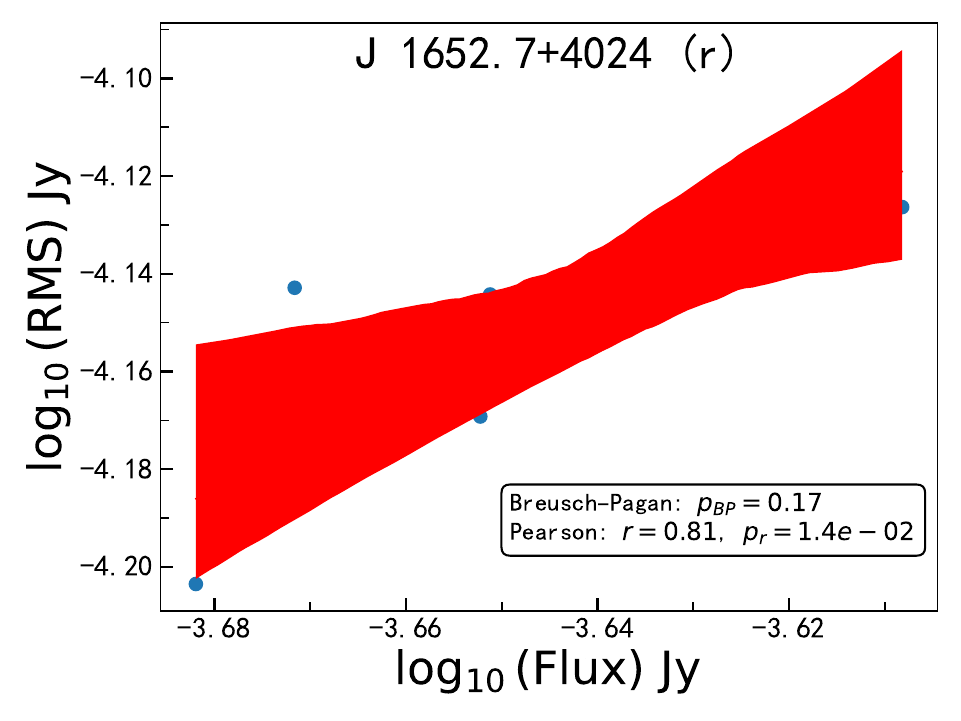}
	\end{subfigure}
	\caption{Same as Figure \ref{fig3}, but the binning is 30 days.}
	\label{fig4}
\end{figure*}

The linear RMS–Flux relation suggests a significant correlation between the variability properties across multiple flux states, which could indicate the presence of nonlinear processes underlying the observed variability. Such a relationship typically implies a flux distribution skewed toward higher flux states and points to multiplicative variability mechanisms, for example the multiplicative coupling of disk and jet perturbations \citep{2020ApJ...891..120}. To test this relation, we binned the light curves in two ways: (1) fixed-size bins of 10 data points, and (2) fixed-duration bins of 30 days. The purpose was to verify whether binning by data points or by time would affect the results. Both binning strategies were designed to ensure that, under limited sampling, a sufficient number of bins could still be retained for RMS–flux fitting. For each segment, the mean log-flux and the variance $S^2$ are computed. Figures \ref{fig3} and \ref{fig4} show the linear relation between the RMS and the corresponding mean log-flux values for the sample sources in 10 data points and 30-day binned segments, respectively. We employed ordinary least squares (OLS) to fit the RMS–flux relation and calculated the Pearson correlation coefficient together with its corresponding p-value. The red lines represent the linear fit to the observational data, while the shaded regions indicate the 95\% confidence intervals, derived from the standard errors of the OLS estimates. To further assess whether the model assumptions were satisfied, we applied the Breusch–Pagan (BP) test to the fitting residuals. The results showed that the BP test p-values for all sources were greater than 0.05, indicating that the adopted fitting method is valid under the basic statistical assumptions. The results indicate that, regardless of the binning method, all four sample sources exhibit a linear RMS–Flux relationship. The weaker correlation observed in J 1503.5+4759 (g-band) is mainly attributed to the limited number of data points (only eight) and the narrow flux range, with six points clustered between $\log_{10}(\mathrm{Flux/Jy})=-3.74$ and $-3.72$. Similar results have also been reported in other literature \citep{2020ApJ...891..120,2021ApJ...923..7,2023RAA..23..115011}.

\subsection{Flux Distribution}

Flux distribution analysis in blazars is a crucial aspect of understanding the physical processes contributing to their emission mechanisms and the origin of their variability \citep{2020ApJ...891..120, 2021ApJ...923..7, 2012AA..548..A123}. By examining the long-term flux distribution, one can gain insights into the nature of high-energy emission processes and the underlying mechanisms driving the observed variability. Two popular distributions are often adopted: a normal distribution and a log-normal distribution. By testing the flux distribution against these probability density functions (PDFs), researchers can gain insights into the nature of the high-energy emission processes and the nature of the variability processes in blazars. For instance, if the flux distribution of a blazar fits a log-normal PDF, it suggests that the variability is likely due to multiplicative processes, such as perturbations in the jet or disk \citep{2012AA..548..A123}. This is often seen in systems where the variability is driven by a series of cascading events, each influencing the next, resulting in a skewed distribution toward higher flux levels. On the other hand, a normal distribution fit would indicate additive processes, where various independent components contribute to the observed flux.

\begin{figure*}
	\centering
	\begin{subfigure}[b]{0.24\textwidth}
		\includegraphics[width=\textwidth]{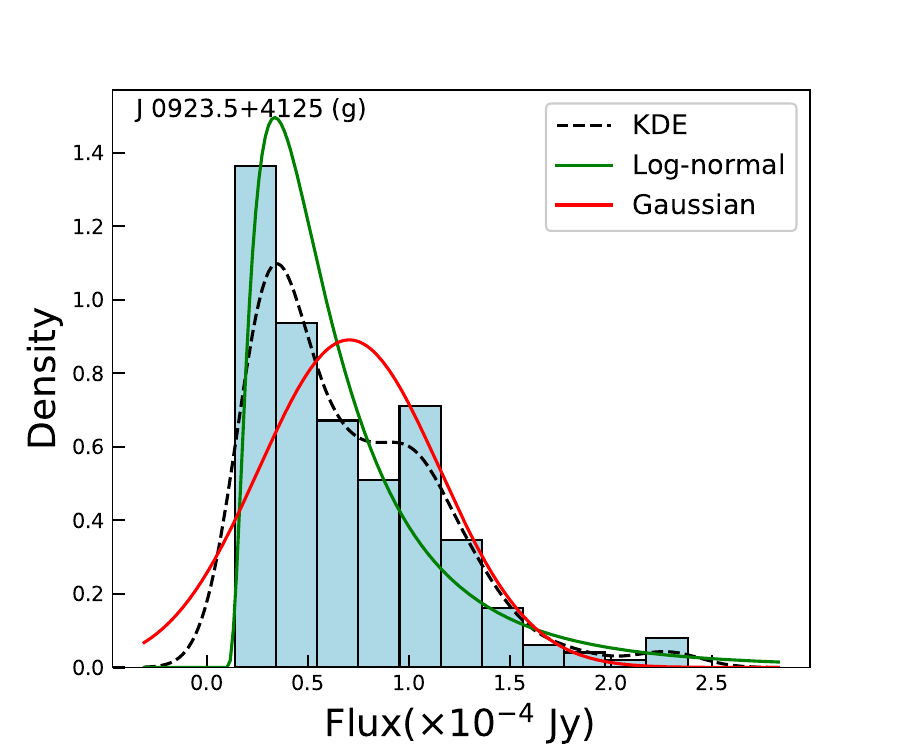}
	\end{subfigure}
	\hfill
	\begin{subfigure}[b]{0.24\textwidth}
		\includegraphics[width=\textwidth]{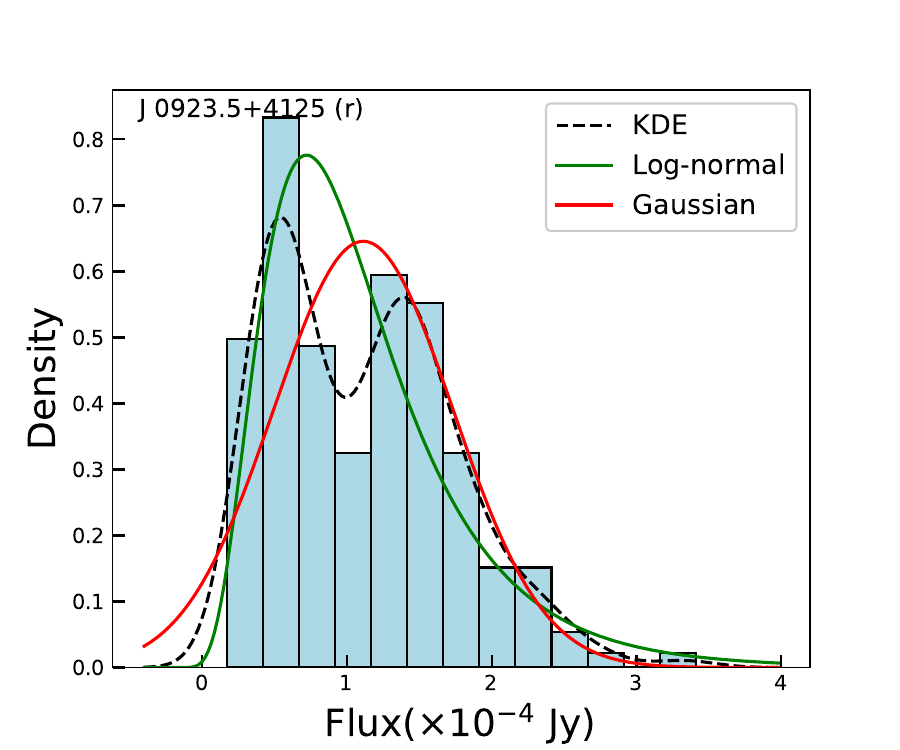}
	\end{subfigure}
	\hfill
	\begin{subfigure}[b]{0.24\textwidth}
		\includegraphics[width=\textwidth]{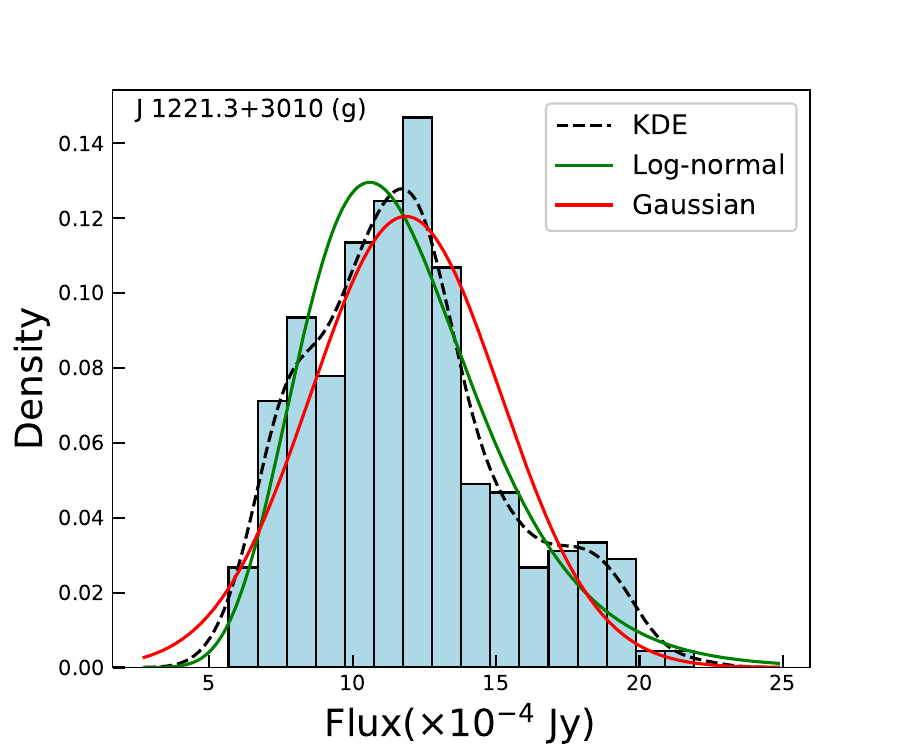}
	\end{subfigure}
	\hfill
	\begin{subfigure}[b]{0.24\textwidth}
		\includegraphics[width=\textwidth]{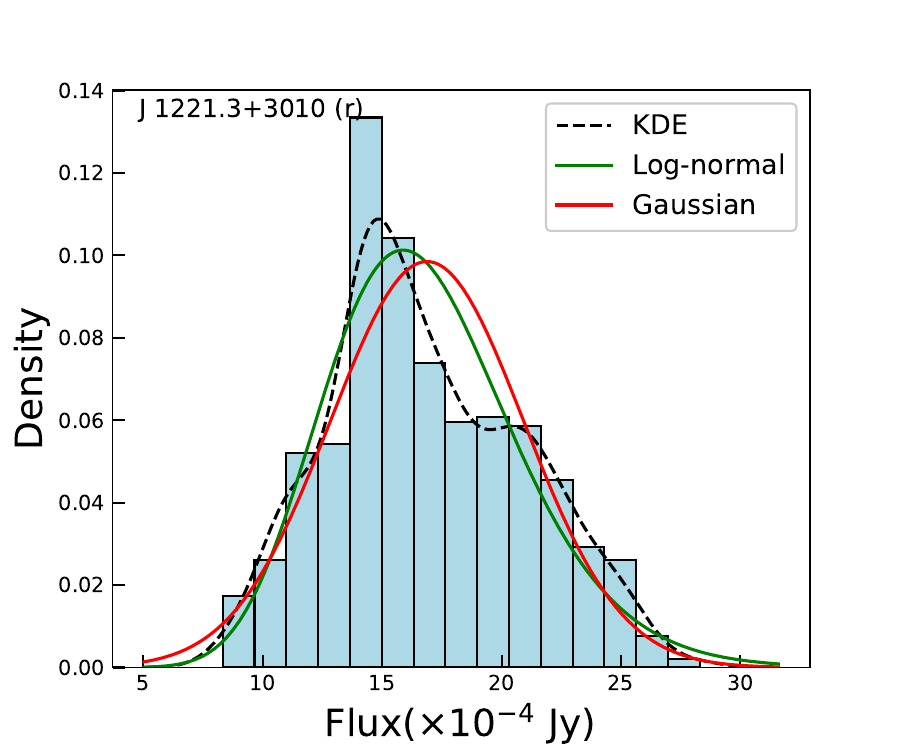}
	\end{subfigure}
	\vspace{0.1cm}
	\begin{subfigure}[b]{0.24\textwidth}
		\includegraphics[width=\textwidth]{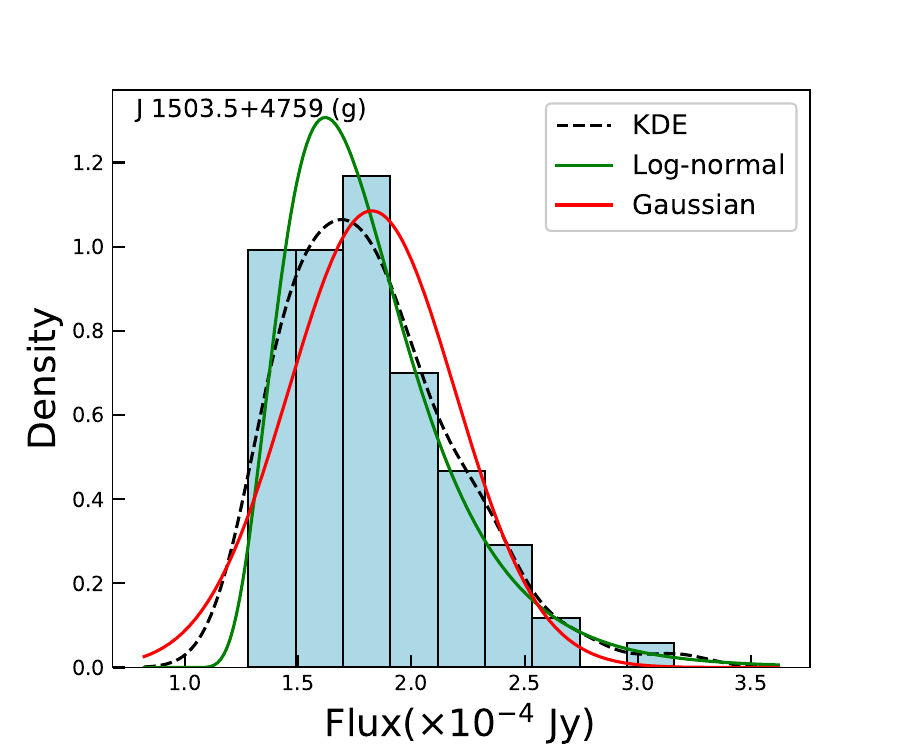}
	\end{subfigure}
	\hfill
	\begin{subfigure}[b]{0.24\textwidth}
		\includegraphics[width=\textwidth]{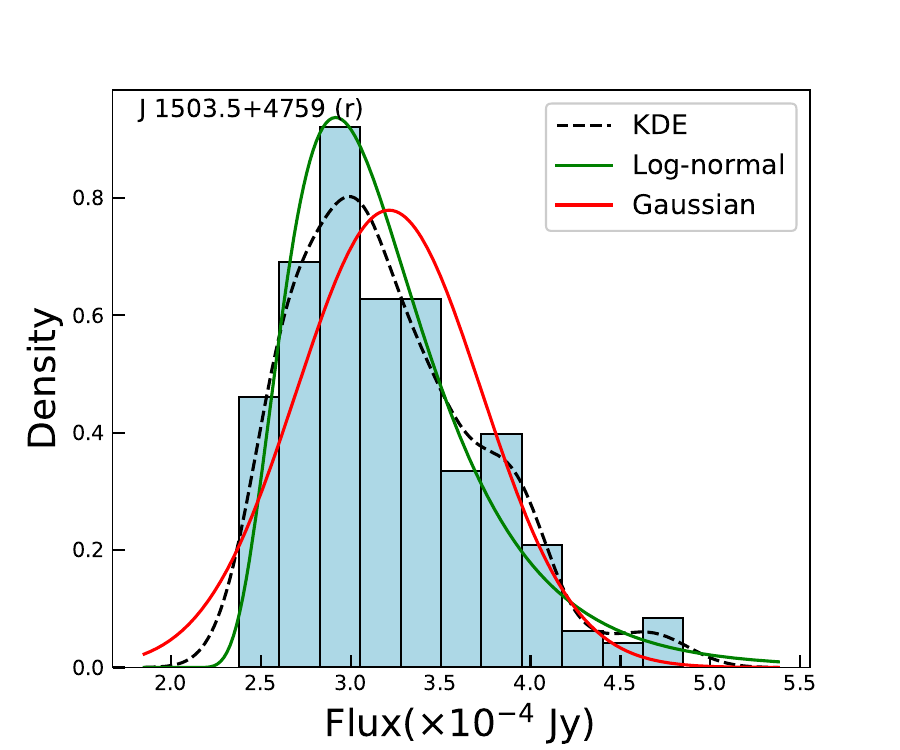}
	\end{subfigure}
	\hfill
	\begin{subfigure}[b]{0.24\textwidth}
		\includegraphics[width=\textwidth]{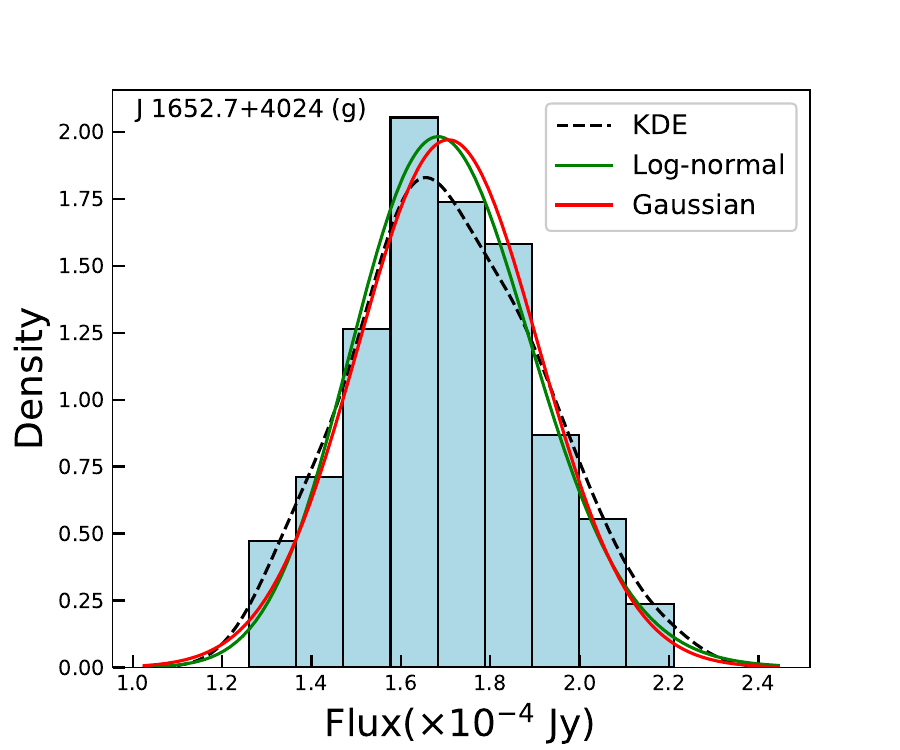}
	\end{subfigure}
	\hfill
	\begin{subfigure}[b]{0.24\textwidth}
		\includegraphics[width=\textwidth]{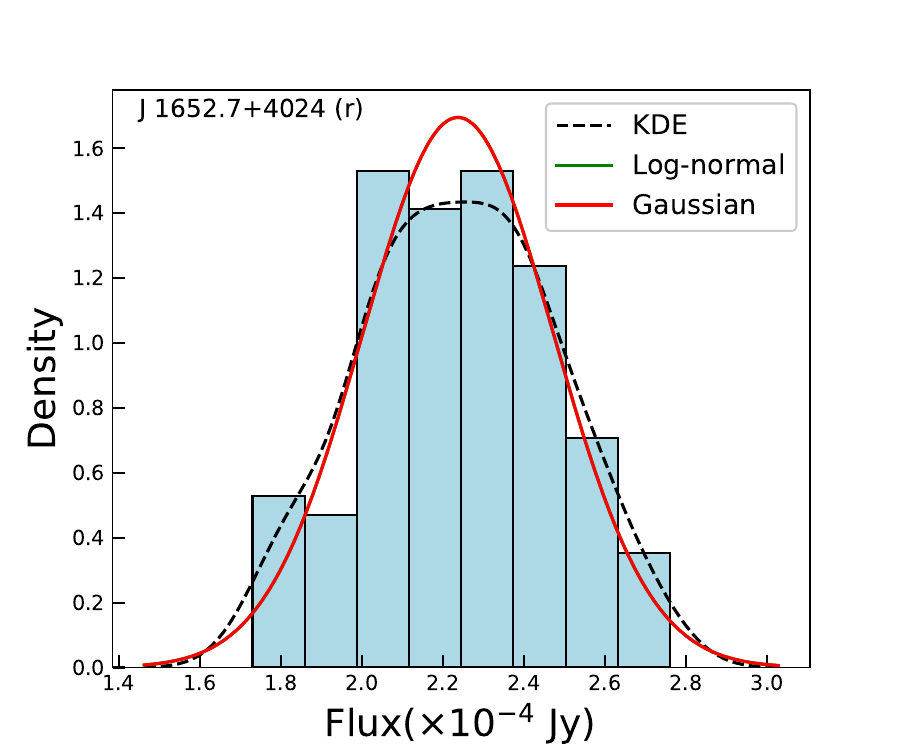}
	\end{subfigure}
	\caption{The best-fit normal (red solid curves) and log-normal (green solid curves) function to the flux distribution (hatched blue histograms) of the blazars. The black dotted curves represent the kernel density estimation (KDE) results.}
	\label{fig5}
\end{figure*}

This approach of fitting normal and log-normal distributions to flux histograms allows for a statistical probe of the PDF of long-term observation. The flux distribution histograms for the four sources in the g- and r-bands are presented in Figure \ref{fig5}. We adopted the automatic binning method implemented in \texttt{numpy}, which combines the Sturges \citep{1926JAmS...21...65S}, Freedman-Diaconis \citep{1981ZWiG...57..453F}, and automatically selects the larger number of bins. This approach avoids overly coarse binning when the sample size is small or the distribution is close to normal, while capturing more details in skewed or long-tailed distributions. The flux distribution histograms were fitted using the two aforementioned PDF models, and the fitting statistics for the log-normal and normal PDFs are presented in Table \ref{tab2}. In addition, we applied the Maximum Likelihood Estimation (MLE) method to directly analyze the data and used the Bayesian Information Criterion (BIC) for model comparison. The values of $\Delta$ BIC ($={\rm BIC}_{\rm normal} -{\rm BIC}_{\rm log-normal}$) are presented in Table \ref{tab2}. When $\Delta$ BIC $\textgreater$ 8–10, it indicates that the flux distribution is more inclined to follow a log-normal model. According to the $\Delta$ BIC, we found that, for most sources, the flux distribution is better fitted by a log-normal PDF in both the g- and r-bands, except for J 1652.7+4024. For J 1652.7+4024 in the r-band, both log-normal and normal PDFs can adequately fit the flux distribution histogram. Additionally, the $p$-value \citep{2014MolPsy..19..1336} from the Shapiro–Wilk (S-W) test and D’Agostino-Pearson normality test \citep{1973Biometrika..60..613D} (using \texttt{scipy.stats.normaltest}) also indicate that the flux distribution of most sources tends to follow a log-normal distribution. 

\setlength{\tabcolsep}{4pt}
\begin{table*}
	\caption{Log-normal and normal distribution fit parameters for the flux distribution. The Mean, standard deviation, Shapiro--Wilk test $p$-value, and D’Agostino--Pearson normality test $p$-value.}
	\noindent\hspace*{-8.5em}
	\label{tab2}
	\begin{tabular}{lcccccccccc}
		\toprule
		Name (band) & \multicolumn{4}{c}{Normal Fit} & & \multicolumn{4}{c}{Log-normal Fit} & $\Delta \rm BIC$ \\
		\cmidrule(lr){2-5} \cmidrule(lr){7-10}
		& Mean & $\sigma$ & S-W test(p) & Normality(p) & & Mean & $\sigma$  & S-W test(p) & Normality(p) & \\
		\midrule
		J 0923.5+4125 (g) & 0.70 & 0.45 & $\ll0.0001$ & $\ll0.0001$ & & $-0.55$ & 0.64  & $\ll0.0001$ & $\ll0.0001$ & 91.93 \\
		J 0923.5+4125 (r) & 1.12 & 0.62 & $\ll0.0001$ & $\ll0.0001$ & & $-0.06$ & 0.62  & $\ll0.0001$ & $\ll0.0001$ & 45.16\\
		J 1221.3+3010 (g) & 11.85 & 3.32  & $\ll0.0001$ & $\ll0.0001$ & & 2.43 & 0.28 &  $>{0.05}$ & $>{0.05}$ & 32.78\\
		J 1221.3+3010 (r) & 16.88 & 4.05 & $\ll0.0001$ & $\ll0.0001$ & & 2.80 & 0.24 & $\ll0.0001$ & 0.001 & 14.71\\
		J 1503.5+4759 (g) & 1.82 & 0.37  & $>{0.05}$ & 0.0006 & & 0.58 & 0.19  & $>{0.05}$ & $>{0.05}$ & 10.95\\
		J 1503.5+4759 (r) & 3.21 & 0.51 & 0.0009 & $\ll0.0001$ & & 1.16 & 0.15  & 0.0002 & 0.01 & 21.09\\
		J 1652.7+4024 (g) & 1.71 & 0.20  & $>{0.05}$ & $>{0.05}$ & & 0.53 & 0.12   & $>{0.05}$ & $>{0.05}$ & 0.48 \\
		J 1652.7+4024 (r) & 2.24 & 0.24  & $>{0.05}$ & $>{0.05}$ & & 0.80 & 0.11  & $>{0.05}$ & $>{0.05}$ & -1.65 \\
		\bottomrule
	\end{tabular}
\end{table*}

\begin{figure*}[t]
	\centering
	\captionsetup[sub]{justification=centering}
	\begin{subfigure}[t]{0.49\textwidth}
		\includegraphics[width=\linewidth,height=5cm,keepaspectratio]{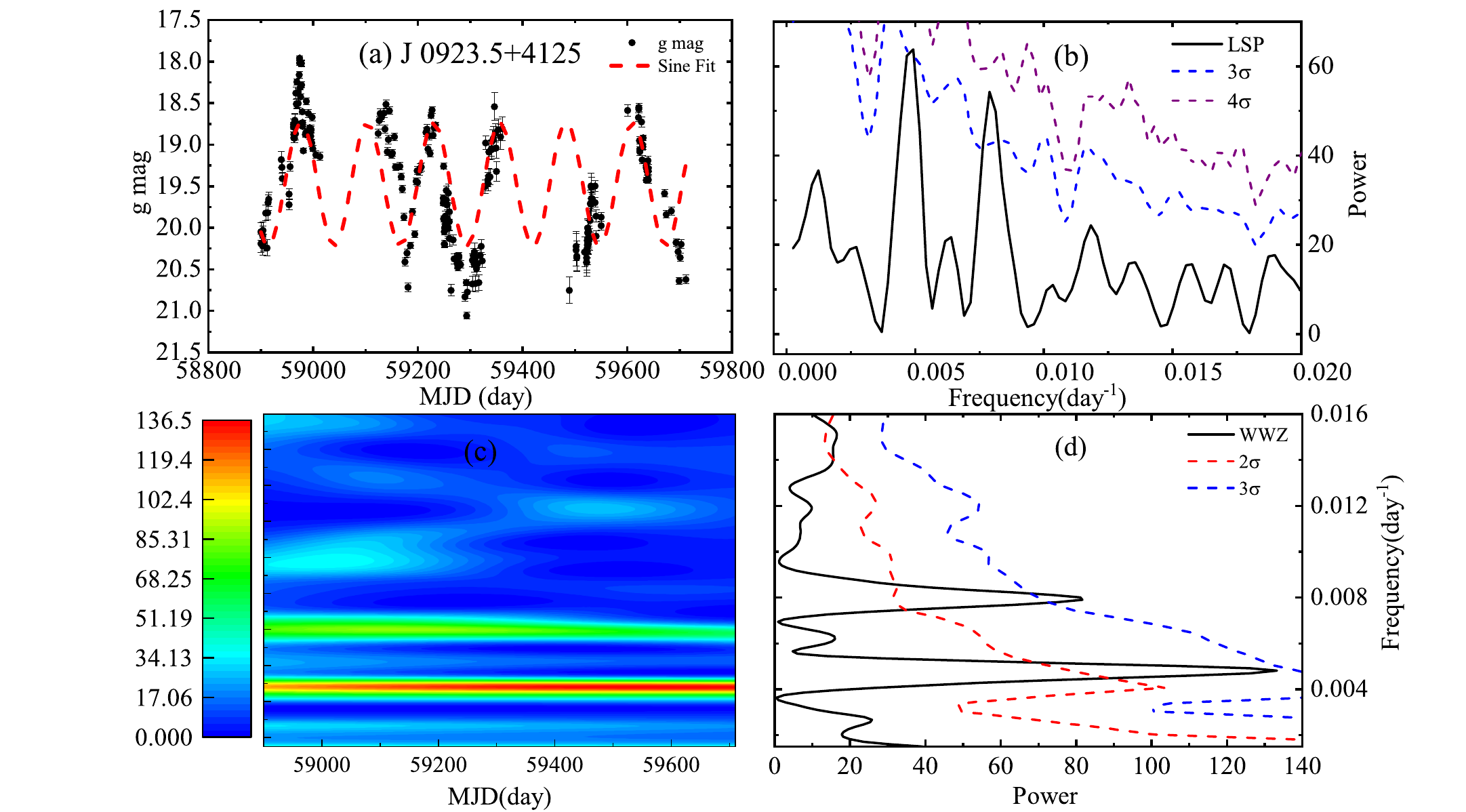}
		
	\end{subfigure}\hfill
	\begin{subfigure}[t]{0.49\textwidth}
		\includegraphics[width=\linewidth,height=5cm,keepaspectratio]{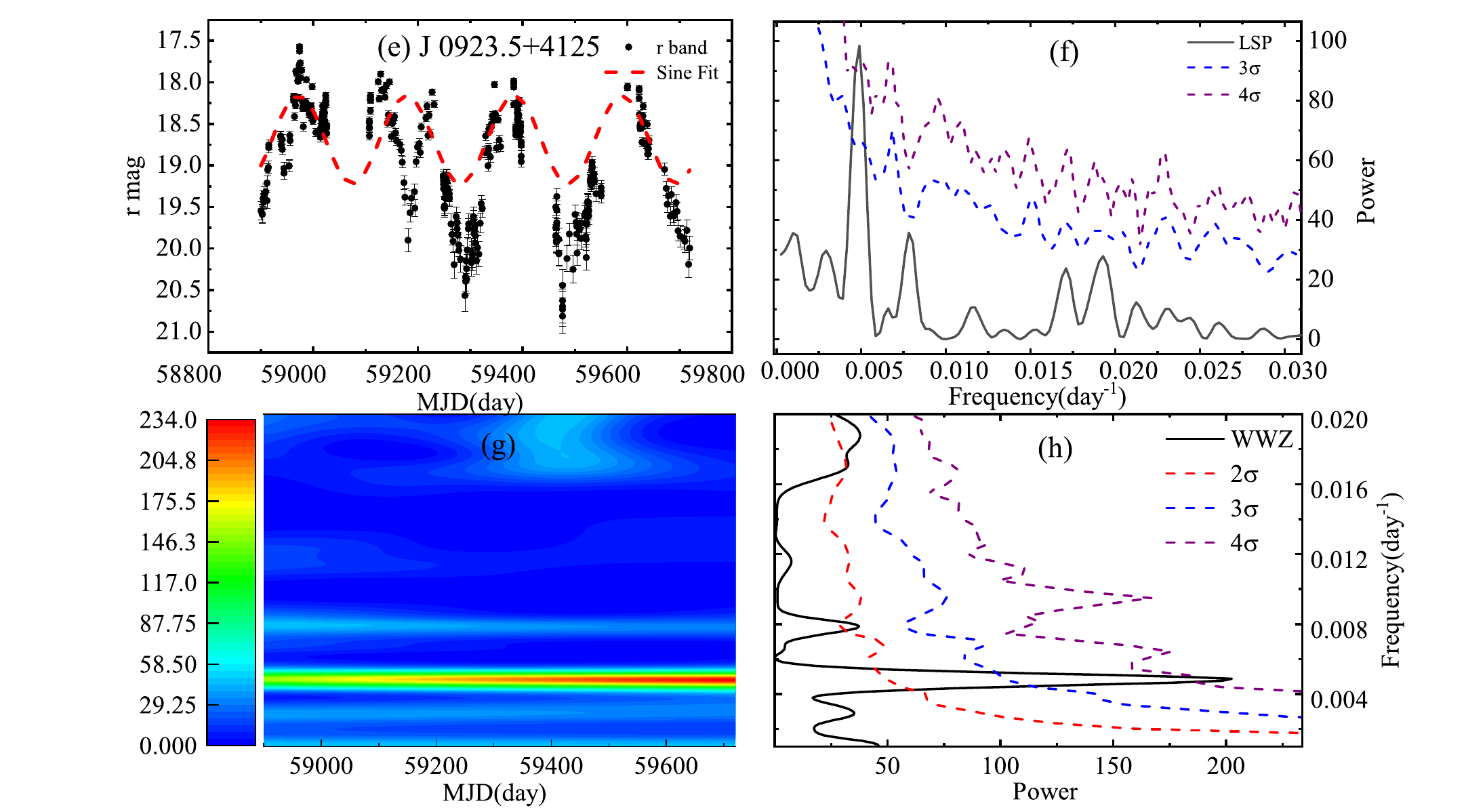}
		
	\end{subfigure}	
	\medskip
	\begin{subfigure}[t]{0.49\textwidth}
		\includegraphics[width=\linewidth,height=5cm,keepaspectratio]{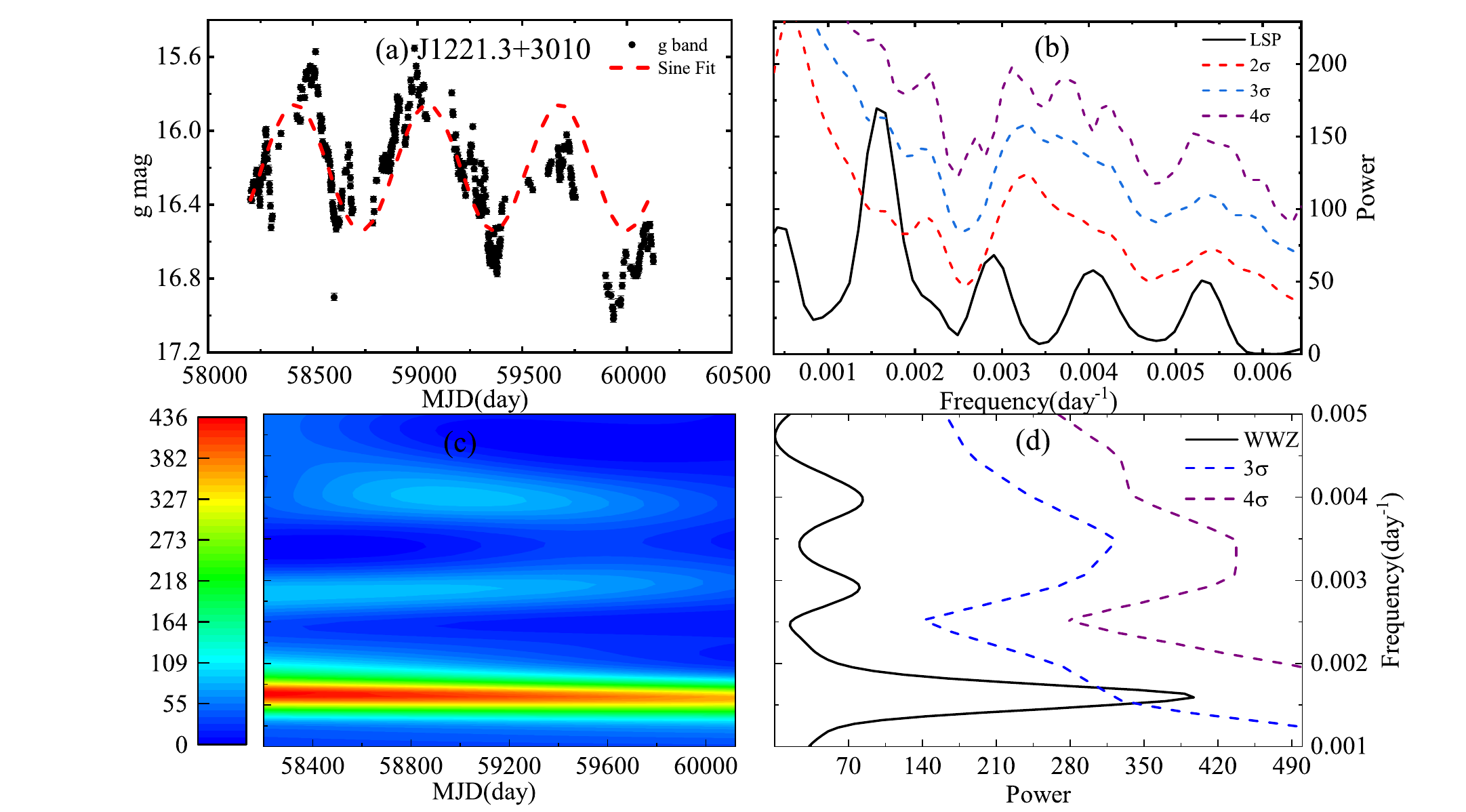}
		
	\end{subfigure}\hfill
	\begin{subfigure}[t]{0.49\textwidth}
		\includegraphics[width=\linewidth,height=5cm,keepaspectratio]{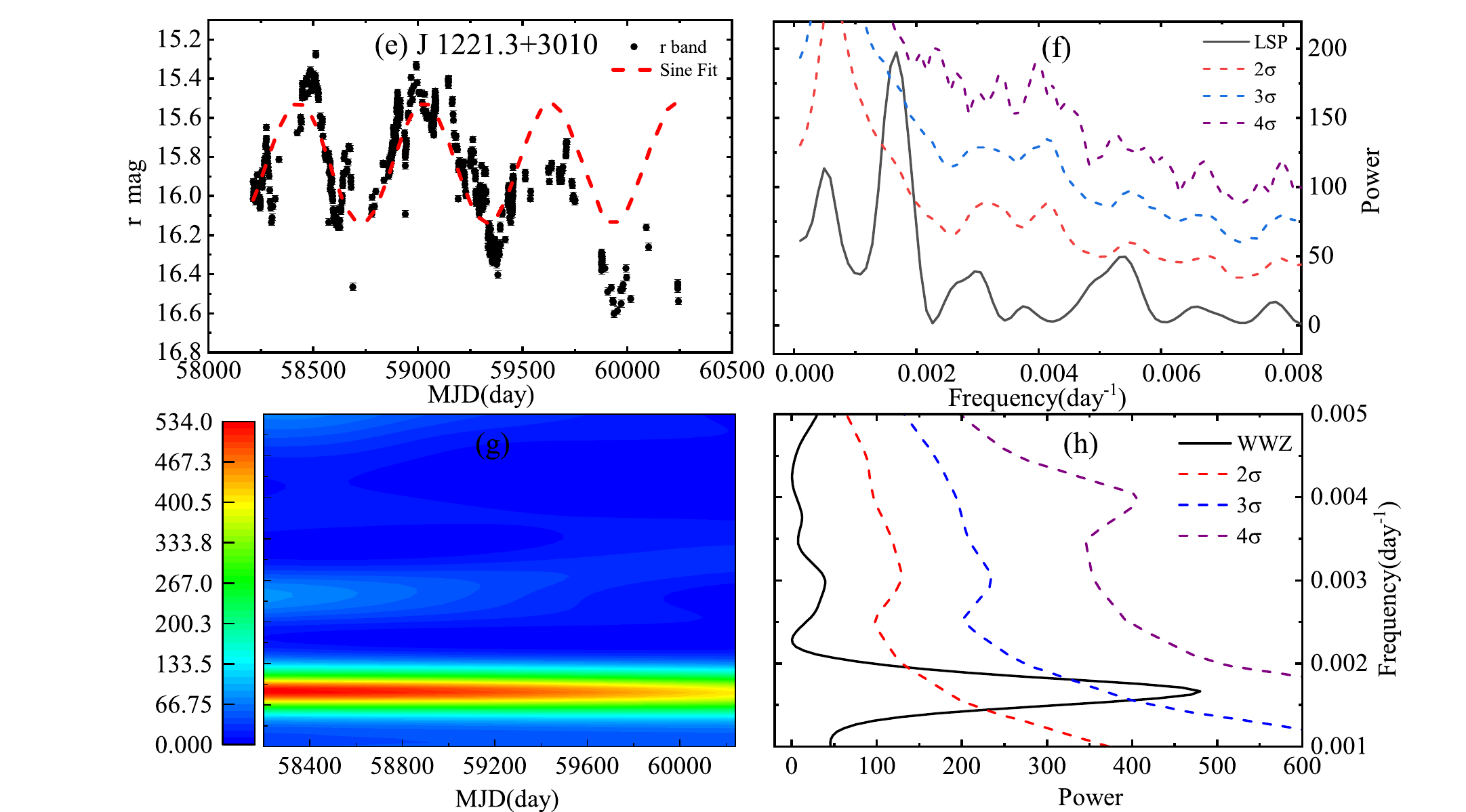}
		
	\end{subfigure}	
	\medskip
	\begin{subfigure}[t]{0.49\textwidth}
		\includegraphics[width=\linewidth,height=5cm,keepaspectratio]{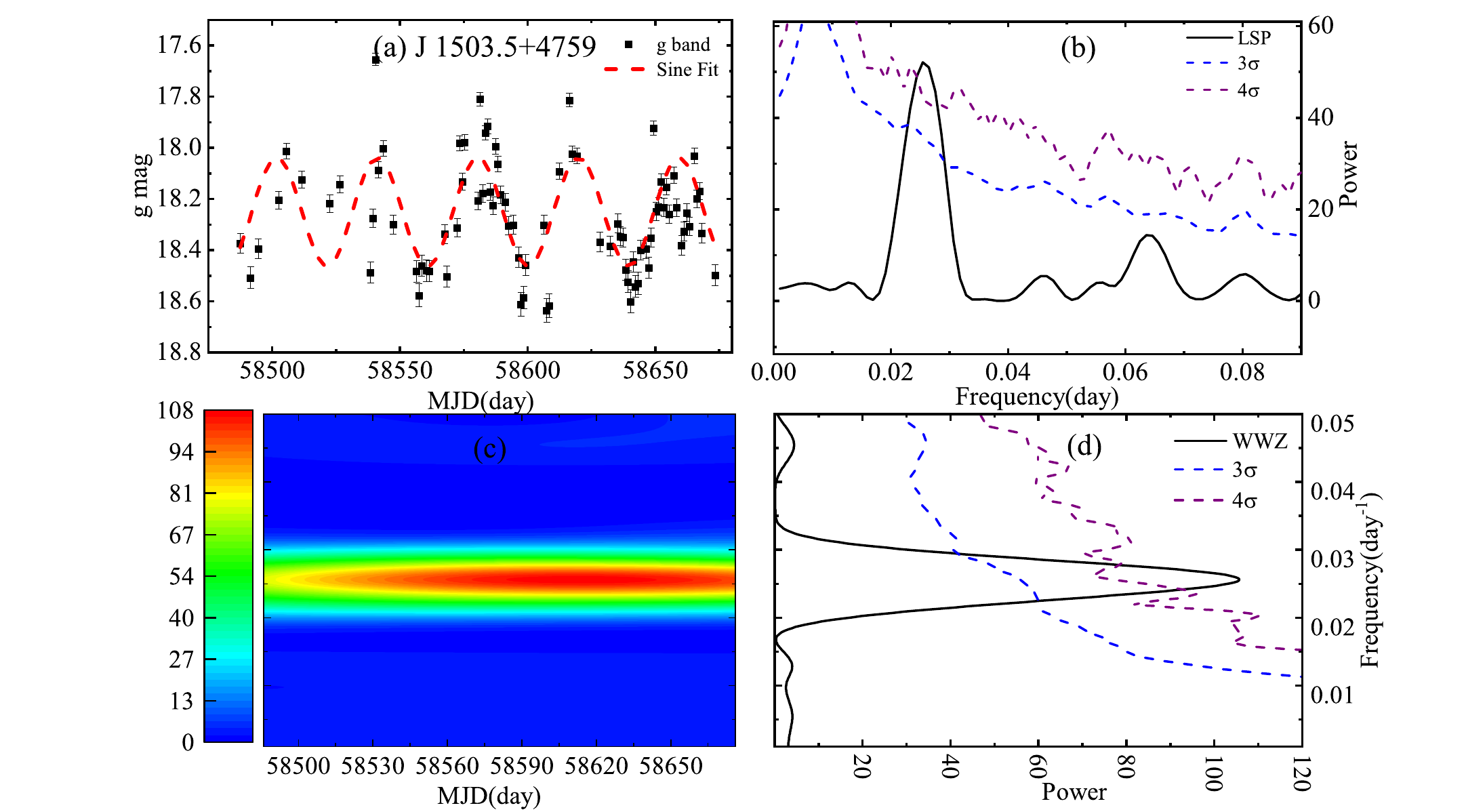}
		
	\end{subfigure}\hfill
	\begin{subfigure}[t]{0.49\textwidth}
		\includegraphics[width=\linewidth,height=5cm,keepaspectratio]{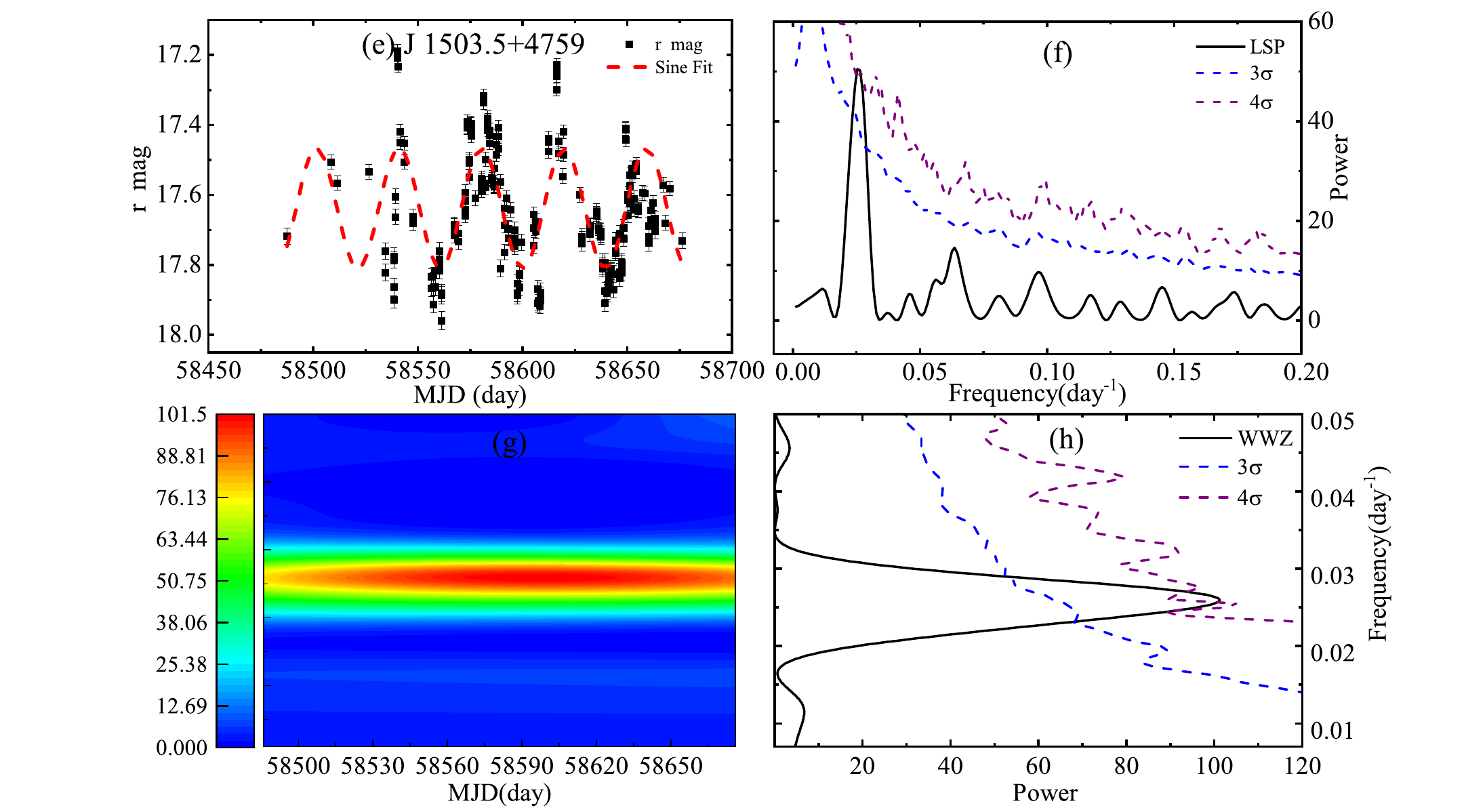}
		
	\end{subfigure}	
	\medskip
	\begin{subfigure}[t]{0.49\textwidth}
		\includegraphics[width=\linewidth,height=5cm,keepaspectratio]{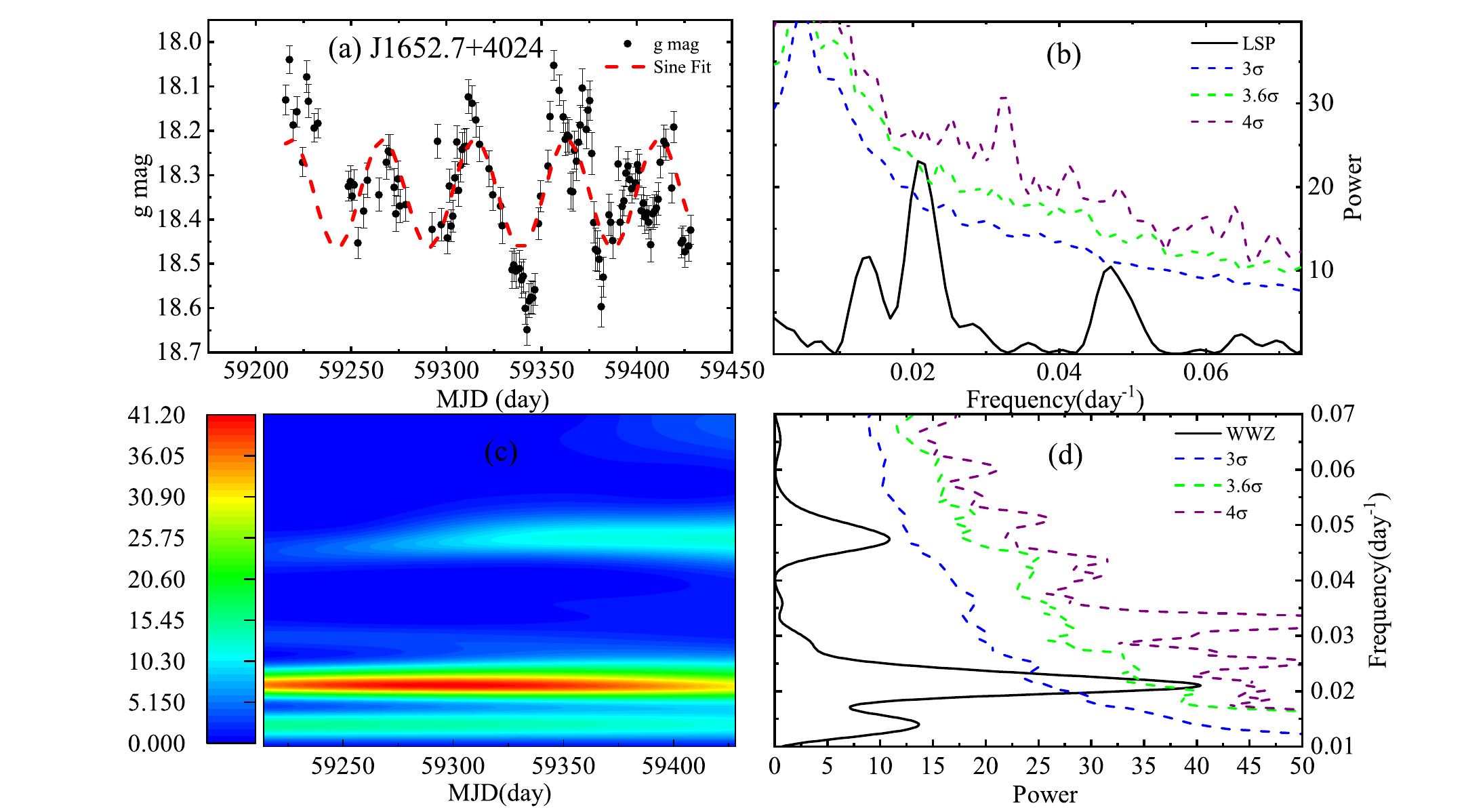}
		
	\end{subfigure}\hfill
	\begin{subfigure}[t]{0.49\textwidth}
		\includegraphics[width=\linewidth,height=5cm,keepaspectratio]{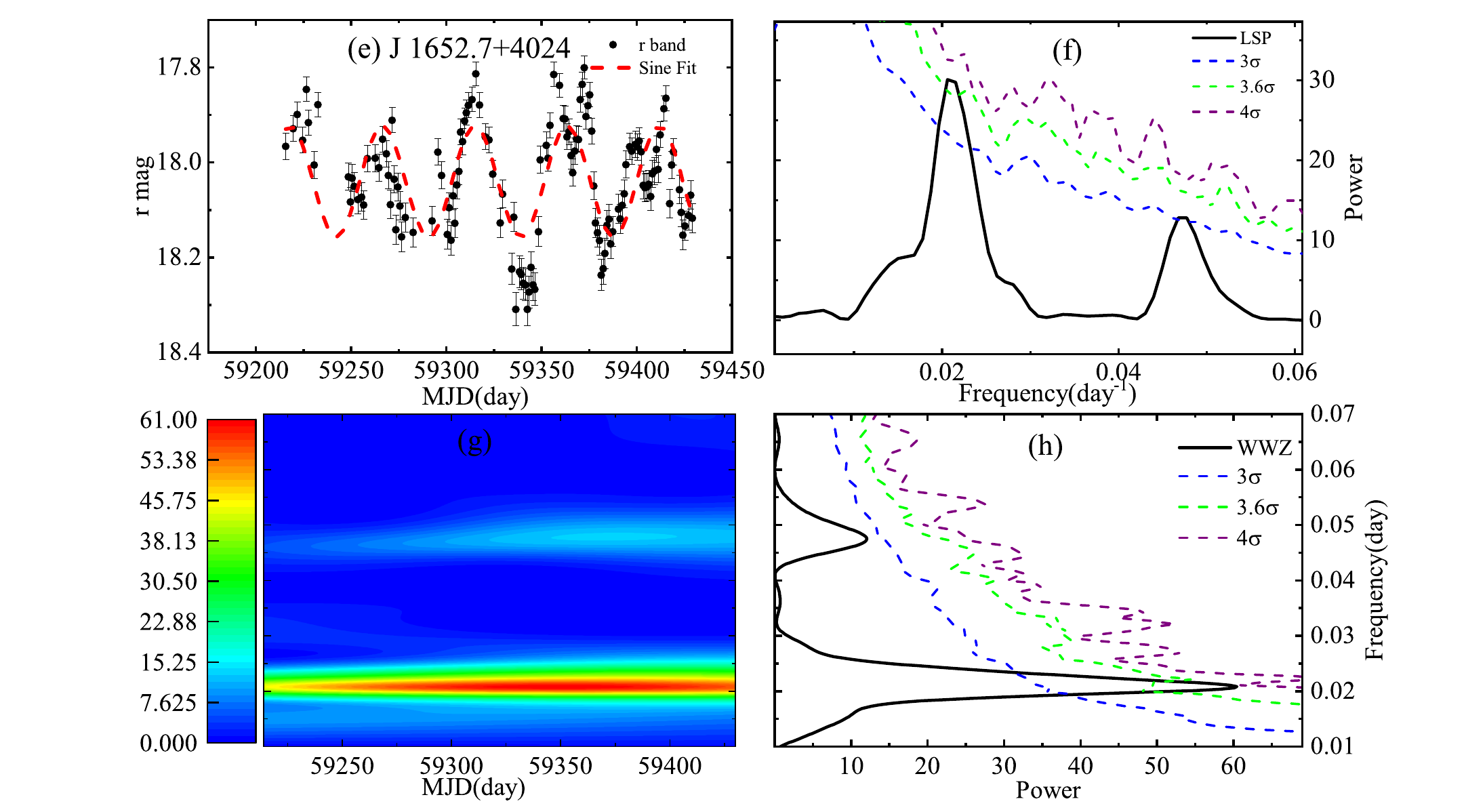}
		
	\end{subfigure}	
	\caption{The WWZ and LSP results for the sample sources. Left panels: (a) the g-band light curve. The red dotted curve and black point represent the fitting result of the sine function and the ZTF g-band magnitude, respectively. (b) the power spectrum of the light curve. The black solid curve is the power spectrum, and the red, blue, green, and purple dashed curves indicate the confidence levels of 95.45\% (2$\sigma$), 99.73\% (3$\sigma$), 99.95\% (3.6$\sigma$) and 99.99\% (4$\sigma$), respectively. (c) the WWZ map of the light curve. (d) the black solid curve shows the time-averaged WWZ. Right panels: similar to the left panels, but for the r-band.}
	\label{fig6}
\end{figure*}

\begin{figure*}
	\centering
	\includegraphics[width=0.5\textwidth,keepaspectratio]{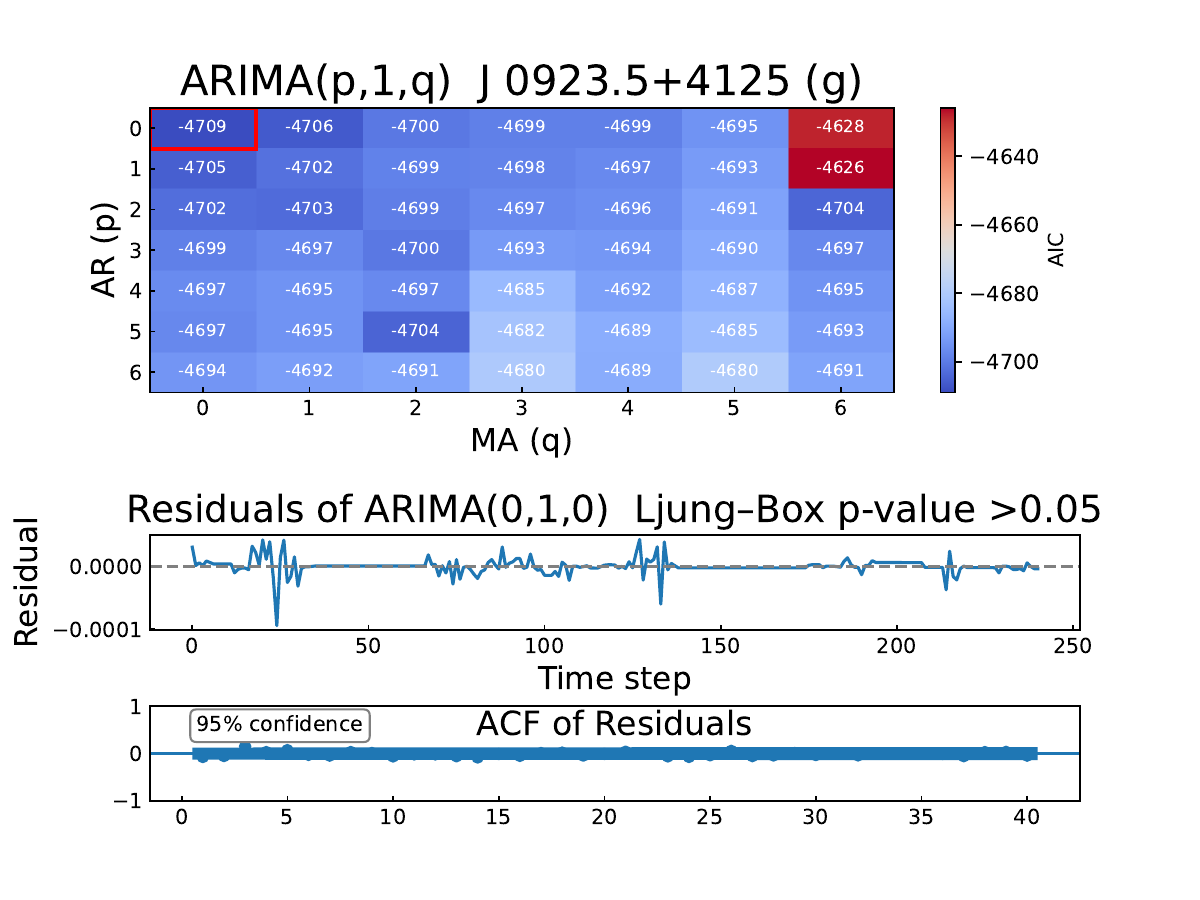}\hfill
	\includegraphics[width=0.5\textwidth,keepaspectratio]{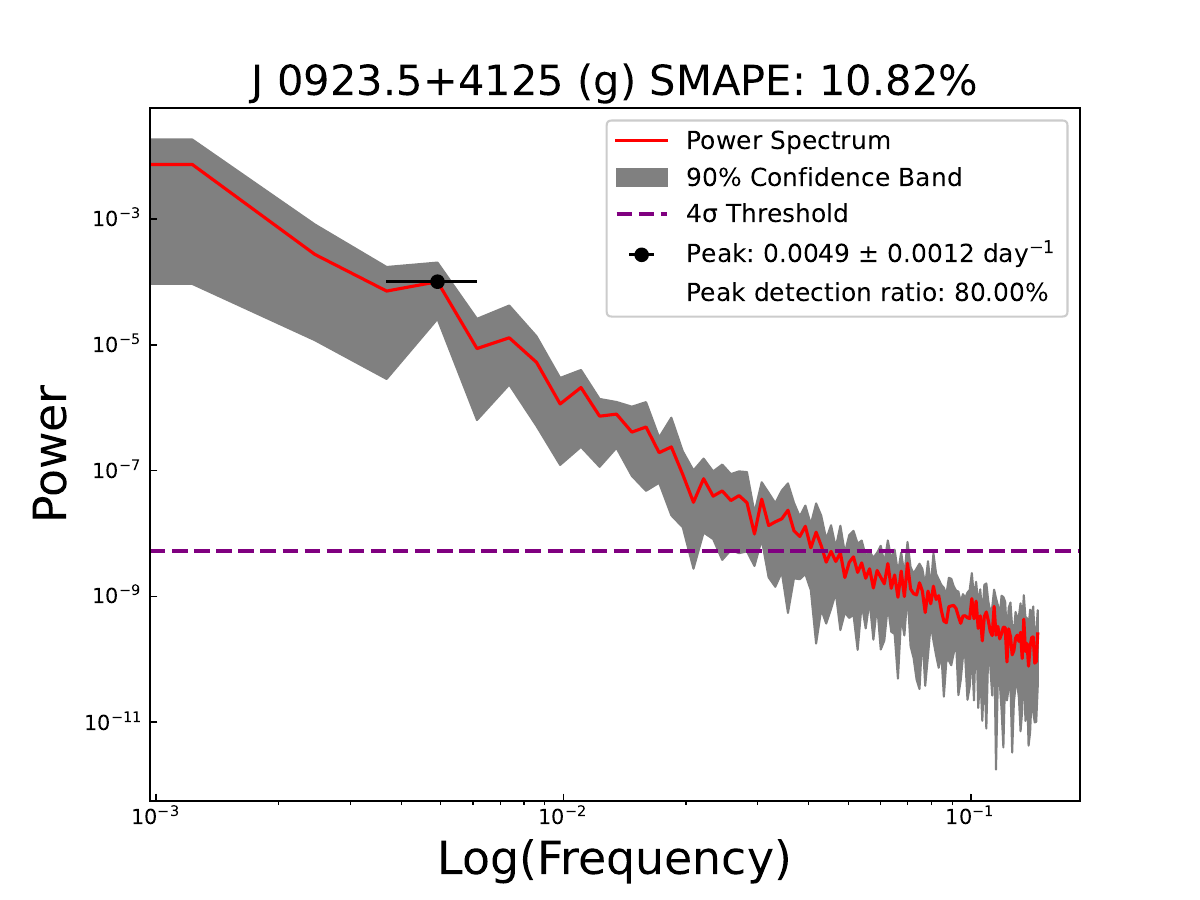}\\[0.5\baselineskip]
	\includegraphics[width=0.5\textwidth,keepaspectratio]{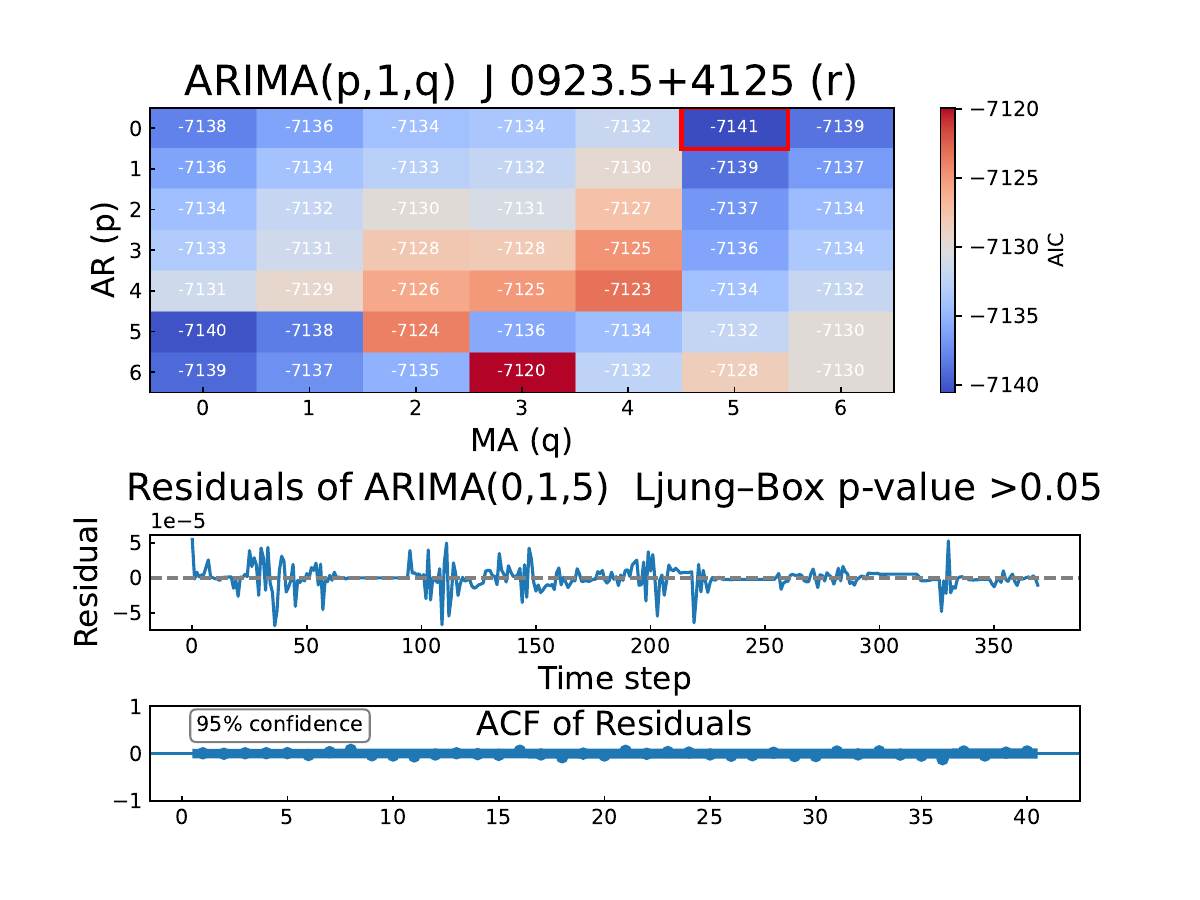}\hfill
	\includegraphics[width=0.5\textwidth,keepaspectratio]{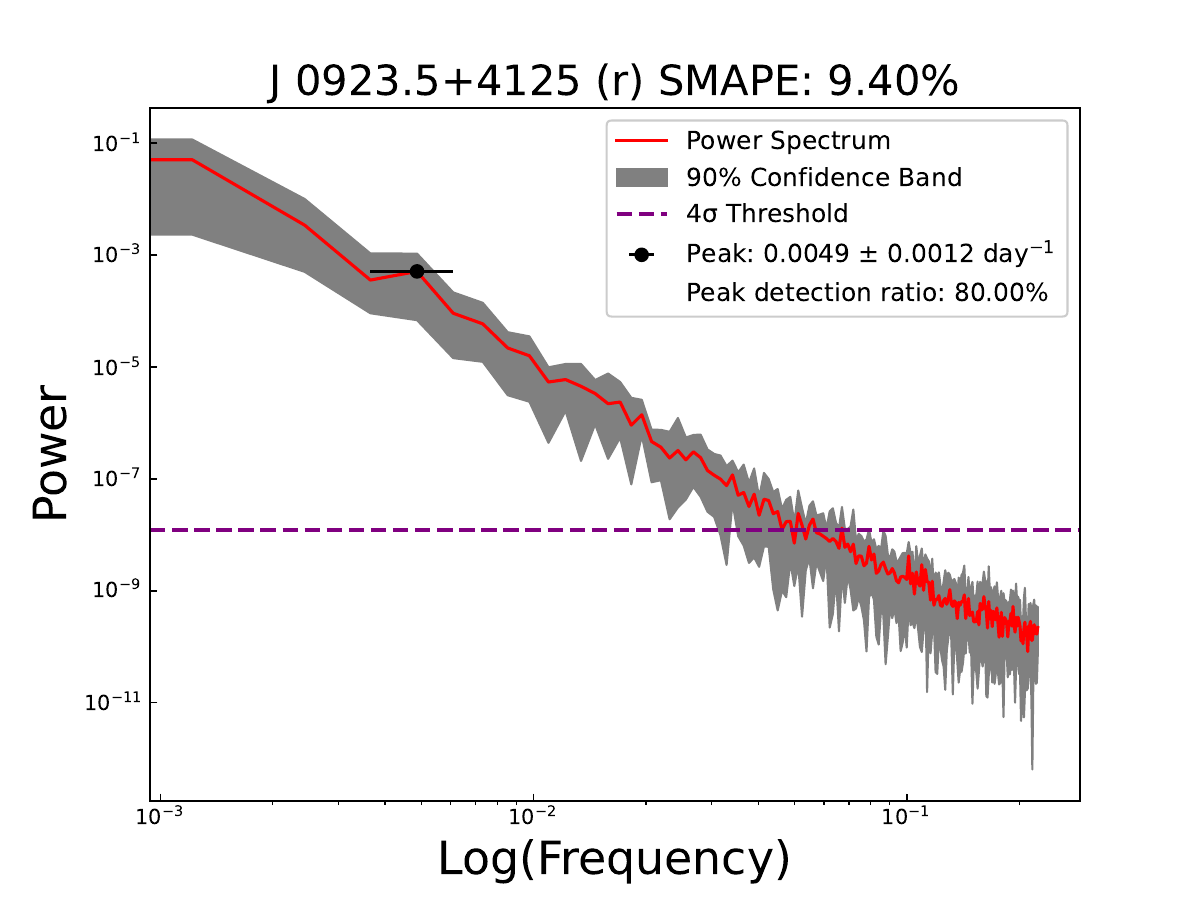}\\[0.5\baselineskip]
	\includegraphics[width=0.5\textwidth,keepaspectratio]{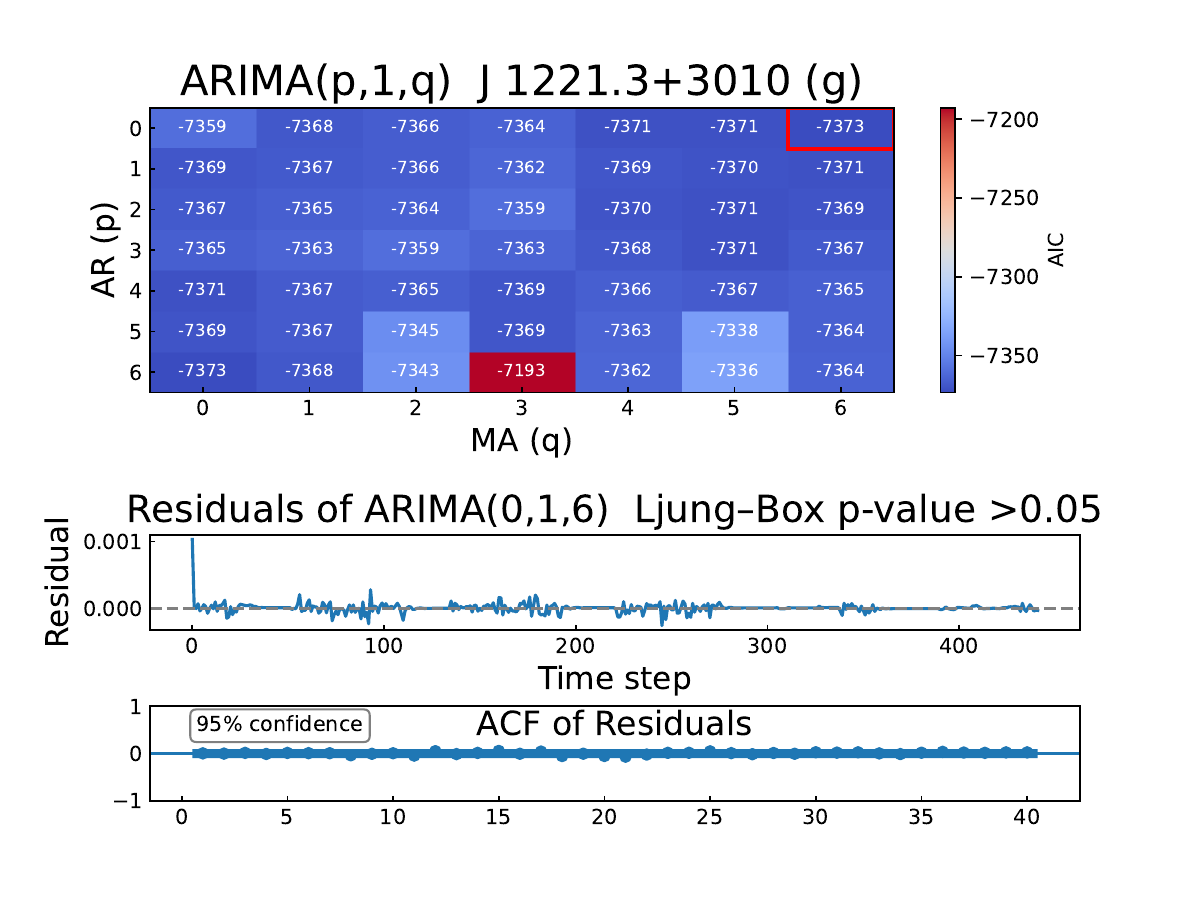}\hfill
	\includegraphics[width=0.5\textwidth,keepaspectratio]{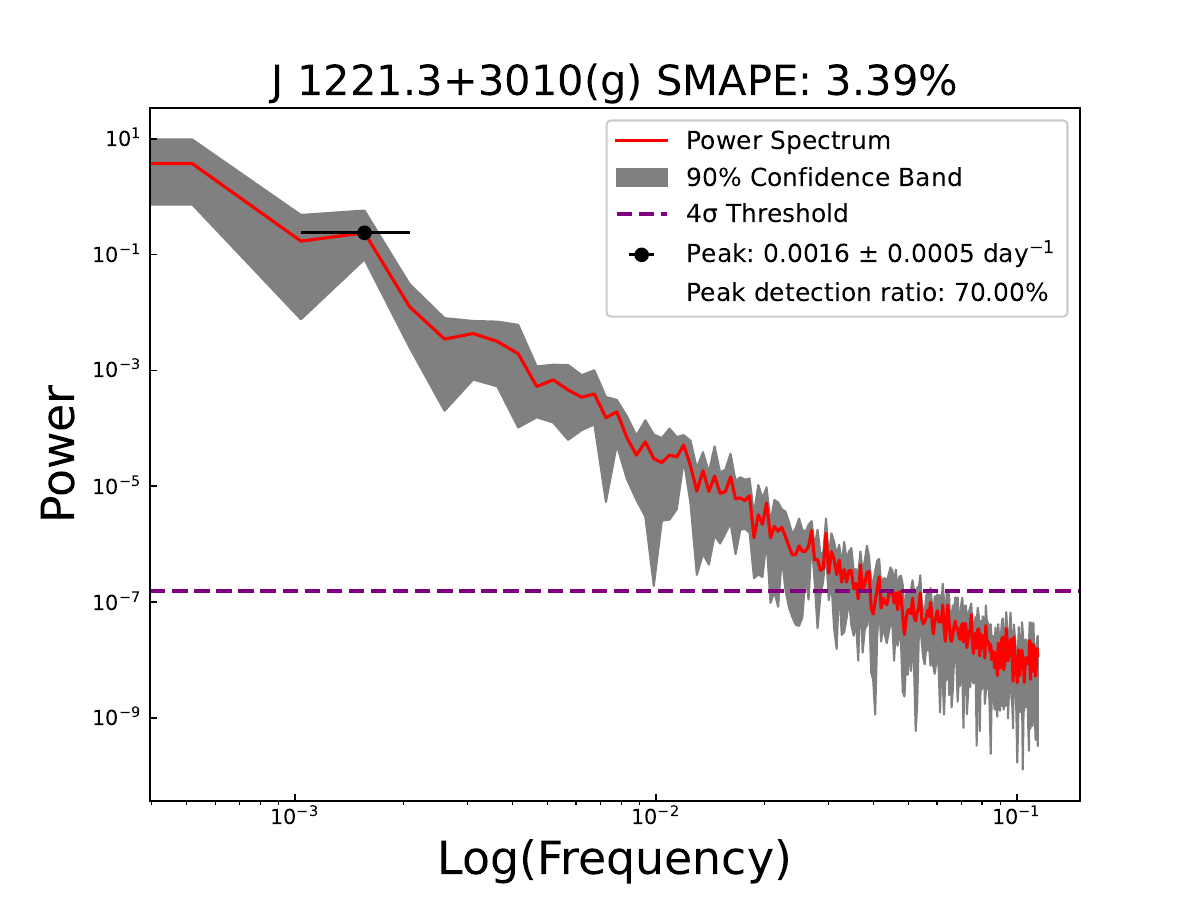}\\[0.5\baselineskip]
	\includegraphics[width=0.5\textwidth,keepaspectratio]{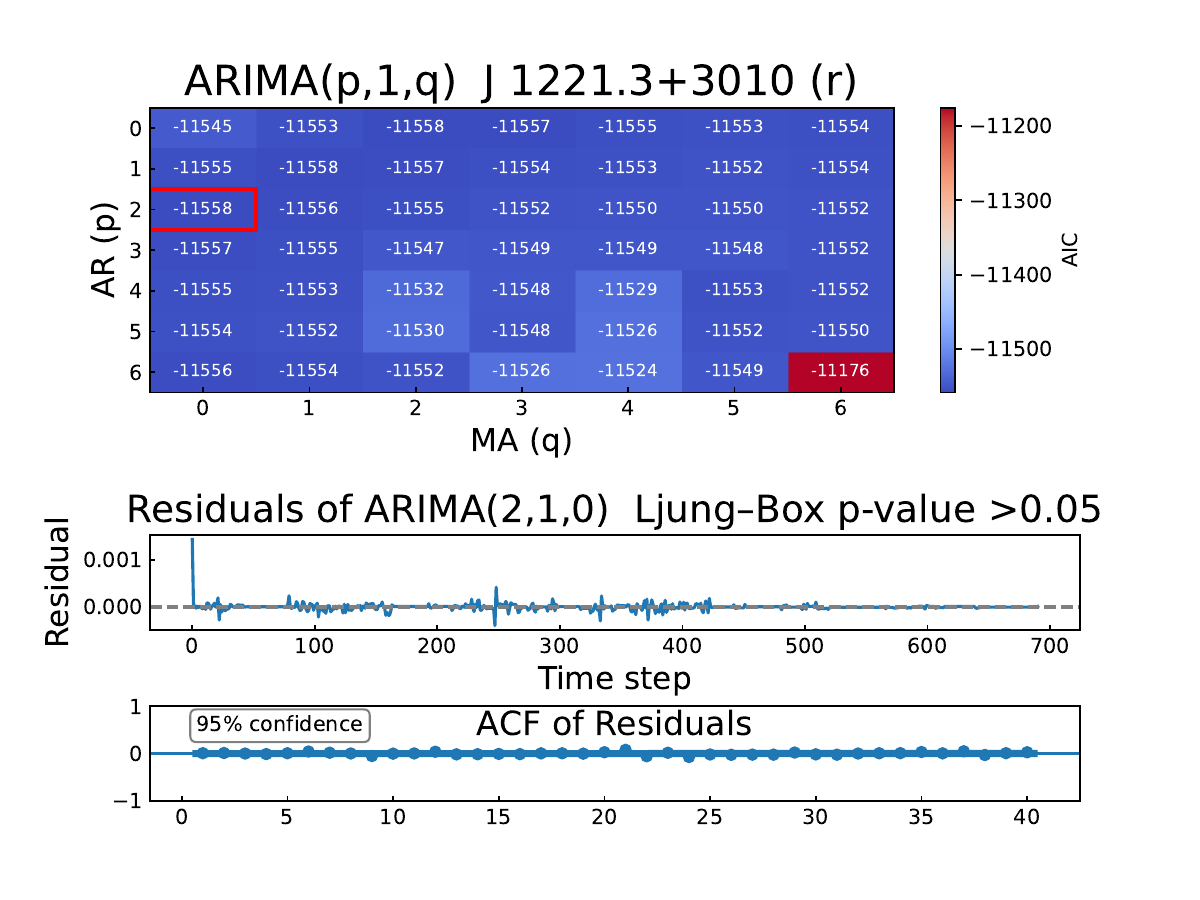}\hfill
	\includegraphics[width=0.5\textwidth,keepaspectratio]{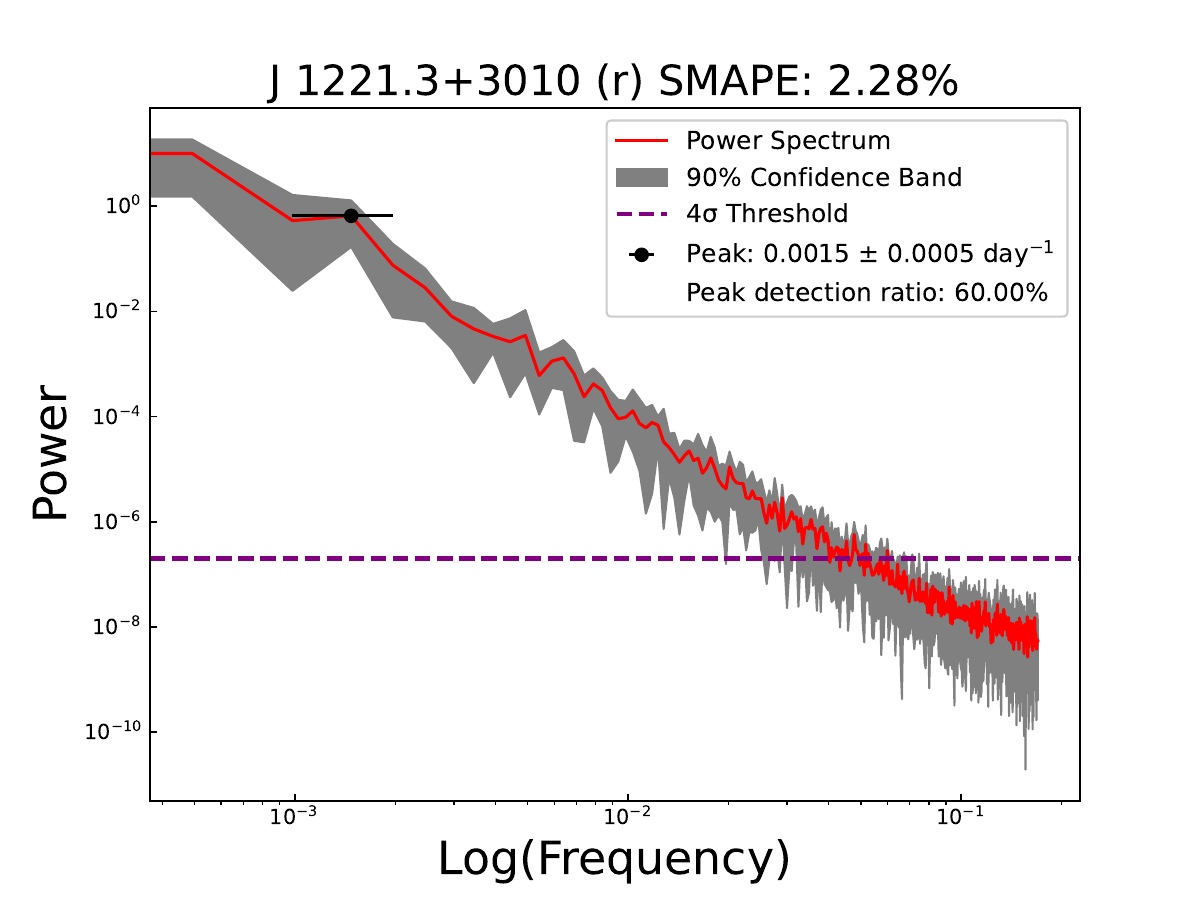}
	\caption{ARIMA model analysis of the sample sources for g- and r-bands.}
	\label{fig7}
\end{figure*}

\begin{figure*}\ContinuedFloat
	\centering
	\includegraphics[width=0.5\textwidth,keepaspectratio]{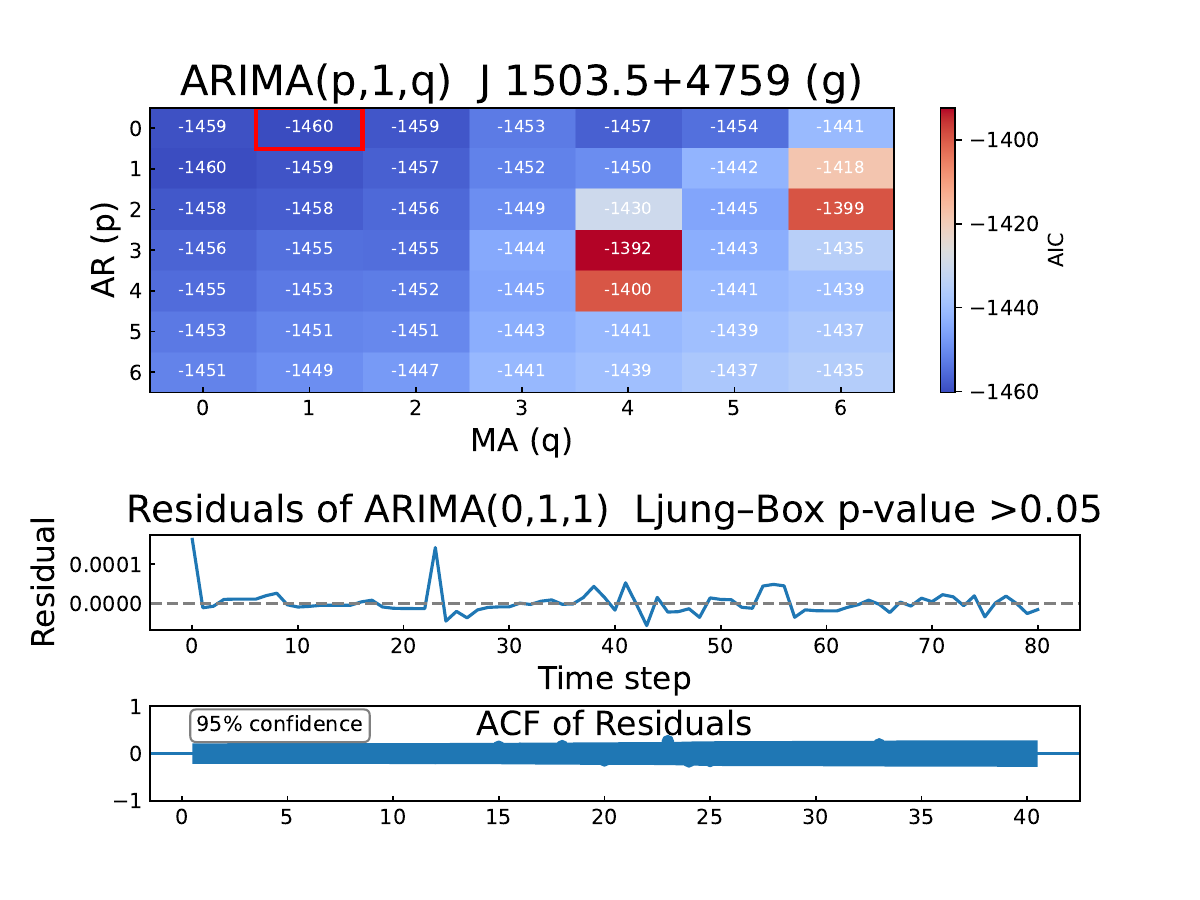}\hfill
	\includegraphics[width=0.5\textwidth,keepaspectratio]{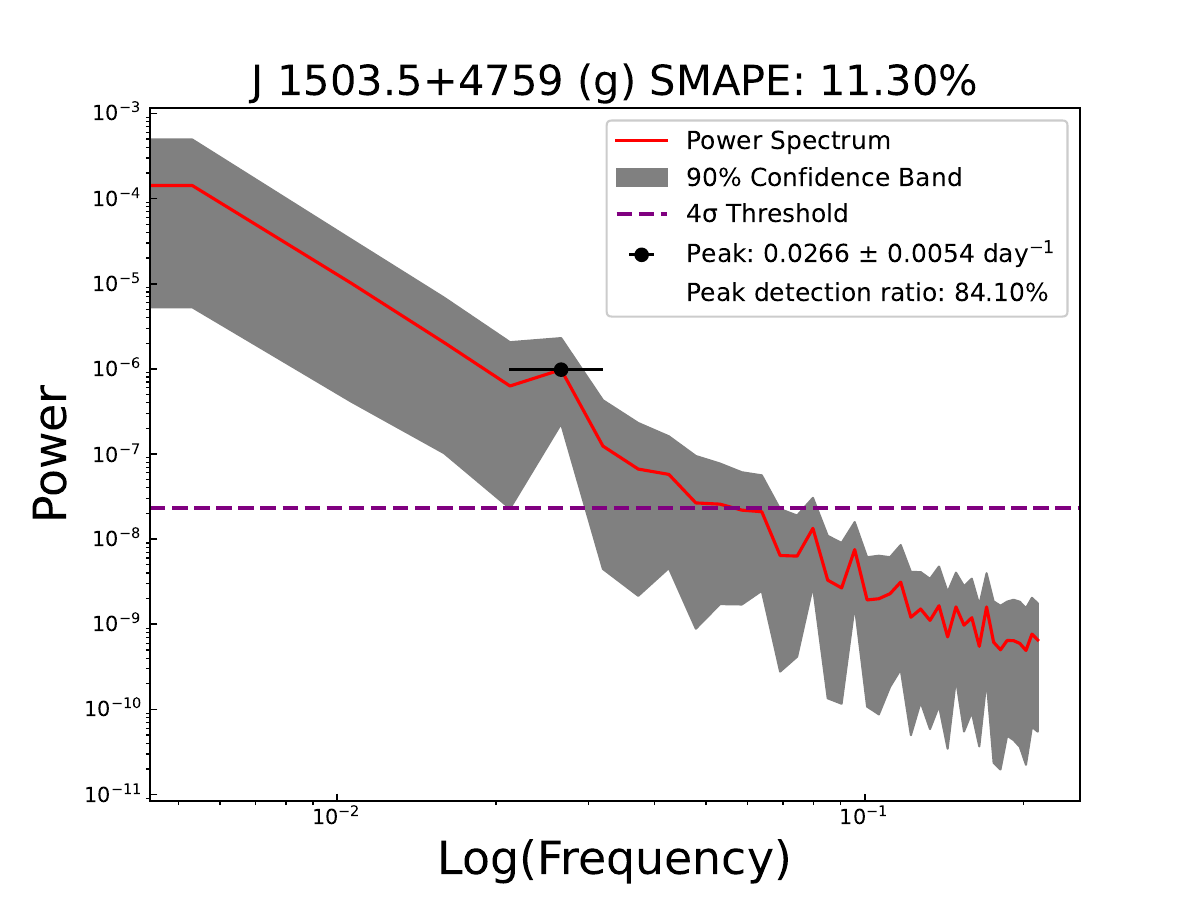}\\[0.5\baselineskip]
	\includegraphics[width=0.5\textwidth,keepaspectratio]{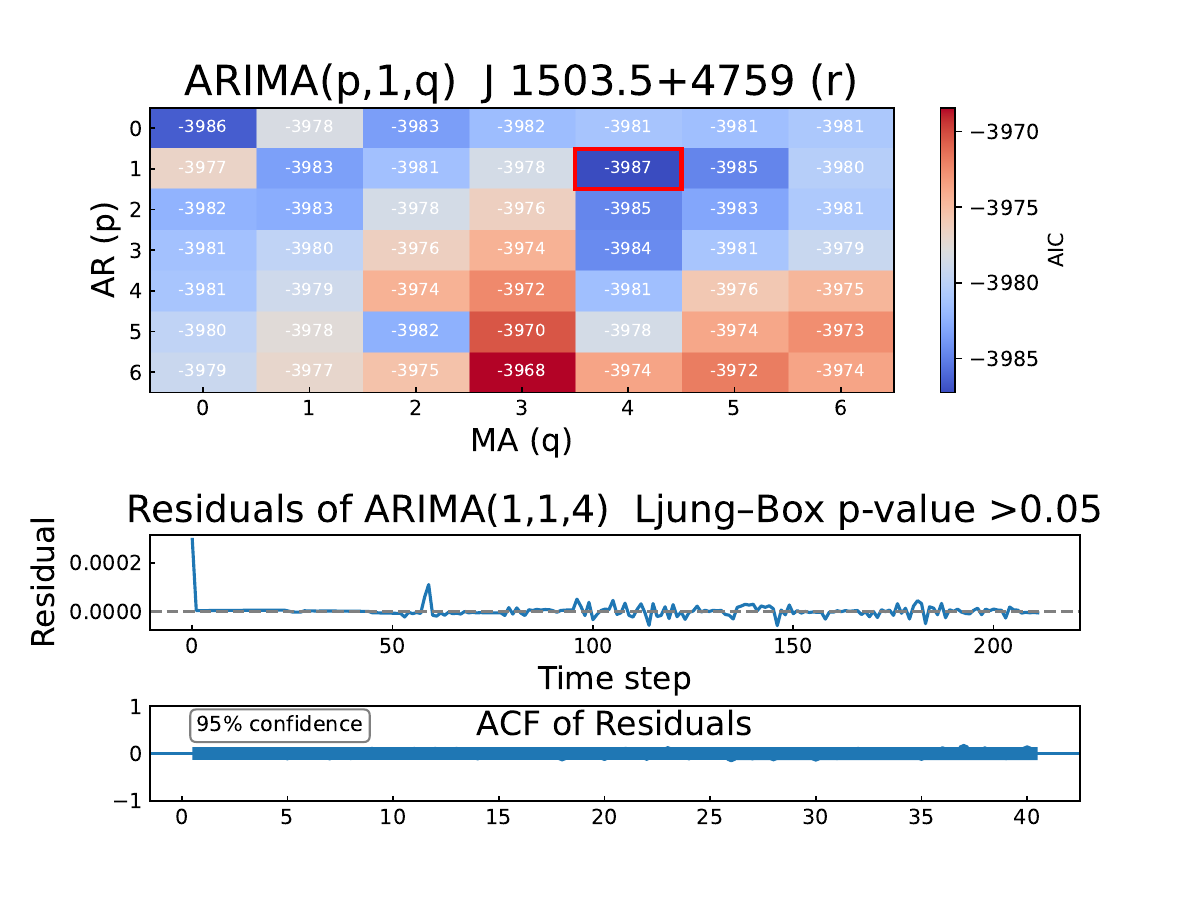}\hfill
	\includegraphics[width=0.5\textwidth,keepaspectratio]{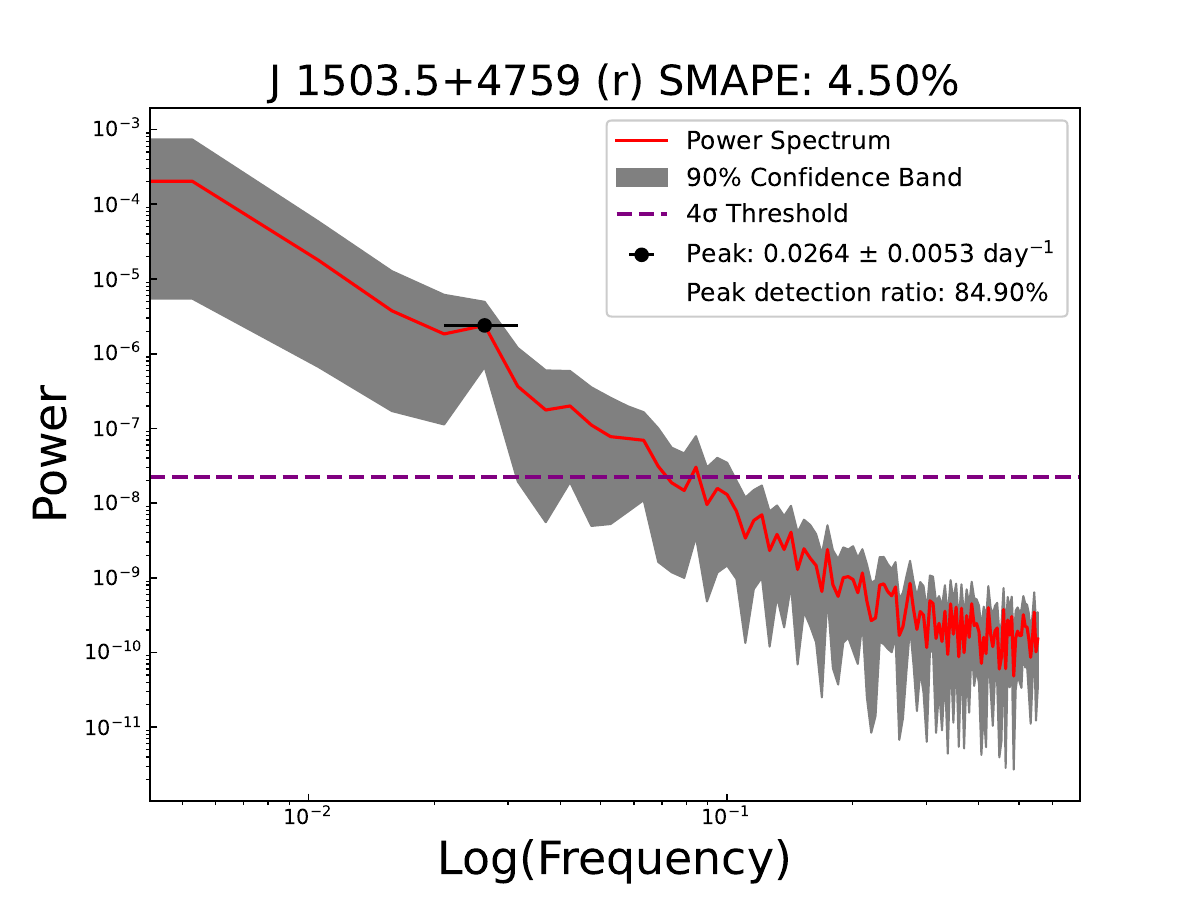}\\[0.5\baselineskip]
	\includegraphics[width=0.5\textwidth,keepaspectratio]{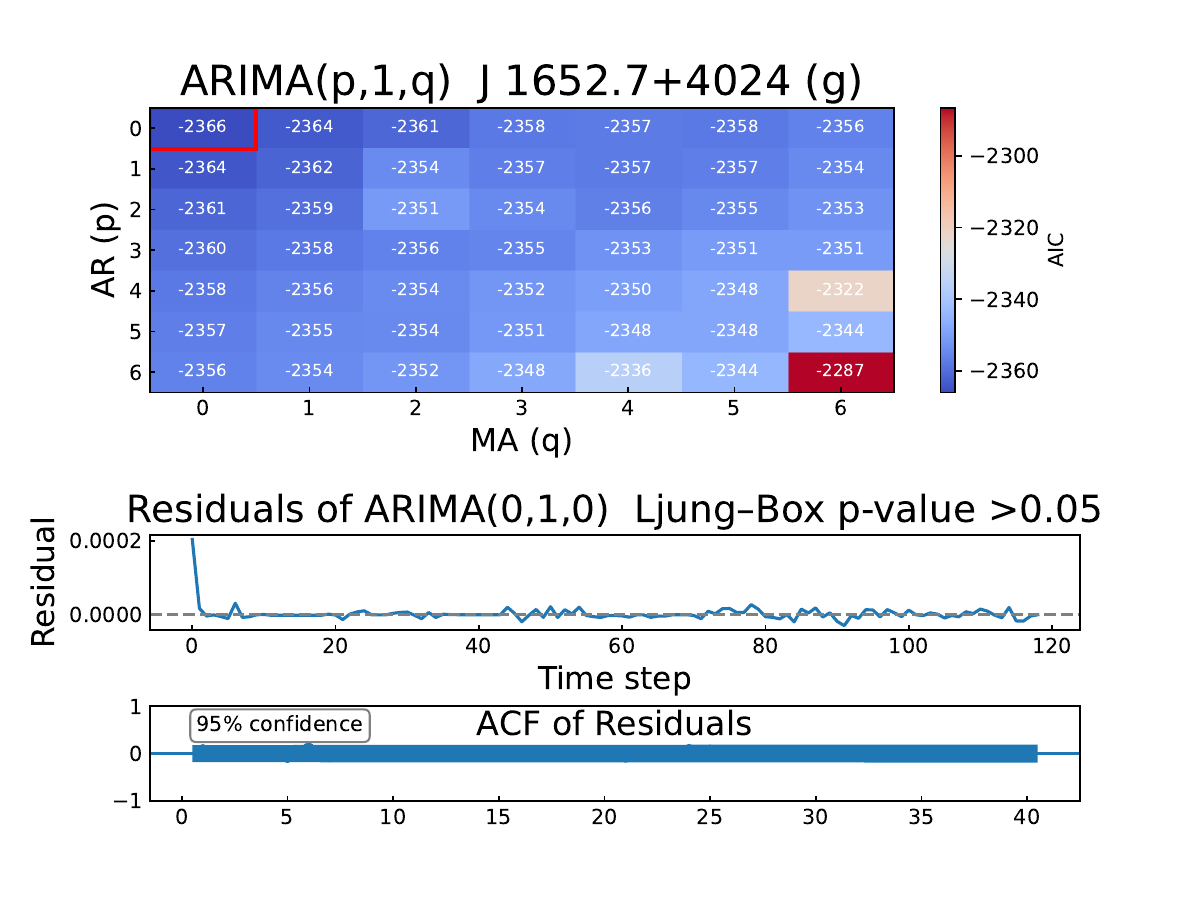}\hfill
	\includegraphics[width=0.5\textwidth,keepaspectratio]{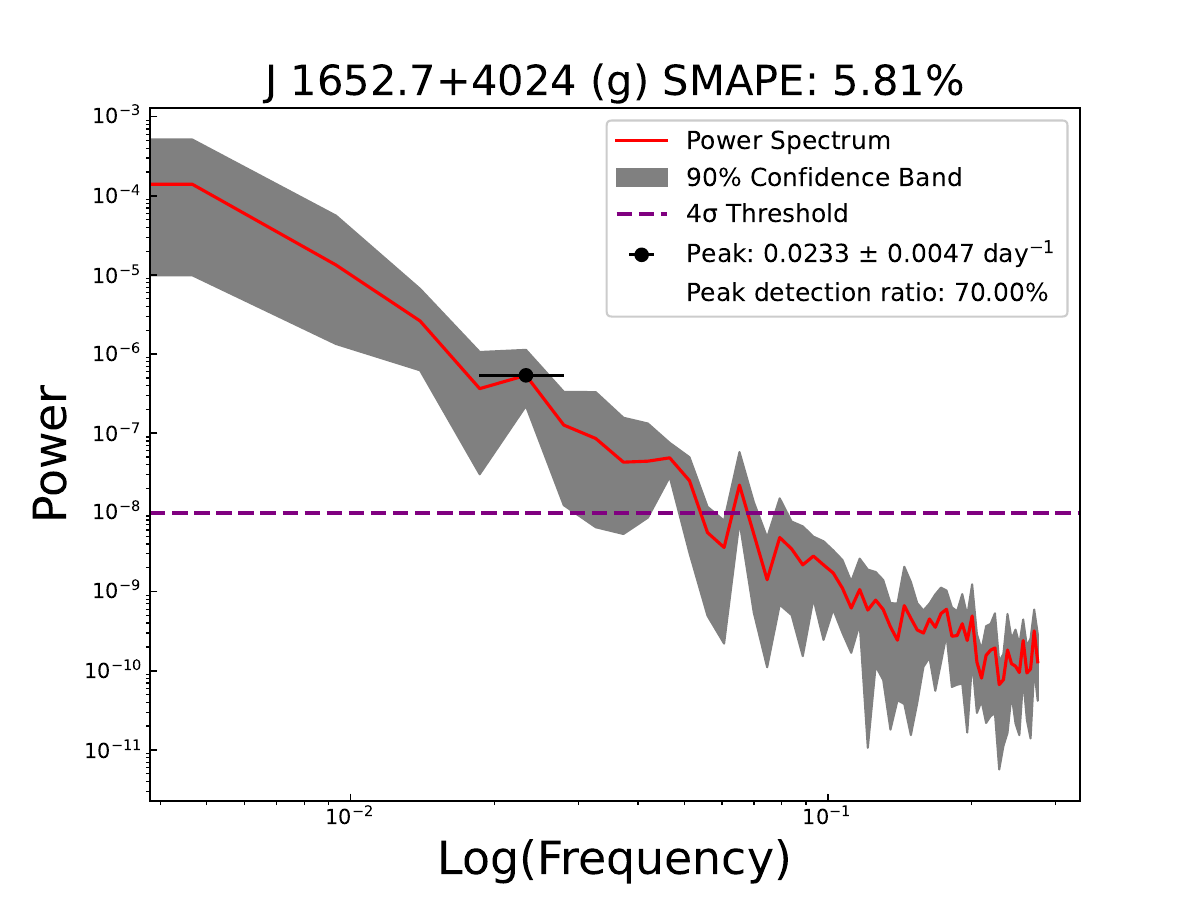}\\[0.5\baselineskip]
	\includegraphics[width=0.5\textwidth,keepaspectratio]{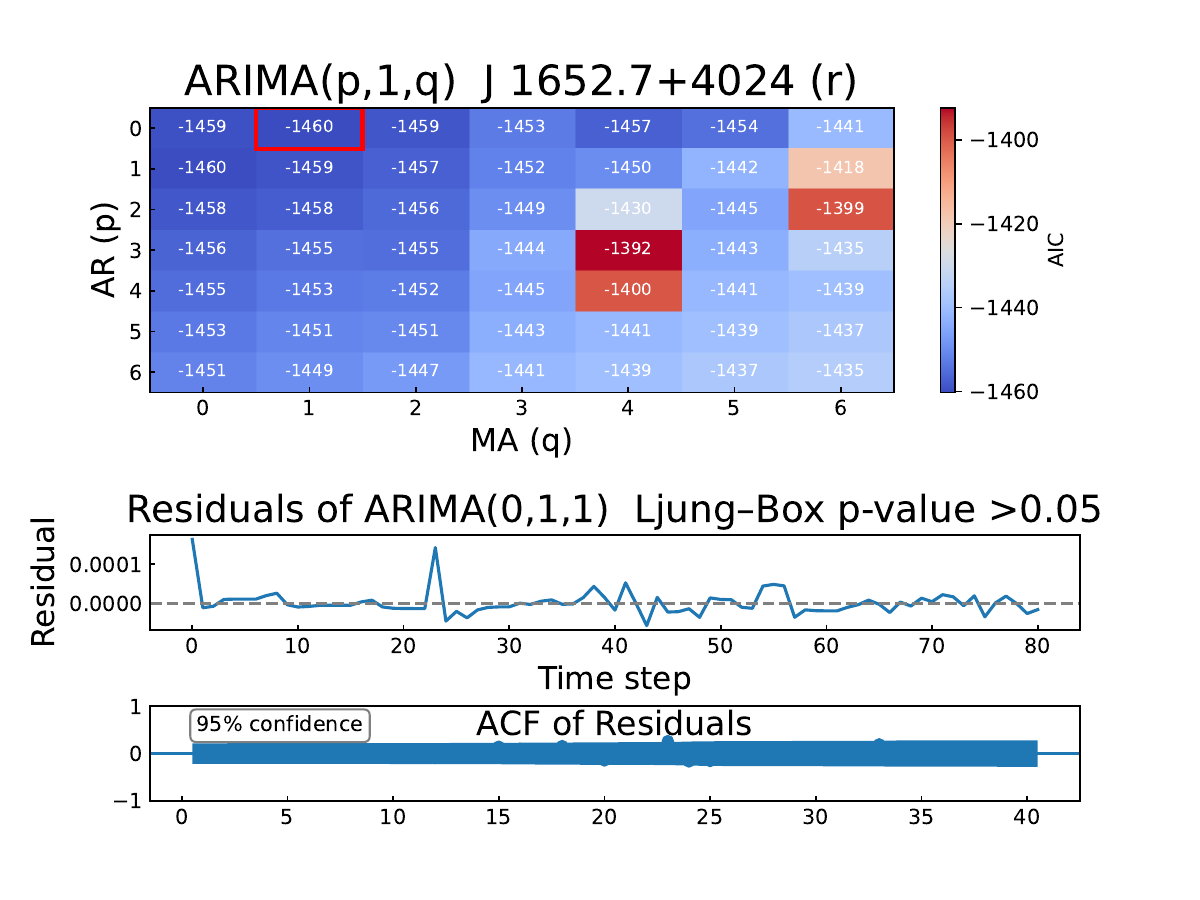}\hfill
	\includegraphics[width=0.5\textwidth,keepaspectratio]{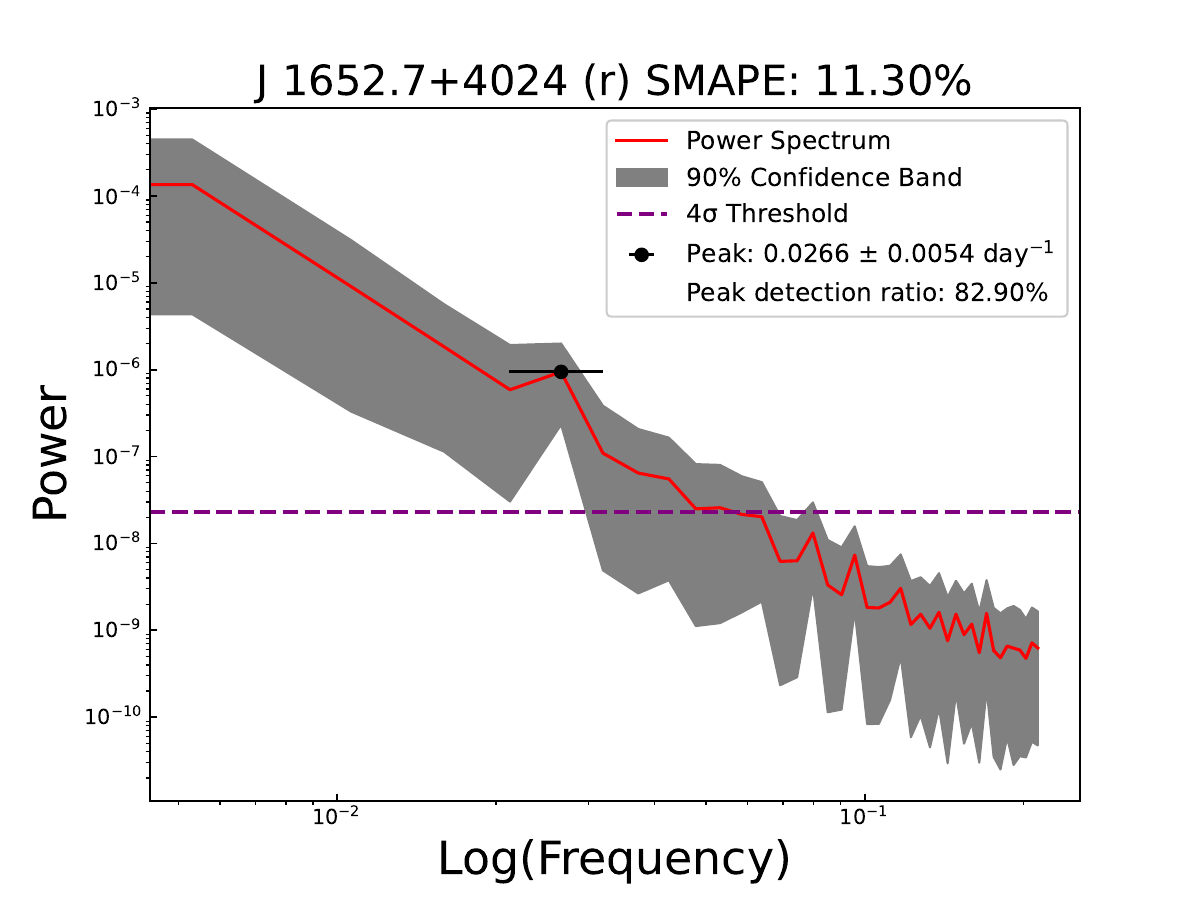}
	\caption{ARIMA model analysis of the sample sources for g- and r-bands. The first and third lists: top panels represent heatmaps of the ARIMA ($p$, $d$, $q$) parameter grid, with the red box indicating the lowest AIC value (optimal ARIMA model); middle panels show the residual time-series plots of the best-fit ARIMA model, annotated with the corresponding Ljung–Box test results; bottom panels are ACF of the residuals, where the blue shaded region represents the 95\% confidence interval. The second and fourth lists: results of the power spectral analysis of the residuals, in which the red curve represents the power spectrum, the gray shaded area indicates the 90\% confidence interval (obtained via MC simulations), and the purple dashed line denotes the confidence threshold (4$\sigma$) used to estimate the overall noise level. The ``peak detection ratio'' represents the percentage of MC simulated light curves whose PSD exhibits a peak at the reported frequency.}
\end{figure*}

\subsection{Approach to QPO Detection}

Here, we employ the weighted wavelet Z-transform (WWZ), the Lomb–Scargle Periodogram (LSP), and autoregressive integrated moving average (ARIMA) methods for QPO detection. The LSP is a renowned algorithm for detecting and characterizing periodicities in time series and has been widely applied in time-domain astronomy. This method efficiently computes the Fourier power spectrum of unevenly sampled observational data to identify potential periodic oscillations. In contrast, the WWZ method searches for oscillatory components across different time segments while simultaneously focusing on both time and frequency domains. Its scalable time window often renders it more sensitive and efficient than traditional Fourier transform techniques. To address the non-stationarity in light curves, we also employed an ARIMA model. This model is less effective at detecting periodic signals compared to the other two methods and is therefore used only as a simple verification tool. After interpolating the irregularly sampled data, we analyzed the residual power spectrum to detect periodic signals. The WWZ method converts the data into the time domain and frequency domain, and convolutes the light curve with the kernel related to time and frequency \citep{1996AJ....112.1709F}. The WWZ map is given by
\begin{equation}
	W[\omega, \tau: x(t)]=\omega ^{1/2} \int x(t)f^*[\omega(t-\tau)]dt,
\end{equation}
and $f$, i.e., the complex conjugate of $f^{*}$, is as follows
\begin{equation}
	f[\omega(t-\tau)]=\mathrm{exp}[i\omega(t-\tau)-c\omega^2(t-\tau)^2],
\end{equation}
where $\tau$ is the time-shift, and $\omega$ is the frequency. We adopted a frequency range of $10^{-4}$--$0.01~\mathrm{day^{-1}}$, which enables the detection of QPOs on time-scales from several months to years. A frequency step of $10^{-5}~\mathrm{day^{-1}}$ and a decay constant of $c=0.001$ were then employed to strike a balance between frequency and time resolution. Similar to other QPO analysis methods, the WWZ technique is adept at identifying potential periodic signals embedded within the original data. Notably, this method offers a unique advantage, i.e., during the analysis process, we only consider instances where power is concentrated across the entire duration of the observation period.

The LSP is one of the most widely used and effective techniques for periodicity searches, particularly due to its ability to handle irregularly sampled data, making it well suited for astronomical observations. The LSP window method facilitates the transformation of datasets into the frequency domain and quantifies the contribution of each frequency to the overall signal \citep{1976Ap&SS..39..447L,2018ApJS..236...16V}. For a set of irregularly sampled time-series data $x(t_{j}), ~j=1,~ 2,~ 3..., ~K$, the power spectral density can be represented as \citep{2008A&A...489..349D}

\begin{equation}
	\begin{aligned}
		P_{x}(\omega_{j})
		&= \frac{1}{2}\biggl\{
		\frac{\Bigl[\sum_{j=1}^{K} (x(t_{j})-\bar{x})\,
			\cos\!\bigl(\omega_{j}(t_{j}-\tau)\bigr)\Bigr]^2}
		{\sum_{j=1}^{K} \cos^{2}\!\bigl(\omega_{j}(t_{j}-\tau)\bigr)}
		\\[4pt]
		&\quad+\;
		\frac{\Bigl[\sum_{j=1}^{K} (x(t_{j})-\bar{x})\,
			\sin\!\bigl(\omega_{j}(t_{j}-\tau)\bigr)\Bigr]^2}
		{\sum_{j=1}^{K} \sin^{2}\!\bigl(\omega_{j}(t_{j}-\tau)\bigr)}
		\biggr\},
	\end{aligned}
\end{equation}
where $\bar{x}$ is the mean of the time series, $\omega_{j}=2\pi f_{j}$, $f_{j}$ is the frequency of the QPO, $P_{x}(\omega_{j})$ is the power spectrum, and $\tau$ is the time series phase correction, which can be obtained by

\begin{equation}
	\tan\!\bigl(2\omega_{j}\tau\bigr)
	= \frac{\sum_{j=1}^{K}\sin\!\bigl(2\omega_{j} t_{j}\bigr)}
	{\sum_{j=1}^{K}\cos\!\bigl(2\omega_{j} t_{j}\bigr)} \, .
\end{equation}

In Figure \ref{fig6}, we used the LSP and WWZ methods to analyze the light curves of sources in the g- and r-bands. The power spectrum color scale diagrams from the LSP and WWZ analyses are shown. Moreover, the significant periods that resulted from the LSP and WWZ methods are listed in Table \ref{tab3}. The uncertainties of the observed QPO periods were estimated by fitting a Gaussian profile to the power peak in the LSP or WWZ spectrum, and adopting the Half Width at Half Maximum (HWHM) as a measure of the period error. Because both LSP and WWZ can be affected by spectral leakage and aliasing under uneven sampling, we computed the spectral window function for each source to evaluate these effects \citep{2023MNRAS.526.5172B}. Specifically, we preserved the original observing time sequence but replaced all flux values with a constant unity, and then performed a Fourier transform of the resulting time series. The resulting peaks reflect only the periodicities induced by the temporal sampling, and the dominant one was identified as the spectral window peak period. To assess the significance of the detected QPOs, we generated $10^4$ white-noise light curves using the same time sampling, and applied LSP analysis to each. The dominant period in each realization was recorded and compared with the spectral window peak period. Among the four blazar samples, only J~1221.3+3010 showed a $\sim0.18\%$ probability that the simulated dominant period coincides with the $1\sigma$ range of the estimated QPO period, while in all other cases, the simulated peak periods lie outside the $1\sigma$ intervals. This indicates that the reported QPOs are unlikely to be artifacts caused by the temporal sampling.

\setlength{\tabcolsep}{3pt}
\begin{table}
	\caption{QPO behaviours of the four sample sources from g- and r-bands.}
	\noindent\hspace*{-8.5em}
	\label{tab3}
	\begin{tabular}{lccccc c ccccc}
		\hline\hline
		Name & \multicolumn{5}{c}{g-band} & & \multicolumn{5}{c}{r-band} \\
		\cline{2-6} \cline{8-12}
		& WWZ($\sigma$) & LSP($\sigma$) & $n$ & Local sig & Global sig &  & WWZ($\sigma$) & LSP($\sigma$) & $n$ & Local sig & Global sig \\
		\hline
		J 0923.5+4125 & 208$\pm30$(3) & 203$\pm18$(3) & 780  & 99.8650\% & 34.8641\% & & 206$\pm20$(4) & 205$\pm17$(4) & 930  & 99.9937\% & 94.3092\%  \\
		J 1221.3+3010 & 629$\pm67$(3) & 645$\pm58$(3) & 1990 & 99.8650\% & 6.7995\%  & & 602$\pm63$(3) & 598$\pm82$(3) & 1730 & 99.8650\% & 9.6610\%  \\
		J 1503.5+4759 & 39$\pm6$(4)  & 39$\pm5$(4)  & 130  & 99.9937\% & 99.1843\% & & 38$\pm7$(4)  & 38$\pm4$(4)  & 130  & 99.9937\% & 99.1843\% \\
		J 1652.7+4024 & 48$\pm4$(3.6)& 48$\pm4$(3.6)& 150  & 99.9841\% & 97.6430\% & & 48$\pm5$(4)  & 48$\pm4$(3.6)& 180  & 99.9889\% & 98.3487\%  \\
		\hline

	\end{tabular}
\end{table}

The periodicity in the radiation process can be revealed by modeling the light curve with a stochastic model \citep{2020A&A...642A.129}. Typically, blazar light curves are non-stationary, whereas the autoregressive moving-average (ARMA) model is applicable only to stationary time series. To address this issue, a more advanced ARIMA model is used. Before performing periodicity analysis using this model, the irregular sampling of astronomical data should first be addressed by interpolating the raw data. To construct the uniformly sampled series required by the ARIMA model, we generated an evenly spaced time grid with the same number of points as the original data and applied the linear interpolation (via the function \texttt{numpy.interp}) to both the flux and their uncertainties. This method assumes a linear variation of the flux between adjacent time points, resulting in a continuous and regularly sampled signal. This model can transform light curves through successive differencing \citep{2020ApJS..250....1T}. As shown in Figure \ref{fig7}, the optimal ARIMA configuration is obtained by selecting the parameter set that minimizes the Akaike Information Criterion (AIC), defined as
\begin{equation}
	{\rm AIC}=2k-2\ln(L),
\end{equation}
where $k$ is the number of model parameters, and $L$ is the maximum likelihood. Specifically, we explore all combinations of $(p, 1, q)$ (when $d=1$, the Augmented Dickey–Fuller test $p$-values for all sample sources are $\ll 1\times10^{-5}$, indicating that the data have already been rendered stationary. For each combination, we fit an ARIMA model and compute the corresponding AIC, with the optimal model corresponding to the minimum AIC. To determine the fit goodness of the optimal ARIMA model, we calculated the Symmetric Mean Absolute Percentage Error (SMAPE). The SMAPE is described by
\begin{equation}
	{\rm SMAPE} = \frac{2}{n} \sum_{i=1}^{n} \frac{|Y_{\text{true}, i} - Y_{\text{pred}, i}|}{|Y_{\text{true}, i}| + |Y_{\text{pred}, i}|} \times 100\%,
\end{equation}
where $Y_{\text{true}, i}$ is the true value and $Y_{\text{pred}, i}$ is the model prediction value. Furthermore, to verify whether the selected ARIMA model provides an adequate fit, we performed a residual analysis on the AIC-optimal ARIMA model. We extracted the residuals from the model and plotted their time series to examine whether they fluctuate around zero and whether there are any apparent trends or periodic structures. We also conducted the Ljung–Box test for each residual sequence to quantitatively assess autocorrelation \citep{1978Biometrika...65..297L}. A $p$-value $>0.05$ indicates that the residuals do not exhibit statistically significant autocorrelation, i.e., they are approximately white noise. Additionally, we plotted the autocorrelation function (ACF) of the residuals to visually assess their autocorrelation at each lag; if all coefficients lie within the 95\% confidence bands, the residuals can be considered approximately white noise. As shown in Figure \ref{fig7}, the residuals fluctuate around zero without any visible periodic structure, indicating that the ARIMA model sufficiently captures the underlying signal and leaves only white-noise residuals. The corresponding Ljung–Box test also yields $>$0.05. Therefore, we conclude that the best ARIMA model is providing a good fit to the data.

To account for the effects of differencing on the Power Spectral Density (PSD) analysis of the residuals, we applied a theoretical correction to the PSD obtained from the differenced time series. In the ARIMA, first-order differencing is defined as
\begin{equation}
	\Delta y(t) = y_{t}-y_{t-1},
\end{equation}
where $y$ denotes a time series. The Fast Fourier Transform (FFT) of a time series is given by
\begin{equation}
	Y(f)=\sum_{-\infty }^{+\infty } y_{t} e^{-i2\pi ft}.
\end{equation}
For the differenced series $\Delta y_{t}$, its FFT can be expressed as
\begin{equation}
	\Delta Y(f) = Y(f)\,\bigl(1 - e^{-i2\pi f}\bigr),
\end{equation}
where $1 - e^{-i2\pi f}$ represents the frequency response function of the differencing operator. Since the PSD is given by the squared magnitude of the FFT, the corresponding power response (i.e., the correction factor) is
\begin{equation}
	\lvert H(f)\rvert^2 = 2[1 - \cos(2\pi f)].
\end{equation}

By dividing the PSD by this correction factor, the subsequent PSD analysis more accurately reflects the true characteristics of the original signal. Additionally, to account for the effects of non-uniform sampling, measurement errors, and interpolation on the analysis of the residual PSD, Monte Carlo (MC) simulations (with 1000 iterations) were employed to incorporate these uncertainties. The procedure is as follows: First, the ARIMA model is fitted by the MLE method, yielding the estimates of the autoregressive coefficients, moving average coefficients, and residual variance along with their uncertainties (the square roots of the diagonal elements of the information matrix). Subsequently, these uncertainties are used to generate new ARIMA parameters and noise distributions, thereby producing a uniformly sampled simulated time series with similar characteristics. Specifically, the uncertainties are employed to perform parameter sampling via a normal distribution to obtain new model parameters, which are then used to construct a differenced sequence that is cumulatively summed to restore the complete simulated light curve, ultimately resulting in the simulated uniformly sampled time series. Second, samples equivalent in number to the original observations were randomly drawn from the uniformly simulated series to mimic the actual non-uniform sampling. Third, Gaussian noise derived from the original measurement errors was added to these samples, along with an additional 5\% interpolation perturbation to simulate interpolation errors. Fourth, the resampled data were linearly interpolated to reconstruct a uniformly sampled time series, thereby enabling the computation of residuals by comparing the interpolated data to the ARIMA model predictions. Fifth, the residuals were transformed into the frequency domain by FFT to compute the PSD. As shown in Figure \ref{fig7}, for each simulation the lower and upper bounds of the 90\% confidence interval at each frequency point were defined using the 5th and 95th percentiles (shaded region). Simultaneously, in the PSD obtained from the FFT of the residuals, a 4$\sigma$ confidence line was drawn under the white-noise assumption, where the threshold was derived from the upper quantile of the $\chi^2$ distribution. Finally, the legend indicates the percentage of the 1000 MC light curves in which a PSD peak was observed at the reported frequency. The ARIMA residual analysis is \emph{consistent} with a periodic component, but given interpolation and modeling assumptions, we regard this as supportive. Overall, a power spectrum analysis of the ARIMA model residuals, as shown in Figure \ref{fig7}, is performed to detect periodic signals not explained by the model, consistent with analyses conducted using the other two methods. The interpolation, random resampling, and differencing steps in this method may render the results more conservative, but they also ensure the robustness of signal detection against noise and sampling effects.

\subsection{Local Significance Estimation and Results}\label{result}

In light curves, detected QPOs may not necessarily originate from genuine physical processes. Statistical analyses indicate that the light curves of AGNs typically exhibit red-noise characteristics, which can give rise to apparent periodic signals. Moreover, observational data are often affected by factors such as weather and seasonal variations, leading to irregular sampling that may result in spurious QPO detections. Therefore, the local confidence level is computed to indirectly test the null hypothesis that the detected QPO is produced by red noise. To estimate the local significance level of the PSD peaks, we employed the following method of simulating $10^4$ light curves. This approach consists of three main steps: First, computing the PSD of the original light curve and obtaining the model parameters using the MLE; Second, based on these model parameters, we simulate light curves using the MC method, with the original data's time series as the foundation. The sampling cadence of the simulated light curves is consistent with that of the original light curve, and measurement errors are not taken into account; Finally, the PSD is computed for each simulated light curve, and the distribution of PSD data points is assessed at each frequency to construct a significance curve, thereby evaluating the authenticity of the periodic signals. Blazars typically exhibit variability characteristics of red noise type in the time-frequency domain. When simulating the PSD of light curves, three different red noise models are commonly employed:

\noindent (i) the bending power law
\begin{equation}
	P(f)=\frac{Af^{-\alpha}}{1+(f/f_{bend})^{\beta - \alpha}} +C,
\end{equation}
(ii) the power law
\begin{equation}
	P(f)=Af^{-\alpha} +C,
\end{equation}
and (iii) the broken power law
\begin{equation}
	P(f) = A \begin{cases}
		\displaystyle \left(\frac{f}{f_{bend}}\right)^{-\alpha} + C,
		& f \leq f_{bend} \\[6pt]
		\displaystyle \left(\frac{f}{f_{bend}}\right)^{-\beta} + C,
		& f_{bend} < f \leq f_{\max},
	\end{cases}
\end{equation}
where the model parameters $A$, $\alpha$, $f_{bend}$, $\beta$, and $C$ are the normalization, low-frequency slope, bend frequency, high-frequency slope, and Poisson noise, respectively \citep{2013MNRAS.433..907E}. In general, the pure power-law model does not adequately characterize the PSD at both low and high frequencies. Therefore, the bending power-law model is adopted as the preferred fitting model \citep{2012A&A...544A..80G}. The AIC values for the three PSD models are listed in Table \ref{tab4}. The results indicate that, in most cases, the bending power law model appears to offer a more appropriate fit. To fit the PSD of the light curve, we adopted the bending power-law model and performed the fitting using the DELightcurve Simulation code\footnote{\url{https://github.com/samconnolly/DELightcurveSimulation}}, a widely recognized method \citep{2023ApJ...943..157L, 2024ApJ...976...51G}. The fitting process utilized the Basin-Hopping algorithm in combination with the Nelder-Mead minimization algorithm, both implemented in the Python package SciPy. The fitting method is based on minimizing the objective function derived from the MLE. This approach ensures robust parameter estimation and improves the reliability of PSD characterization.

\begin{table*}
	\centering
	\caption{AIC results for the three model of power law density.}
	\label{tab4}
	\begin{tabular}{lccc}
		\toprule
		4FGL Name (band) & Bending Power Law & Broken Power Law & Power Law \\
		\midrule
		J 0923.5+4125(g)  & -924.11 & -922.70 & -897.73\\
		J 0923.5+4125(r)  & -988.85 & -952.17 & -955.99\\
		J 1221.3+3010(g)  & -1739.36 & -1661.21 & -1671.90\\
		J 1221.3+3010(r)  & -3145.45 & -2994.99 & -3006.11\\
		J 1503.5+4759(g)  & -502.83 & -499.15 & -503.22\\
		J 1503.5+4759(r)  & -1413.47 & -1386.27 & -1400.95\\
		J 1652.7+4024(g)  & -902.06 &  -890.50 & -899.28\\
		J 1652.7+4024(r)  & -963.18 &  -937.94 & -947.98\\
		\bottomrule
	\end{tabular}
\end{table*}

In periodicity analysis, it is crucial to consider both local and global significance to ensure the scientific robustness and reliability of the results. Local significance refers to the periodicity detected at a single frequency in the light curve, while global significance accounts for corrections across the entire frequency range, considering all independent trials. This correction is designed to address the ``look-elsewhere effect'' \citep{2010EPJC...70..525G}, which compares the probability of observing an excess at a fixed frequency to the probability of observing an excess across the entire analyzed frequency range. Without such correction, multiple trials can significantly increase the false positive rate, leading to misleading conclusions. Furthermore, $p_{g}$ can be used to test the null hypothesis that the detected QPO is produced by red noise. The relationship between these two can be expressed as \citep{2022arXiv221101894P}
\begin{equation}
	p_{g}=1-(1-p_{l})^{n},
\end{equation}
where $1-p_g$ represents the global significance, $1-p_l$ is the local significance, and $n$ is the test factor. Generally, since we do not know which sources will exhibit periodic behavior and the prior frequency for each source, $n$ should include the following information: sample size and the number of independent frequencies in the periodogram. In this work, the sources we studied all exhibit certain periodic behavior (see section \ref{Data Source and Sample Composition}). Therefore, the value of $n$ equals the number of independent frequencies. The independent frequencies $n$ can be calculated by $m(f_{max}-f_{min})/\delta f$  \citep{2018ApJS..236...16V}, where $m$ is the number of samples ($m=10$). For non-uniformly sampled time series,  we approximately set $f_{max}\approx 1/(a$ days) and $f_{min}\approx 1/(b$ days), where $a$ is close to the data collection pace ($\sim$10), $b$ is half the total span of the light curve data, and $\delta f\approx 1/2b$ is the frequency resolution determined by the length of the light curve data. Through the afore mentioned procedure, we place the calculation results including the values of QPO, local significance, test factor, and global significance into Table \ref{tab3}. The magnitude of the global confidence level may be dependent on the sample size, so global confidence levels cannot be compared in absolute terms but must be evaluated in conjunction with the number of samples. The probability that none of the 10 samples exhibit a false positive is given by $(1-p_l)^n$, resulting in a global confidence level of 99.18$\%$.  The global $p$-value $p_{g} \sim0.008$ ($<0.05$) indicates the rejection of the null hypothesis that the detected QPO is produced by red noise. Thus, from a statistical perspective, in these 10 samples an average of only 0.08 sources are expected to be misclassified as true signals (i.e., false positives), and since the expected number is less than 1, the detection method is highly reliable. Consequently, the $\sim$48-day feature in J 1652.7+4024 remains a candidate periodicity. In the cases of J 0923.5+4125 and J 1221.3+3010, the lower global confidence levels may be attributed to an excessively large frequency resolution $\delta f$ in their long light curves (ranging from 800 to 2000 days).

\section{Discussion}\label{Discussion}

\subsection{Variability Properties}

$F_{\rm var}$ is employed to quantify the observed variability in the analysis of variability. Higher $F_{\rm var}$ values typically indicate more intense variability activity. The values listed in Table \ref{tab1} suggest that blazar sources exhibit significant activity in the optical band. Many physical models have been proposed to explain variability in blazars, which can be broadly classified into external and internal mechanisms. The external mechanisms include interstellar scintillation and microlensing effects, while the internal mechanisms involve relativistic jet activities and instabilities in the accretion disk \citep{1995ARA&A..33..163W}. Optical intraday and short-term variability is generally dominated by the internal processes, as the synchrotron emission from the relativistic jet overwhelms the thermal radiation from the accretion disk. In this context, the “shock-in-jet” model is widely adopted to interpret optical variability in blazars \citep{2008Natur..452..966M, 1985ApJ...298..114M}. Furthermore, small variability may arise from turbulence behind shock or localized perturbations in the jet \citep{2015MNRAS.450..541A}. When the source is in a low state, instabilities in the accretion disk may also contribute to the observed variability, since the jet emission is relatively weak under such conditions \citep{1993ApJ...406..420M}. The CCF analysis indicates that the g and r-bands emission are co-spatial. Moreover, apart from J 1221.3+3010, which exhibits the RWB behavior, that is, brightening accompanied by spectral softening, the remaining three sources display BWB trend indicative of spectral hardening. Statistically, the BWB trend is more prominent in BL Lacs, whereas the RWB trend occurs more frequently in FSRQs. The shock-in-jet model has been widely applied to interpret the BWB behavior observed in BL Lacs \citep{2015MNRAS.450..541A, 2022MNRAS.510.1791N, 2023MNRAS.520.4118C}. In this framework, a shock propagating down the relativistic jet encounters regions with enhanced electron density and produces synchrotron radiation at different wavelengths from various distances downstream of the shock front. Because higher-energy photons are emitted earlier and closer to the shock front, time-dependent color variations naturally arise \citep{2015MNRAS.450..541A}. Negi et al. (2022) reported that many RWB BL Lacs are significantly affected by host-galaxy contamination \citep{2022MNRAS.510.1791N}; thus, the RWB behavior detected in J 1221.3+3010 (BL Lac) may be partially influenced by its host galaxy. Negi et al. (2022) also found that brighter FSRQs tend to exhibit the BWB trend, while fainter ones more frequently display RWB behavior \citep{2022MNRAS.510.1791N}.

\subsection{RMS-Flux relation and flux distribution}

The RMS-Flux relation is commonly observed in astrophysical systems. A linear RMS–Flux relationship, along with a log-normal flux distribution, has been documented in various blazars across multiple frequency bands, including X-ray \citep{2009A&A...503..797G}, optical \citep{2021ApJ...923..7}, and $\gamma$-ray \citep{2020ApJ...891..120}. This linear relationship may suggest underlying nonlinear processes that drive the observed variability, leading to a flux distribution skewed towards higher values, similar to a log-normal PDF \citep{2005MNRAS..359..345}. In the $\gamma$-ray band, Wang et al. (2023) discovered that J 0923.5+4125, J 1221.3+3010, and J 1503.5+4759 exhibit a strong leading RMS–Flux relation, and the flux distribution tends towards a log-normal distribution \citep{2023RAA..23..115011} . To examine the flux distribution in the optical light curves of the four blazars in our sample, both normal and log-normal PDFs were fitted to the flux histograms. The fitting results and $\Delta$ BIC indicate that the observed flux histograms align more closely with a log-normal distribution, showing a heavy tail skewed toward higher flux.

In accretion disk models, the log-normal flux distribution and linear RMS–Flux relation may be associated with viscosity fluctuations at different radii that propagate outward, influencing mass accretion rates on larger scales \citep{1997MNRAS.292..679L}. 
For a standard Shakura–Sunyaev thin disk, the viscous timescale can be estimated as \citep{2010MNRAS.402.2087V}
\[
t_{\rm vis} \approx \alpha^{-1} \left( \frac{H}{R} \right)^{-2} t_{\rm dyn},
\]
where \( t_{\rm dyn} = \left( R^{3} / GM \right)^{1/2} \) is the Keplerian dynamical timescale, and \( H/R \) is the disk aspect ratio. 
If we express the radius as \( R = r R_{\rm g} \), where \( R_{\rm g} = GM / c^{2} \) is the gravitational radius, 
then the viscous timescale can be written as
\begin{equation}
	t_{\rm vis} \approx 4.93 \times 10^{-6}\ {\rm s}\,
	\left( \frac{M}{M_\odot} \right) r^{3/2}\, \alpha^{-1}\, \left( \frac{H}{R} \right)^{-2},
	\label{eq:tvis}
\end{equation}
where \( M \) is the black hole mass, \( r \) is the radius normalized to the gravitational radius, and \( \alpha \) is the dimensionless viscosity parameter. Adopting typical blazar parameters,
\(M_{\rm BH}=10^{8}\,M_\odot\), \(\alpha=0.1\), \(H/R=0.01\),
and \(r=10^{2}\), Eq.~(\ref{eq:tvis}) yields a viscous timescale of
\(t_{\rm vis}\sim10^{3}\ {\rm yr}\). It is longer than the observed optical variability, indicating that viscous disk fluctuations cannot directly drive the optical variability. Instead, the variability likely originates in the relativistic jet. In relativistic jet models, highly skewed flux distributions may arise in a ``mini-jets in a jet'' scenario, where jets dominated by Poynting flux create conditions for the production of isotropically distributed mini-jets \citep{2009MNRAS.395L..29G}. Emission from identical, independent, but randomly oriented mini-jets follows a Pareto distribution, leading to both normal and log-normal distributions while maintaining a linear RMS–Flux relation \citep{2012AA..548..A123}. Moreover, we found that the double-peaked structure of J 0923.5+4125 (r-band) persists under different binning schemes, indicating that it is not an artifact of the binning method. This bimodality has previously been interpreted as the superposition of a quiescent state and a flaring state \citep{2017MNRAS.467.4565L}.

\subsection{QPOs}

We have identified QPOs candidates in both the g- and r-band light curves of all four sources, with peaks that exceed the local noise estimates in the PSDs but whose global significances are reduced after accounting for trials. The detected QPO candidates exhibit local confidence levels above $3\sigma$, even exceeding $4\sigma$. However, to account for the potential increase in false positive rates caused by multiple trials across the frequency range, we calculated the global significance based on the ``look-elsewhere'' effect. As shown in Table \ref{tab3}, the global significances of J 1503.5+4759 and J 1652.7+4024 are 99.18\% and $\sim$98\%, respectively. For J 1503.5+4759, the global $p$-values for the g- and r- bands are only 0.008 ($<$ 0.05). J 1652.7+4024 has $p_{g}$ in the g- and r-bands equal to 0.024 and 0.017, respectively. Hence, we reject the null hypothesis for J 1503.5+4759 and J 1652.7+4024 and propose that both sources could be QPO candidates. They indicate a relatively high level of confidence in the signal. Further verification is still required. The lower global significance may indicate that the detected QPO signals is more likely due to random fluctuations or background noise rather than a true physical signal. Alternatively, it could result from issues such as sampling in the light curves. For such sources, longer-term and higher-quality observational data are required to further validate the authenticity of the periodic signal.

There is currently no conclusive theoretical explanation for QPOs. For instance, magnetic reconnection events within equidistant magnetic islands inside the jet and the helical motion of plasma blobs along the jet may lead to transient QPOs. When QPOs occur on timescales of weeks to months, they may result from emission originating in localized regions exhibiting kink instabilities \citep{2021MNRAS.506.1862}. Non-ballistic hydrodynamical Newtonian-driven jet precession may account for variations with periods longer than one year \citep{2004ApJ...615L...5}. For year-scale QPOs, the Keplerian orbital motion of SMBBHs or gravitational torque induced by a companion is often employed as an explanation \citep{2020ApJ...891..120, 2021ApJ...923..7}. Principally, the origins of QPOs can be attributed to three major scenarios.

First, in an SMBBHs system, the misalignment between the orbital plane of the secondary SMBH and the accretion disk of the primary SMBH, or the warping of the accretion disk, can induce precession of the jet from the primary BH. This precession leads to a periodic variation in the angle between the jet and the observer's line of sight on the orbital timescale. The observed timescale can be interpreted as the Keplerian period of the secondary BH orbiting the central BH, given by $T=2\pi d^{3/2} (GM) ^{-1/2}$, where $M$ is the SMBH mass, and $d$ is the semimajor axis. We have found that the SMBH masses of J 1221.3+3010 and J 1503.5+4759 are $\sim 10^{9}$ and $\sim 10^{7.97}~M_{\odot}$, respectively \citep{2023MNRAS..526..4079C}. Assuming the SMBH masses of J 0923.5+4125 and J 1652.7+4024 are typical AGN mass, $10^9~M_{\odot}$, the separations between two BHs in the four blazars are estimated to be $3.3\times 10^{-3}$, $7.0\times 10^{-3}$, $4.7\times 10^{-3}$, and $1.2\times 10^{-3}$ pc, respectively. Such sub-parsec separations represent relatively stable configurations in SMBBHs evolution. If the quasi-periodic behavior originates from such systems, these objects may be potential targets for future gravitational wave detectors.

Second, QPOs could arise from instabilities within the accretion disk. Assuming circular orbits of a bright hotspot, for BH masses of the four sources, the radius of the Keplerian orbit timescale can be estimated to be a few tens of Schwarzschild radii \citep{2020ApJ...891..120}. Therefore, in this scenario, instabilities in the disk may also produce QPOs, but longer light curves and multi-wavelength observations are required to confirm this. Additionally, the disk in a binary system may also be affected by hydrodynamical instabilities. If the orbital plane of the companion is inclined relative to the accretion disk, the induced torque causes disk precession on a characteristic timescale \citep{2000AA...360..57R}. Thick accretion disks may experience global perturbations, leading to $p$-mode oscillations \citep{2003ApJ...593..980L} with fundamental frequencies similar to observed QPOs \citep{2013MNRAS..434..3487A}.

Third, QPOs might be linked to instabilities within the jets themselves. This explanation associates quasi-periodic flux modulations with the relativistic motion of emission regions along helical paths in magnetized jets \citep{1992AA...255...59C}. As these regions move, relativistic effects cause periodic flux modulation due to changes in the viewing angle. Bhatta noted that in non-ballistic motion with inclination $i\sim 1/\Gamma_{b}$ and bulk Lorentz factor $\Gamma_{b} \sim 5$–15, the observed period can be shortened by a factor $\Gamma_{b}^{2}/(1+z)$ \citep{2021ApJ...923..7}. While red noise in blazar light curves complicates QPO identification, the consistent overlapping signals in g- and r-bands strengthen the case for physical QPOs.

The helical motion of plasma blobs within the jet can produce periodic flux modulations. Jet bending may reflect intrinsic helical structure driven by magnetohydrodynamic instabilities \citep{1999ApJ...524..650}. As blobs traverse helical paths, their viewing angle changes periodically, and Doppler boosting enhances the observed flux when alignment is favorable. Adopting typical parameters $\phi=2^\circ$, $\psi=2^\circ$, and $\Gamma=20$ \citep{2022MNRAS.510.3641}, the observed periods for J 1503.5+4759 and J 1652.7+4024 ($\sim38.5$ and $\sim48$ days) correspond to rest-frame periodicities of $\sim42.74$ yr and $\sim52.61$ yr, respectively. The travel distances per period are then $\sim13.08$ pc and $\sim16.10$ pc, indicating parsec-scale propagation over a few cycles. Observational evidence for such structures exists \citep{2017ApJ...846...98,2024A&A...684A202}, and optical polarization studies support helical magnetic geometries \citep{2008Natur..452..966M}.

\section{Conclusions}\label{Conclusion}

After applying selection criteria of S/N > 15 and more than 100 valid data points, we identified 10 blazars from the ZTF. A periodicity search among these sources revealed that four blazars (J 0923.5+4125, J 1221.3+3010, J 1503.5+4759, and J 1652.7+4024) potentially show tentative evidence for QPOs-like modulation on timescales ranging from months to years. The detected QPOs show local confidence levels above 3$\sigma$, with some exceeding 4$\sigma$. After accounting for the ``look-elsewhere'' effect, the global significance of the QPOs in J 1503.5+4759 and J 1652.7+4024 remains at the 2.7$\sigma$ and 2.5$\sigma$ levels, respectively. We employed several time series analysis methods to analyze the optical light curves of the four sources over the past five years, with the aim of studying their variability characteristics. Similar to the $\gamma$-ray band, the optical emission from the sample blazars also shows pronounced flux modulation. The optical emission from these four sources exhibits linear correlations in both g- and r-bands under two different binning schemes (10 data points and 30 days per bin), indicating that higher flux states are often more variable. Moreover, the flux distribution in the g- and r-bands tends towards a log-normal PDF. Similar to $\gamma$-ray variability, the processes driving optical variability are likely multiplicative nonlinear processes that operate across diverse timescales and flux states. We analyzed these sample sources using three methods: the LSP, the WWZ and ARIMA. Especially, the results indicate that the four blazars may displays tentative QPO behavior in the g-band also exhibit the same behavior in the r-band. In this work, we propose that the possible QPOs of four sources can be interpreted as the Keplerian periods of the secondary BH orbiting the central BH, or as instabilities intrinsic to the accretion disk. Notably, the periods of J 1652.7+4024 and J 1503.5+4759 may originate from plasma blobs spiraling within the curved jet, potentially indicating parsec-scale jets. These specific mechanism require multi-band and more light curve investigation. For these blazars, we will keep investigating the potential for multi-band QPOs and correlations in flux distributions. Additional research will focus on time delays and the scales of AGN structural components.

\section*{Acknowledgements}
This work was supported by the National Key R\&D Program of China under grants 2023YFA1607902 and 2023YFA1607903, the National Natural Science Foundation of China under grants 12173031, 12494572, 12221003, 12494572, 12322303, and 12063007, the Natural Science Foundation of Fujian Province of China (No. 2022J06002), the Fundamental Research Funds for the Central Universities (No. 20720240152), the Fund of National Key Laboratory of Plasma Physics (No. 6142A04240201), and the China Manned Space Program with grant No. CMS-CSST-2025-A13. Based on the observations obtained with the Samuel Oschin 48-inch Telescope at the Palomar Observatory as part of the Zwicky Transient Facility project. ZTF is supported by the National Science Foundation under Grant No. AST-1440341 and a collaboration including Caltech, IPAC, the Weizmann Institute for Science, the Oskar Klein Center at Stockholm University, the University of Maryland, the University of Washington, Deutsches Elektronen-Synchrotron and Humboldt University, Los Alamos National Laboratories, the TANGO Consortium of Taiwan, the University of Wisconsin at Milwaukee, and Lawrence Berkeley National Laboratories. Operations are conducted by COO, IPAC, and UW.

\bibliographystyle{unsrt}

\end{document}